\documentclass[amsmath,amssymb,aps,pre,english,showkeys,onecolumn,superscriptaddress]{revtex4-2}
\usepackage{graphicx}% Include figure files
\usepackage{dcolumn}% Align table columns on decimal point
\usepackage{bm}% bold math
\usepackage{hyperref}% add hypertext capabilities
\hypersetup{
    unicode=false,          % non-Latin characters in AcrobatÕs bookmarks
    pdftoolbar=false,        % show AcrobatÕs toolbar?
    pdfmenubar=true,        % show AcrobatÕs menu?
    pdffitwindow=false,     % window fit to page when opened
    pdfstartview={FitH},    % fits the width of the page to the window
    pdfkeywords={MBL} {AGP} {Mass-deformed SYK}, % list of keywords
    pdfnewwindow=true,      % links in new window
    colorlinks=true,       % false: boxed links; true: colored links
    linkcolor=red,          % color of internal links (change box color with linkbordercolor)
    citecolor=blue,        % color of links to bibliography
    filecolor=magenta,      % color of file links
    urlcolor=blue           % color of external links
}
\usepackage{mathtools}
\usepackage{xcolor}
\usepackage[page]{appendix}

\newcommand{\reffig}[1]{\mbox{Fig.~\ref{#1}}}
\newcommand{\refsec}[1]{\mbox{Sec.~\ref{#1}}}
\newcommand{\appsec}[1]{\mbox{Appendix~\ref{#1}}}

\newcommand{\be}{\begin{equation}}
\newcommand{\ee}{\end{equation}}
\newcommand{\bal}{\begin{align}}
\newcommand{\eal}{\end{align}}
\newcommand{\ba}{\begin{eqnarray}}
\newcommand{\ea}{\end{eqnarray}}
\renewcommand{\Re}{\mathrm{Re}}

\newcommand{\T}{${\mathcal T}\,$}
\newcommand{\Ti}{${\mathcal T}$}
\def\II{\hbox{$1\hskip -1.2pt\vrule depth 0pt height 1.6ex width 0.7pt\vrule depth 0pt height 0.3pt width 0.12em$}}

\newcommand{\pcsadd}{Center for Theoretical Physics of Complex Systems, Institute for Basic Science (IBS), Daejeon, Korea, 34126}
\newcommand{\ustadd}{Basic Science Program, Korea University of Science and Technology (UST), Daejeon 34113, Republic of Korea}
\newcommand{\skcgadd}{P.G. Department of Physics, S.K.C.G. (Auto.) College, Paralakhemundi, Odisha, India, 761200}
\newcommand{\ictpadd}{ICTP South American Institute for Fundamental Research \\
Instituto de F\'{i}sica Te\'{o}rica, UNESP - Univ. Estadual Paulista \\
Rua Dr. Bento Teobaldo Ferraz 271, 01140-070, S\~{a}o Paulo, SP, Brazil}

\begin{document}

\title{The Rosenzweig-Porter model revisited for the three Wigner-Dyson symmetry classes}

\author{Tilen \v{C}ade\v{z}}
    \email{tilencadez@ibs.re.kr}
    \address{\pcsadd}

\author{Dillip Kumar Nandy}
    \email{nandy.pawan@gmail.com}
    \address{\skcgadd}
    \address{\pcsadd}

\author{Dario Rosa}
    \email{dario\_rosa@ictp-saifr.org}
    \address{\ictpadd}
    \address{\pcsadd}
    \address{\ustadd}

\author{Alexei Andreanov}
    \email{aalexei@ibs.re.kr}
    \address{\pcsadd}
    \address{\ustadd}

\author{Barbara Dietz}
    \email{barbara@ibs.re.kr}
    \address{\pcsadd}
    \address{\ustadd}

%\author{Tilen Čadež, Dillip Nandy, Dario Rosa, Alexei Andreanov, Barbara Dietz}

\date{\today}

\begin{abstract}
	Interest in the Rosenzweig-Porter model, a parameter-dependent random-matrix model which interpolates between Poisson and Wigner-Dyson (WD) statistics describing the fluctuation properties of the eigenstates of typical quantum systems with regular and chaotic classical dynamics, respectively, has come up again in recent years in the field of many-body quantum chaos. The reason is that the model exhibits parameter ranges in which the eigenvectors are Anderson-localized, non-ergodic (fractal) and ergodic extended, respectively. The central question is how these phases and their transitions can be distinguished through properties of the eigenvalues and eigenvectors. We present numerical results for \emph{all} symmetry classes of Dyson’s threefold way. We analyzed the fluctuation properties in the eigenvalue spectra, and compared them with existing and new analytical results. Based on these results we propose characteristics of the short- and long-range correlations as measures to explore the transition from Poisson to WD statistics. Furthermore, we performed in-depth studies of the properties of the eigenvectors in terms of the fractal dimensions, the Kullback-Leibler (KL) divergences and the fidelity susceptibility. The ergodic and Anderson transitions take place at the same parameter values and a finite size scaling analysis of the KL divergences at the transitions yields the same critical exponents for all three WD classes, thus  indicating  superuniversality of these transitions.
\end{abstract}	
\bigskip

%\submitto{\NJP}
\maketitle

\section{Introduction}

Random matrix theory (RMT)~\cite{Mehta2004} has been successful in the description of the fluctuation properties in the energy spectra of atomic nuclei~\cite{Porter1965,Brody1981,Haq1982,Guhr1989,Guhr1998,Weidenmueller2009,Mitchell2010,Gomez2011} and, within the field of quantum chaos, of those of quantum systems with a chaotic classical counterpart.
The objective of quantum chaos is to identify signatures of classical chaos in the properties of quantum systems.
However, nuclear many-body systems do not have an obvious classical analogue, even though their spectra exhibit features that are similar to those of quantum systems with integrable, chaotic or mixed integrable-chaotic dynamics~\cite{Guhr2009}.
It was demonstrated in Refs.~\onlinecite{Enders2000,Enders2004} that integrability may be associated with collective excitations, i.e. collective motion of the nucleons, whereas chaoticity corresponds to complex motion.
In fact RMT, was introduced by Wigner to describe the spectral properties of nuclei~\cite{Wigner1951,Wigner1955,Wigner1957,Porter1965,Brody1981,Guhr1989,Weidenmueller2009,Dietz2017,Dietz2018}.
In Refs.~\onlinecite{Berry1977b,Casati1980,Bohigas1984} a link between the spectral properties of quantum systems with a chaotic dynamics and random Hermitian matrices with Gaussian-distributed matrix elements was proposed.
This idea was pursued and led to the Bohigas-Giannoni-Schmit (BGS) conjecture~\cite{Bohigas1984} which states that the spectral properties of typical quantum systems, that belong to either the orthogonal ($\beta =1$) universality class, which applies to integer spin systems with preserved time-reversal (\T) invariance, to the unitary one ($\beta =2$), when \Ti-invariance is violated, or to the symplectic one ($\beta =4$) for half-integer spin systems with preserved \Ti-invariance, agree with those of random matrices from the corresponding Wigner-Dyson (WD) ensembles.
These comprise the Gaussian orthogonal ensemble (GOE), the Gaussian unitary ensemble (GUE), and the Gaussian symplectic ensemble (GSE), respectively~\cite{Mehta2004,Haake2018}.
On the other hand, Berry and Tabor demonstrated, based on the Einstein-Brillouin-Keller quantization~\cite{Berry1977a}, that the fluctuation properties in the eigenvalue sequences of typical integrable systems ($\beta =0$) exhibit Poissonian statistics.
The BGS conjecture was confirmed for single-particle systems theoretically for all three universality classes~\cite{Scharf1988,LesHouches1989,StoeckmannBuch2000,Haake2018} and also experimentally, e.g., with flat, cylindrical microwave resonators~\cite{Stoeckmann1990,Sridhar1991,Graef1992,Deus1995} simulating quantum billiards and microwave networks simulating quantum graphs~\cite{Hul2004,Allgaier2014,Bialous2016}.
It also applies to quantum systems with chaotic classical dynamics and partially violated \Ti-invariance~\cite{Pandey1991,Lenz1992,Bohigas1995,Guhr1996}.
These are described by a RMT model interpolating between the GOE and the GUE for complete \Ti-invariance violation.
Such systems were investigated theoretically in Refs.~\onlinecite{French1985,Pluhar1995,Mitchell2010,Assmann2016,Pluhar1995} and experimentally in microwave billiards~\cite{So1995,Stoffregen1995,Wu1998,Dietz2019b}. 

We report in this work on the analysis of the properties of a random matrix model, the Rosenzweig-Porter model (RP), which is a paradigmatic model for the description of universal properties of typical quantum systems,
whose classical counterpart undergoes a transition from integrable to chaotic dynamics, leading to a transition from Poisson to Wigner-Dyson statistics of their spectral properties and a transition from localized to extended for their eigenvectors.
The RP model was introduced in 1960 to describe phenomena like level repulsion or partial level clustering, exhibited by the energy levels that were obtained from experimental atomic spectra~\cite{rosenzweig1960repulsion}.
Depending on a parameter $\lambda$ it interpolates between random diagonal matrices and random matrices from either of the WD ensembles, denoted $\hat H_0$ and $\hat H^{\beta}$, respectively,
\be
    \label{RPH}
    \hat H^{0\to\beta}(\lambda) =\hat H_0+\Gamma_N\lambda\hat H^{\beta}\, ,\, \beta=1,2,4.
\ee
Here, we choose Gaussian distributed random entries for $\hat H_0$, and $\Gamma_N$ denotes a $N$-dependent scaling parameter which depends on the dimension $N$ of $\hat H^{0\to\beta}(\lambda)$ and ensures that the spectral properties of the unfolded eigenvalues only depend on $\lambda$~\cite{Pandey1981,Pandey1995,Guhr1997,Kunz1998,Guhr1998}, as explained below.
Upon increasing $\lambda$, the relative strength of off-diagonal matrix elements with respect to diagonal ones increases, and the spectral properties experience a transition from Poisson to WD statistics, while the eigenvectors undergo a transition from localized to extended ergodic phase.
Recently, the transition from Poisson to GUE was studied experimentally with a microwave billiard~\cite{Zhang2023b}.

It was shown in Ref.~\onlinecite{kravtsov2015random} that a suitable re-parametrization of $\lambda$ in terms of a power law of the matrix dimension uncovers an additional, intermediate phase, consisting of extended non-ergodic eigenstates that exhibit fractal dimensions, referred to as generalized Rosenzweig-Porter (gRP) model in the following.
Like in the original model~\eqref{RPH} the random matrices $\hat H^{\rm gRP}(\gamma)$ of the gRP model are Gaussian distributed, however the variances of the off-diagonal elements are modified by multiplication with an $N$-dependent prefactor, 
\ba
    \label{eq:HgRP}
    && H^{\rm gRP}_{nm}(\gamma)=H_{nn}\delta_{nm}+\frac{1}{N^{\gamma/2}}H_{nm}(1-\delta_{nm})\\ 
    \label{eq:gRP_variances}
    && \sigma_d^2 = \langle H_{nn}^2 \rangle = \frac{1}{\beta N},\, \sigma_{off}^2 = \left\langle\left(H^{(\xi)}_{nm}\right)^2\right\rangle = \frac{1}{2\beta N},\, \xi=0,\dots,\beta -1,\,\beta=1,2,4,
\ea
where $N,\, \sigma_d^2,\, \sigma_{off}^2$ denote the dimension of $\hat H$ and the diagonal and off-diagonal variances, respectively.
The parameter $\xi$ counts the number of independent components of the off-diagonal matrix elements of $\hat H$ and \(\gamma\) determines the phase diagram.
For $\beta=1$ $\hat H^{\rm gRP}(\gamma)$ is real symmetric, for $\beta =2$ it is complex Hermitian and for $\beta =4$ it is quaternion real and can be written in the quaternion representation.
Assuming that $\hat H^{\beta =4}$ is $2N$-dimensional, it is given in terms of an $N\times N$ matrix whose matrix elements are $2\times 2$ quaternion matrices of the form,
\be
    \hat h_{mn}=h^{(0)}_{mn}\II_2 +\boldsymbol{h}_{mn}\cdot\boldsymbol{\tau},\, n,m=1,\dots ,N.
    \label{quaternion}
\ee
Here, $\II_2$ is the 2-dimensional unit matrix, and $\boldsymbol{\tau}=-i\boldsymbol{\sigma}$ with the components of $\boldsymbol{\sigma}$, $\hat\sigma_i,\, i= 1,2,3$, denoting the three Pauli matrices.
Time-reversal invariance implies that the matrices $\hat h_{nm}$ are quaternion real, $h^{(\mu)}_{mn}=h^{(\mu)\ast}_{mn},\mu=0,\dots ,3$, and Hermiticity yields $h^{(0)}_{mn}=h^{(0)}_{nm}$, $\boldsymbol{h}_{mn}=-\boldsymbol{h}_{nm}$, and thus $\hat h_{nn}=h^{(0)}_{nn}\II_2$.
The eigenvalues of quaternion real matrices are Kramers degenerated so that the number of eigenvalues is reduced to one half of the dimension.

For \(0\leq \gamma < 1\) the properties of the eigenstates of the model Hamiltonian~\eqref{eq:HgRP} coincide with those of random matrices from the WD ensemble~\cite{Brezin1996,kravtsov2015random,bogomolny2018eigenfunction} with corresponding value $\beta$. At \(\gamma = \gamma_E = 1\) an {\emph{ergodic}} phase transition occurs.
Furthermore, it was shown in Refs.~\onlinecite{Pandey1995,Brezin1996,Kunz1998,Soosten2018} that for $\gamma >2$ all eigenstates are localized and at \(\gamma_A = 2\) the Anderson localization transition takes place.
In the parameter range \( \gamma_E < \gamma < \gamma_A\), referred to as non-ergodic extended phase, the eigenstates are delocalized and exhibit single-fractal properties~\cite{kravtsov2015random}.
Due to the existence of this intermediate phase and its connection to the phenomenon of many-body localization \cite{abanin2019colloquium, alet2018many-body, sierant2024manybody}, the gRP model has gained considerable attention in the last few years~\cite{landon2019fixed, facoetti2016from, truong2016eigenvectors, monthus2017multifractality, vonsoosten2019non, bogomolny2018eigenfunction, pino2019from, detomasi2019survival, berkovits2020super, skvortsov2022sensitivity, Hopjan2023, Cadez2023}. 
Several extensions and modifications of the model were also studied, among which are the circular~\cite{Buijsman2022,Buijsman2024} and non-Hermitian~\cite{DeTomasi2022} RP model and the effects of fat-tailed distributions of off-diagonal elements~\cite{khaymovich2020fragile, Kutlin2024} or fractal disorder~\cite{Sarkar2023}.

The values of $\gamma$, $\gamma_E$ and $\gamma_A$, where the transitions to ergodic and localization take place, may be estimated using the rule of thumb criteria for ergodicity and localization for dense matrices outlined in Refs.~\onlinecite{bogomolny2018power, khaymovich2020fragile}.
They are based on the following sums over moments of $\vert H_{nm}\vert$, 
\begin{align}
    \label{eq:S_q}
    S_q(N) = \frac{1}{N \, A^q} \sum_{n,m = 1}^N \langle |H_{nm}|^q \rangle,
\end{align}
with \(A = \sqrt{\langle |H_{nn}|^2 \rangle}\), \(q = 1,2\) and \(N\) denoting the dimension of the matrix. 
The criteria are:

\begin{itemize}
    \item The property \(\lim_{N \to \infty} S_1(N) < \infty\) implies that the eigenstates are localized and spectral statistics agrees with Poisson statistics (Anderson localization criterion).
    \item The property \(\lim_{N \to \infty} S_2(N) \to \infty\) implies that the eigenstates are ergodically distributed over the whole available space and spectral statistics agrees with WD statistics (ergodicity criterion).
    \item The property \(\lim_{N \to \infty} S_1(N) \to \infty\) and \(\lim_{N \to \infty} S_2(N) < \infty\) indicates -- but does not necessarily imply -- that the states are extended but non-ergodic. 
    \item Furthermore, a sufficient condition for complete ergodicity is fulfilled~\cite{khaymovich2020fragile} if 
	\(\lim_{N \to \infty} S_1(N) \to \infty\), \(\lim_{N \to \infty} S_2(N) \to \infty\) and \(\lim_{N \to \infty} \bar{S}(N) \to \infty\), with 
    \begin{align}
        \label{eq:barS}
        \bar{S}(N) = \frac{\bigl(\sum_{m} \langle |H_{nm}|^2 \rangle_{\mathrm{t}}\bigr)^2}{S_2(N)},
    \end{align}
    and $\langle .\rangle_t$ denoting the typical value which is given by \( \langle |H_{nm}|^2 \rangle_{\mathrm{t}} = \exp{\bigl[ \langle \ln(|H_{nm}|^2) \rangle\bigr] } \). 
\end{itemize}
For the gRP model we obtain with the definition of the variance $\sigma^2$ of the Gaussian distributions in~\eqref{eq:gRP_variances} \(S_1(N) =  \sqrt{2/\pi} \bigl[ 1 + 1/\sqrt{2} (N-1) N^{-\gamma/2}\bigr]\), \( S_2(N) = 1 + 1/2 (N-1) N^{-\gamma}\) and \( \langle |H_{nm}|^2 \rangle_{\mathrm{t}}= \sigma^2/[2\exp(\gamma_{EM}]\), with $\gamma_{EM}$ denoting the Euler-Mascheroni constant.
This yields in the limit \(N \to \infty\) the values $\gamma_E=1$ and $\gamma_A=2$ for the transition from ergodic to non-ergodic and non-ergodic to localized phase, respectively.

We extend the numerical studies of the Rosenzweig-Porter model to the transition from Poisson to GSE and present results for the spectral properties of the three WD ensembles in~\refsec{RMTSpectr}.
They have been studied thoroughly for the transition from Poisson to GOE, e.g. in Ref.~\onlinecite{berkovits2020super} and for that from Poisson to GUE even analytical results exist~\cite{Lenz1992,Guhr1996a,Guhr1997,Kunz1998,Frahm1998}. These have been tested experimentally and checked with low-dimensional random matrices in Ref.~\onlinecite{Zhang2023b}.
In this work we test them with high-dimensional matrices and derive a Wigner-surmise like analytical expression for the ratio distribution for that transition; see~\appsec{RatioAnalyt} for details.
The ratio distribution has the advantage that it is dimensionless, so that unfolding of the eigenvalues of $\hat H^{0\to\beta}(\gamma)$ to a uniform spectral density is not required.
The average ratios are commonly used as a measure for the size of chaoticity and ergodicity~\cite{pino2019from}.
We propose the position of the maximum of the nearest-neighbor spacing distribution, the position of the minimum of the form factor, the deviation of the number variance from that of the corresponding WD ensemble, and the slope of the power spectrum in the asymptotic limit as measures for the transition from WD behavior to Poisson statistics.
For the long-range correlations the associated measures reveal deviations from the corresponding WD statistics when increasing $\gamma$ beyond $\gamma=1$ and saturate at the value corresponding to Poisson statistics for $\gamma\gtrsim 2.1$.
Yet, to identify the region $1 <\gamma <2$ as a fractal, i.e., non-ergodic phase, the properties of the associated eigenvectors need to be analyzed.
An in-depth analysis of commonly used statistical measures is presented for all three WD ensembles in~\refsec{RMTWFs}.

\section{Analysis of spectral properties and comparison with available analytical results for the RP model}
\label{RMTSpectr}

To study the properties of the eigenvalues $E_\mu$ and the eigenvector components $\psi_{\mu}(i),\, i=1,\dots,N$ of the gRP Hamiltonian~\eqref{eq:HgRP}, we solve the eigenvalue problem, \(\hat H |\psi_{\mu}\rangle = E_{\mu} |\psi_{\mu} \rangle,\, \mu=1,\dots,N\), where the eigenvectors are given in terms of the computational basis \( |\psi_{\mu}\rangle = \sum_i \psi_{\mu}(i) | i \rangle\), by full exact diagonalization for numerous values of $\gamma\in [0.0,3.5]$ for $\beta =1,2,4$ and $N=2^n$ with $n$ varying from 9 to 16.
In the following subsections we present our results on fluctuation properties in the eigenvalue spectrum, localization properties of the eigenvectors, and the fidelity susceptibility which depends on a combination of the $E_\mu$ and the $\psi_{\mu}(i)$. 

The RP model~\eqref{RPH} has been succesfully employed to describe the spectral properties of typical quantum systems whose classical counterpart exhibits a dynamics between regular and chaotic behavior. Furthermore, analytical results for the spectral properties were obtained based on the Hamiltonian~\eqref{RPH}. However, as mentioned above, for the description of the transition from ergodicity to localization of the eigenvectors of typical quantum systems the parametrization~\eqref{eq:HgRP} is more suitable. Accordingly, the functional dependence of the parameter $\lambda$ on $\gamma$ needs to be determined. In~\eqref{eq:HgRP} $\hat H$ and in~\eqref{RPH} $\hat H^{\beta}$ are drawn from the Gaussian ensemble denoted by $\beta$ with variance $\sigma^2=\frac{(1+\delta_{nm})}{2\beta N}$. Thus the $N$-dimensional gRP Hamiltonian~\eqref{eq:HgRP} can be cast to its original form~\eqref{RPH} by replacing $\hat H_0$ by the difference of two diagonal matrices, $\hat H^{(1)}-\frac{1}{N^{\gamma/2}}\hat H^{(2)}$, whose matrix elements are Gaussian distributed with variances $\left\langle\left(H^{(1)}_{nn}\right)^2\right\rangle = \left\langle\left(H^{(2)}_{nn}\right)^2\right\rangle=\frac{1}{\beta N}$, yielding a random diagonal matrix with Gaussian distributed matrix elements  with variance $\left\langle\left(H_{0nn}\right)^2\right\rangle = \frac{1}{\beta N}\left(1+\frac{1}{N^{\gamma}}\right)$.
Accordingly we may expect that $\lambda\propto\frac{1}{N^{\gamma/2}}$. 

For the study of spectral properties we performed for all three values of $\beta$ numerical simulations for random matrices from the gRP Hamiltonian~\eqref{eq:HgRP} with $N=2^{16}$. For the computation of truly universal spectral properties, the dependence of the mean spectral density on energy and the dimension $N$ of the gRP Hamiltonian, which is a system-specific property, needs to be removed. This is achieved by unfolding the sorted eigenvalues $E_1\leq E_2\leq\cdots \leq E_N$ to mean spacing unity. For the WD ensembles the eigenvalues can be unfolded with the integrated semicircle law,
\begin{equation}
    N(E)=\frac{N}{\pi}\left[E\sqrt{1-E^2}+\frac{\pi}{2}+\arcsin\left(E\right)\right],
\end{equation}
with $N(E)$ denoting the number of eigenvalues below $E$. For non-zero values of $\gamma$ we unfolded to average spacing unity with a combination of the semicircle law and a polynomial of 5th order. Here, we excluded the lowest and largest 7500 eigenvalues, corresponding to $\approx 23 \%$ of the total number $N$. Unless otherwise stated, the spectral properties were analyzed for random-matrix ensembles consisting of 5-10 realizations. For the calculation of the statistical measures we performed in addition to the ensemble average a spectral average over the whole spectrum.

Similarly, to achieve truly universal spectral properties of the RP Hamiltonian $\hat H^{0\to\beta}(\lambda)$ in~\eqref{RPH}, the parameter $\lambda$ needs to be unfolded. This is achieved by an appropriate choice of the scaling factor $\Gamma_N$. One possibility is to set $\Gamma_N=1/D_N$, with $D_N$ denoting the mean spacing of the eigenvalues of $\hat H^{\beta}$~\cite{Pandey1981,Pandey1995,Guhr1997,Kunz1998,Guhr1998}. Another possibility considered in~\cite{Frahm1998} is to replace $\Gamma_N\lambda$ by the ratio of the variance of the diagonal matrix elements of $\hat H^{\beta}$ and the average spacing of the entries of $\hat H_0$. Note that analytical results are obtained for the spectral measures in the limit $N\to\infty$ which can be performed only after proper unfolding the eigenvalues and $\lambda$~\cite{Mehta2004,Guhr1998}. In the present case the variances of the diagonal elements of $\hat H_0$ and $\hat H^{\beta}$ are chosen similar for $N\gg 1$, so that an unfolding of $\gamma$ is not needed. Indeed, the spectral properties do not change when increasing the dimension from $N=2^{16}$ to $N=100000$. Thus, to determine the functional dependence of $\lambda$ on $\gamma$, we may evaluate available analytical expressions for short- and long-range correlation functions of the eigenvalues of the model Hamiltonian~\eqref{RPH} and fit to them those obtained for fixed $N$ from random-matrix simulations with the Hamiltonian~\eqref{eq:HgRP}. For $\lambda\to\infty$ the random matrix $\hat H^{0\to\beta}(\lambda)$ approaches the WD ensemble with corresponding $\beta$, however, its spectral properties already coincide with WD statistics for $\lambda\gtrsim 2.5$. 

We analyzed the nearest-neighbor spacing distribution $P(s)$, the distribution of ratios of consecutive eigenvalue spacings $P(r)$, and the two-point cluster function $Y_2(\epsilon^\prime,\epsilon^{\prime\prime})=1-R_2(\epsilon^\prime,\epsilon^{\prime\prime})$, with $R_2(\epsilon^\prime,\epsilon^{\prime\prime})$ denoting the spectral two-point correlation function, $R_2(\epsilon^\prime,\epsilon^{\prime\prime})=\langle\sum_{i\ne j}\delta(\epsilon^\prime-\epsilon_i)\delta(\epsilon^{\prime\prime}-\epsilon_j)\rangle$ for unfolded eigenvalues $\epsilon_i$ and $\epsilon_j$. Both correlation functions depend only on the distance $\vert\epsilon^\prime-\epsilon^{\prime\prime}\vert$, that is the length of the energy interval bordered by $\epsilon^\prime$ and $\epsilon^{\prime\prime}$. Furthermore, we computed the number variance $\Sigma^2 (L)=\langle (N(L)-\langle N(L)\rangle)^2\rangle$ with $N(L)$ denoting the number of unfolded eigenvalues in an energy interval of length $L$ and $\langle N(L)\rangle=L$, the spectral form factor $K(\tau)=1-b(\tau)$ with $b(\tau)=\int_{-\infty}^\infty Y_{2}(x)e^{-i2\pi x\tau}dx$, and the power spectrum, which is not commonly used, yet is sensitive to small perturbations. It is defined as
\be
  \label{eq:power}  s\left(\tau=\frac{l}{N}\right)=\left\langle\left\vert\frac{1}{\sqrt{N}}\sum_{q=0}^{N-1} \delta_q\exp\left(-2\pi i\frac{l}{N} q\right)\right\vert^2\right\rangle,\, l=1,\dots N
\ee
with $\delta_q=\langle\epsilon_{q+i}-\epsilon_i\rangle_i-q$ and $\epsilon_j$ denoting the unfolded eigenvalues~\cite{Relano2002,Faleiro2004}.

\subsection{Short-range correlations}

\begin{figure}
    \centering
    \includegraphics[width=0.49\columnwidth]{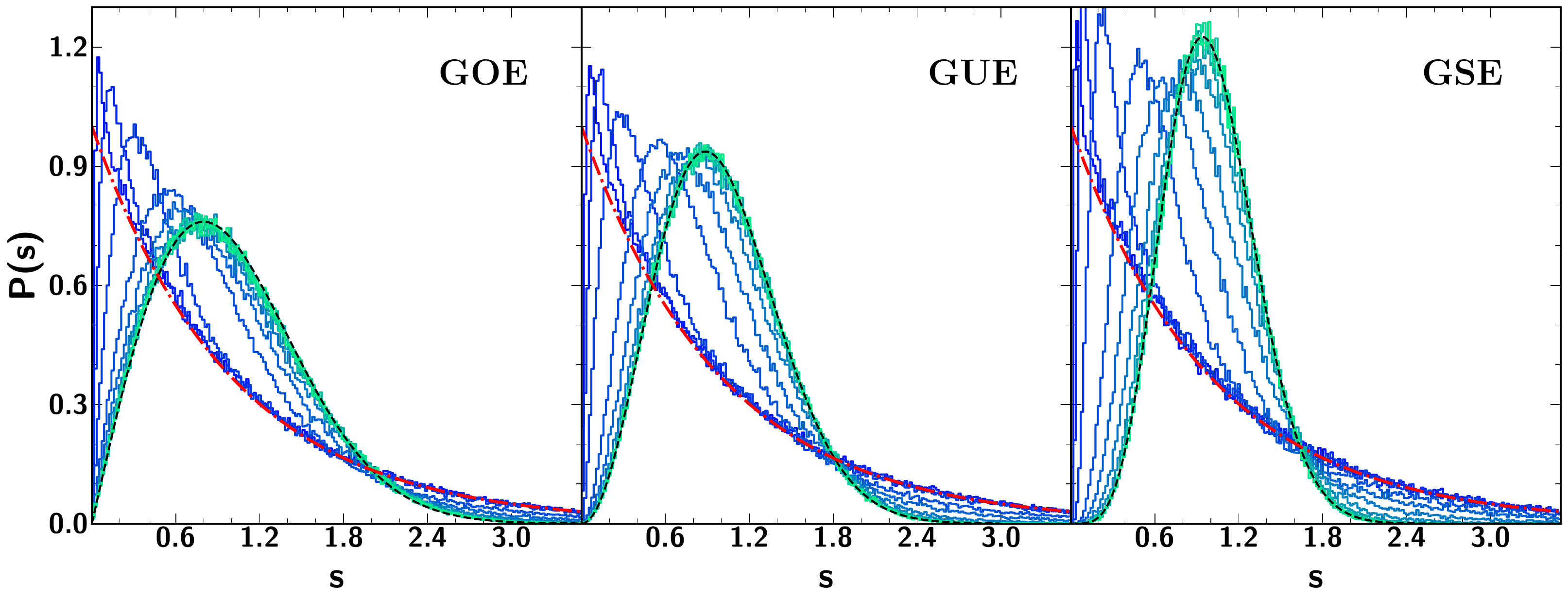}
    \includegraphics[width=0.49\columnwidth]{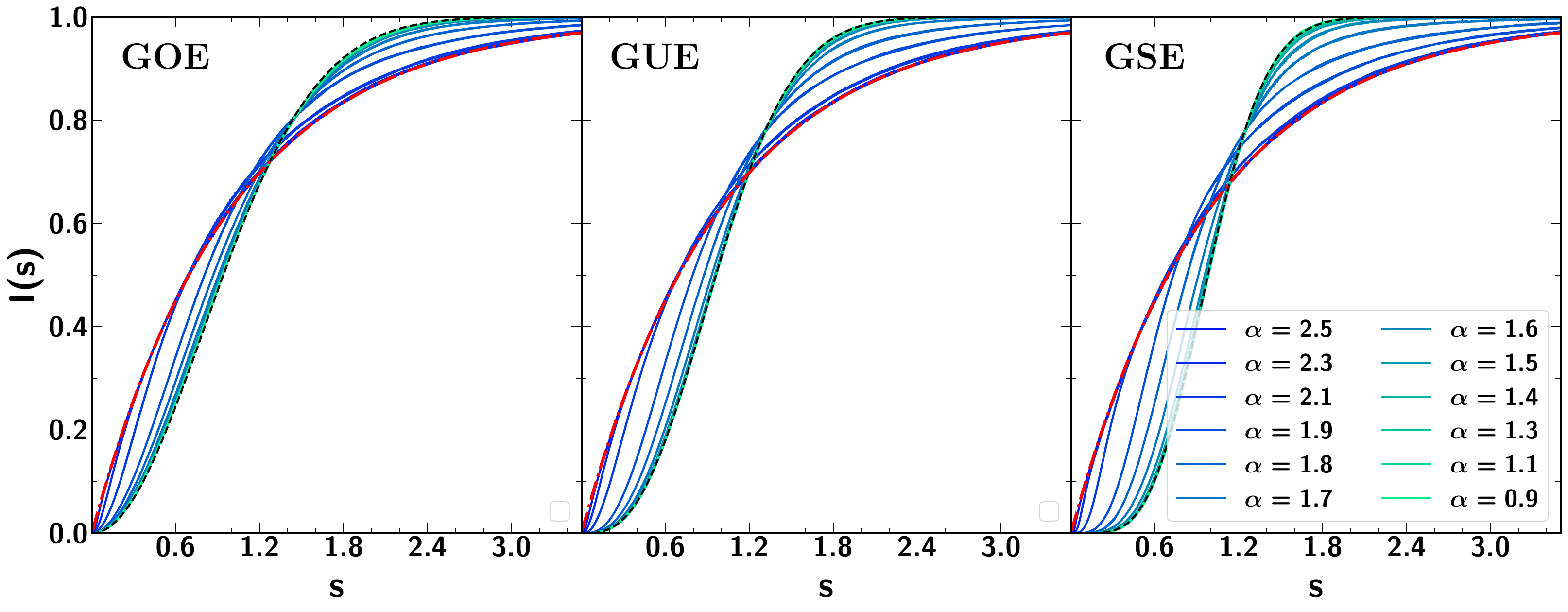}
    \caption{
            Left: Nearest-neighbor spacing distributions $P(s)$ obtained from random-matrix simulations for the gRP model for $\beta =1$ (left), $\beta =2$ (middle) and $\beta =4$ (right) for the values of $\gamma$ given in the insets of the right panel. With increasing $\gamma$ $P(s)$ experiences a transition from WD to Poisson statistics. Actually, for $\gamma=0.9$ the curves lie on top of the WD result (black dashed line), and for $\gamma=2.5$ it is close to the result for Poissonian random numbers (red dash-dotted line).
        Right: Same as left for the cumulative nearest-neighbor spacing distribution.
    }
\label{fig:NNDs}
\end{figure}

In~\reffig{fig:NNDs} we show the nearest-neighbor spacing distributions and its cumulative distribution for several values of $\gamma$. 
Wigner-surmise like approximations have been derived for the nearest-neighbor spacing distributions based on $2\times 2$-dimensional random matrices of the form~\eqref{RPH} for $\beta=1,2$ and based on $2\times 2$ matrices in quaternion basis of the form~\eqref{RPH} for $\beta =4$.
For $\beta =1$ it has been derived in Refs.~\onlinecite{Lenz1991,Haake1991,Kota2014},
\be
    P_{0\to 1}(s)=\frac{su_{\lambda }^2}{\lambda}\exp{\left( -\frac{u_\lambda ^2s^2}
    {4\lambda^2}\right)}\int_0^\infty d\xi e^{-\xi^2-2\xi\lambda}
    I_0\left(\frac{s\xi u_\lambda}{\lambda}\right)
    \label{PSGOE}
\ee
where $u_\lambda=\sqrt{\pi}U(-\frac{1}{2},0,\lambda^2)$, with $U(a,c,x)$ denoting the Tricomi function,
\be
    U(-\frac{1}{2},0,\lambda^2)=\frac{1}{\sqrt{\pi}}e^\frac{\lambda^2}{2}
    \int_0^\frac{\pi}{2}d\Theta \cos \left(\frac{\lambda^2}{2}\tan\Theta - \Theta\right)
\ee
and $I_0(x)$ is the modified Bessel function
\be
    I_0(x)=\frac{1}{\pi}\int_0^\pi d\Theta\cosh (x\cos\Theta)\ .
\ee
This distribution interpolates between Poisson for $\lambda =0$ and the Wigner surmise for $\beta =1$ in the limit $\lambda\to\infty$.
Note that the limit $\lambda\to 0$ has to be taken such that $\lambda <s$.
For finite values of $\lambda$ the distribution decays exponentially for $s\gg \langle s\rangle$ with $\langle s\rangle$ denoting the average spacing, that is, the distribution~\eqref{PSGOE} exhibits the characteristic features of intermediate statistics~\cite{Bogomolny1999}.
In Ref.~\cite{Lenz1992} a Wigner-surmise like expression was derived for $\beta =2$ based on the RP model~\eqref{RPH} with $N=2$, 
\ba
    \label{PSGUE}
%        &&P_{0\to 2}(s)=Cs^2e^{-D^2s^2}\int_0^\infty dxe^{-\frac{x^2}{4\lambda^2}-x}\frac{\sinh z}{z},\,\\
%        &&\nonumber D(\lambda)=\frac{1}{\sqrt{\pi}}+\frac{1}{2\lambda}e^{\lambda^2}[1-\Phi(\lambda)]-\frac{\lambda}{2}{\rm Ei}\left(\lambda^2\right)\\
%        &&\nonumber +\frac{2\lambda^2}{\sqrt{\pi}}{_2{F}_2}\left(\frac{1}{2},1;\frac{3}{2},\frac{3}{2};\lambda^2\right),\\
%        &&\nonumber C(\lambda)=\frac{4D^3(\lambda)}{\sqrt{\pi}},\, z=\frac{xDs}{\lambda},
    &&P_{0\to 2}(s)=Cs^2e^{-D^2s^2}\int_0^\infty dxe^{-\frac{x^2}{4\lambda^2}-x}\frac{\sinh z}{z},\,\\
    &&\nonumber D(\lambda)=\frac{1}{\sqrt{\pi}}+\frac{1}{2\lambda}e^{\lambda^2}[1-\Phi(\lambda)]-\frac{\lambda}{2}{\rm Ei}\left(\lambda^2\right),\,
    +\frac{2\lambda^2}{\sqrt{\pi}}{_2{F}_2}\left(\frac{1}{2},1;\frac{3}{2},\frac{3}{2};\lambda^2\right),\,
    C(\lambda)=\frac{4D^3(\lambda)}{\sqrt{\pi}},\, z=\frac{xDs}{\lambda},
\ea
where $\Phi(x)$ denotes the error function, Ei$(x)$ the exponential integral, and ${_2{F}_2}(\alpha_1,\alpha_2;\beta_1,\beta_2;x)$ the generalized hypergeometric error function~\cite{Abramowitz2013,Gradshteyn2007}.
This distribution was rederived in Ref.~\cite{Kota1999} and is quoted in Ref.~\cite{Schierenberg2012}, where also a Wigner-surmise like expression was derived for $\beta =4$ based on the RP model~\eqref{RPH} with $N=4$, corresponding to a $2\times 2$ dimensional matrix in the quaternion basis,
\ba
    \label{PSGSE}
	&&P_{0\to 4}(s)=D\frac{\lambda}{2\sqrt{\pi}}s_0e^{-\frac{s_0^2}{4}}\int_0^\infty dxe^{-x^2-2\lambda x}\frac{z\cosh(z)-\sinh z}{x^3},\\
%        &&\nonumber D(\lambda)=\frac{\lambda}{4\sqrt{\pi}}\int_0^\infty dx\frac{(4x^3+2x)e^{-x^2}+\sqrt{\pi}(4x^4+4x^2-1)\rm{erf}(x)}{x^3},\\
%        &&\nonumber s_0=2Ds,\, z=s_0x,
	&&\nonumber D(\lambda)=\frac{\lambda}{\sqrt{\pi}}\int_0^\infty dx\frac{(4x^3+2x)e^{-x^2}+\sqrt{\pi}(4x^4+4x^2-1)\Phi(x)}{x^3}e^{-2\lambda x},\,
    s_0=2Ds,\, z=s_0x.
\ea

\begin{figure}
    \centering
    \includegraphics[width=0.4\columnwidth]{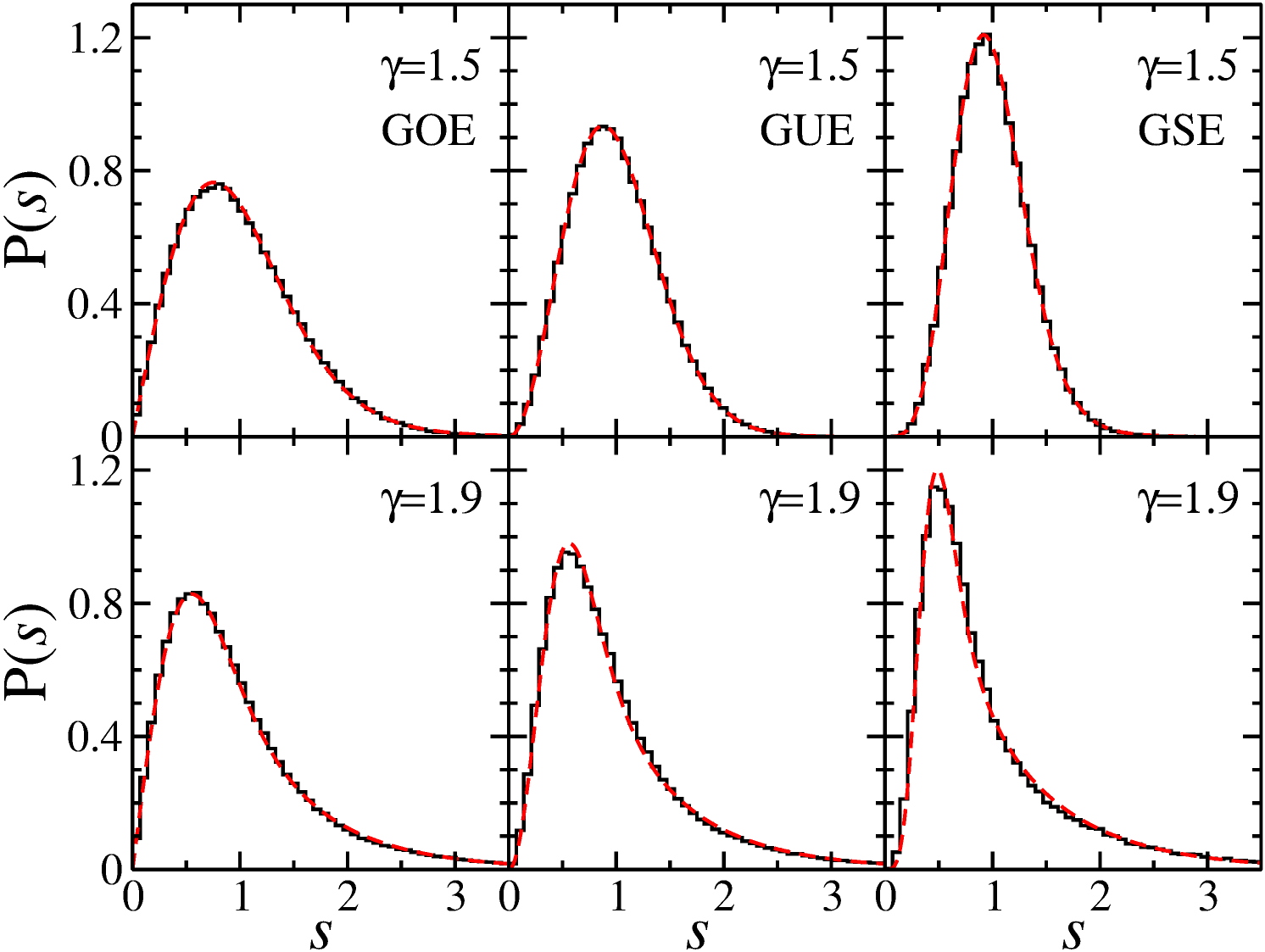}
    \includegraphics[width=0.4\columnwidth]{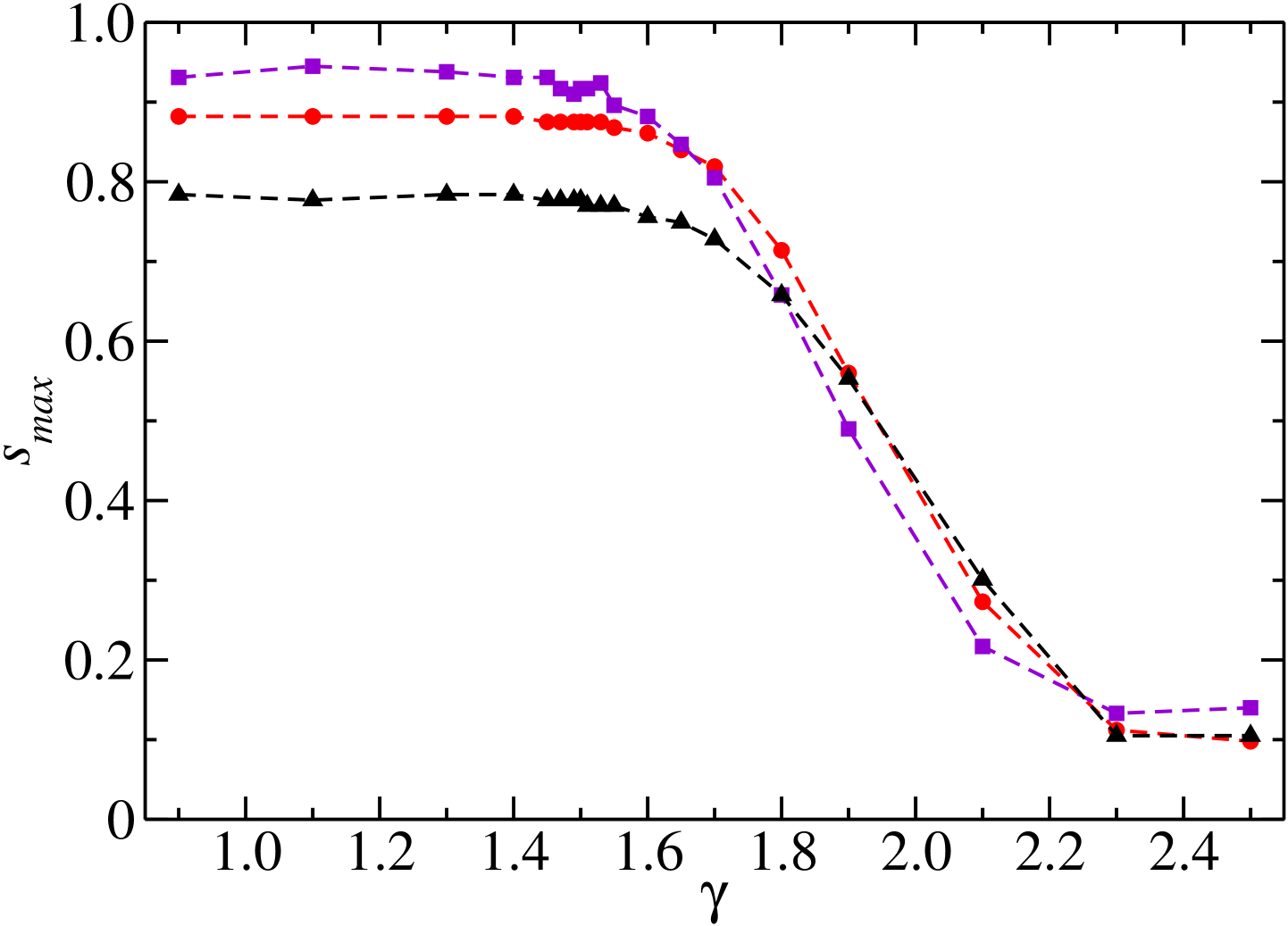}
	\caption{
	    Left: Examples for the nearest-neighbor spacing distributions obtained from random-matrix simulations for the gRP model (black histograms) for the transition from Poisson to GOE (left column), GUE (middle column) and GSE (right column), respectively.
        They are compared to the curves $P_{0\to\beta}(s)$ given in Eqs.~(\ref{PSGOE}-\ref{PSGSE}) best fitting them (red dashed line). 
	    Right: Position of the maximum of the best-fitting $P_{0\to\beta}(s)$ as function of the transition parameter $\gamma$ for $\beta =1$ (black triangle), $\beta=2$ (red circles) and $\beta =4$ (purple squares).
    }
   \label{fig:NNDGSE}
\end{figure}

\begin{figure}
    \centering
    \includegraphics[width=0.4\columnwidth]{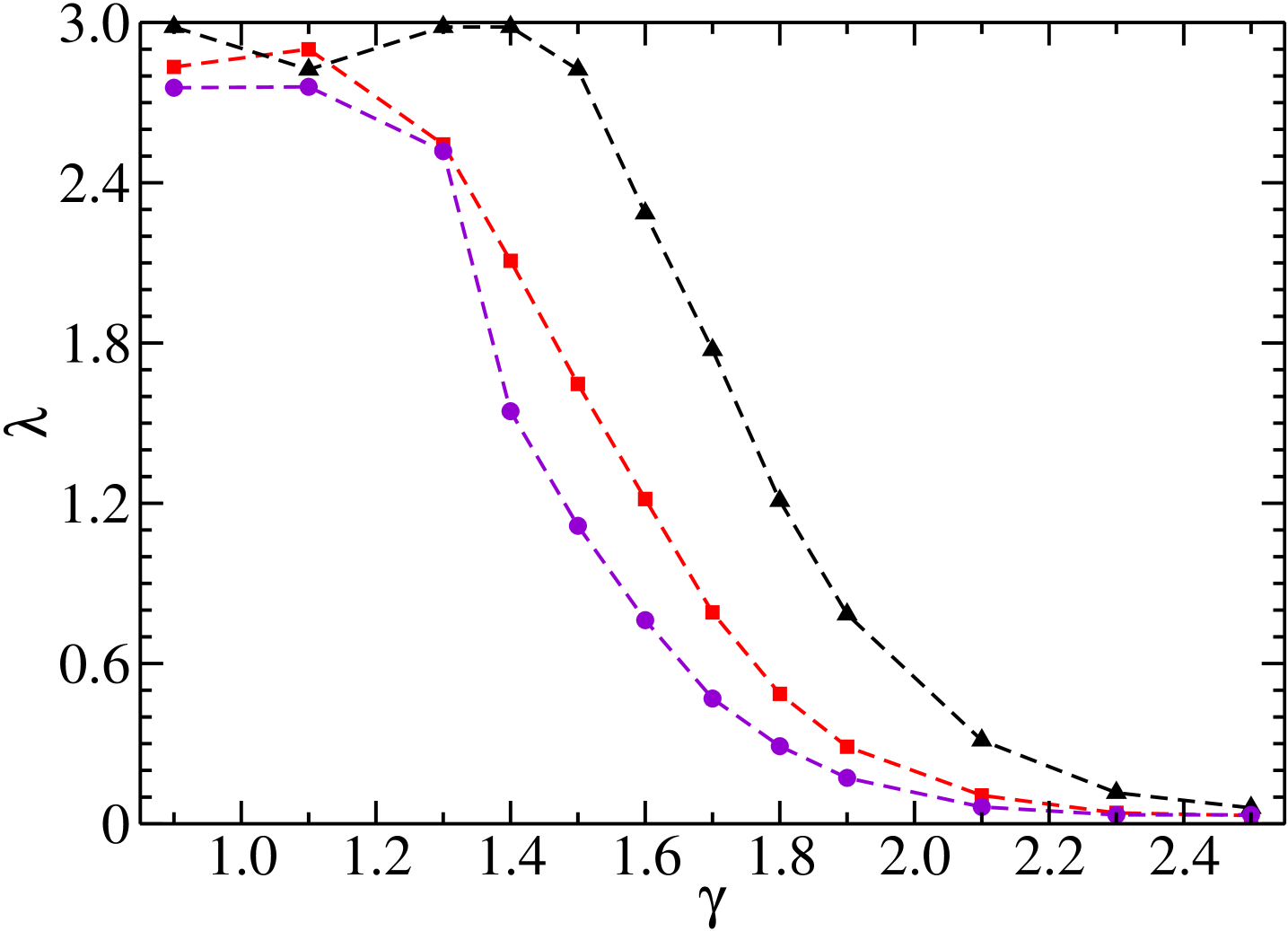}
    \includegraphics[width=0.4\columnwidth]{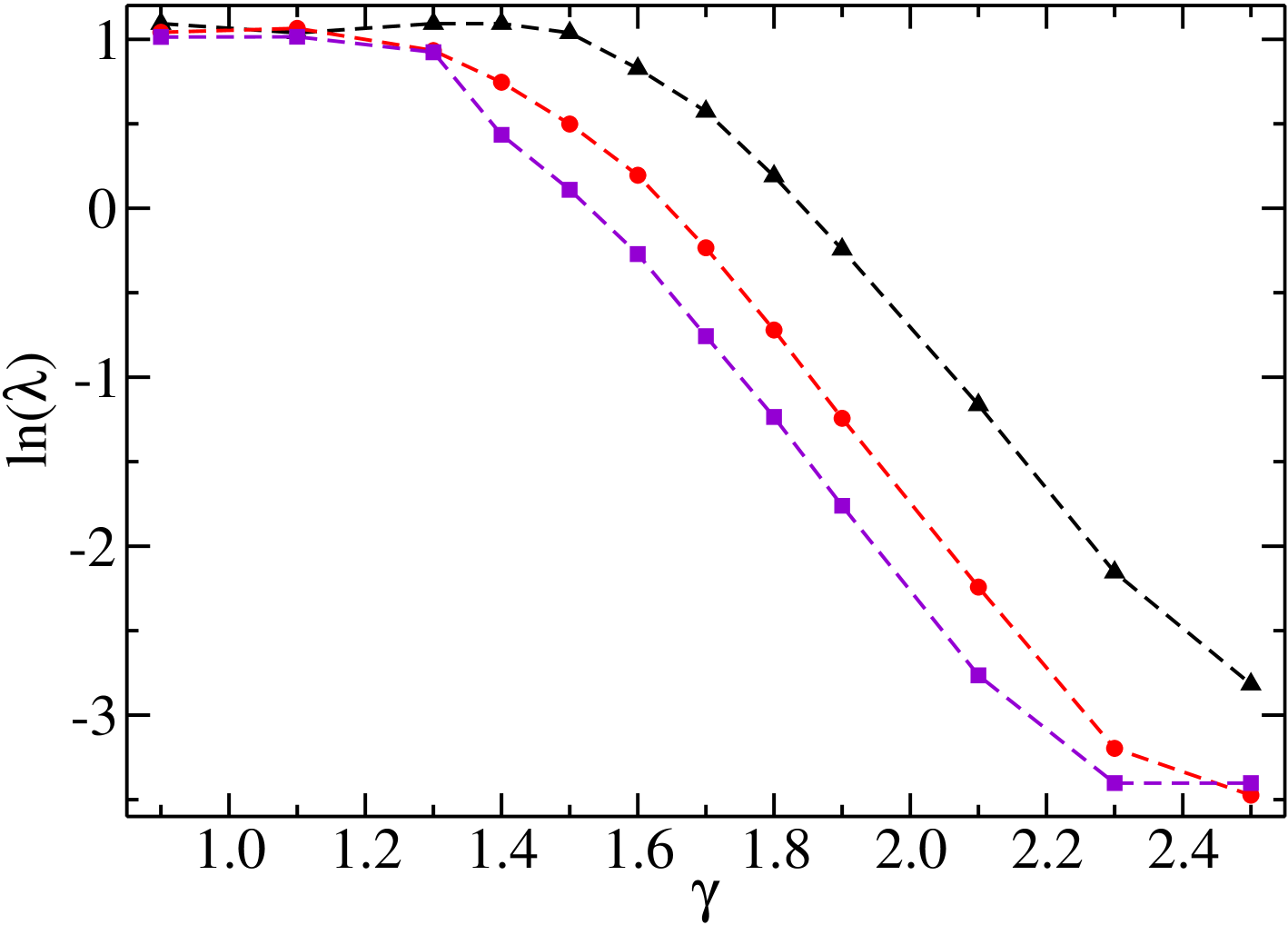}
    \caption{
	    Left: Values of $\lambda$ obtained from the fit of $P_{0\to\beta}(s)$ to the numerical results for $\beta =1$ (black triangles), $\beta=2$ (red circles) and $\beta=4$ (purple squares) as function of $\gamma$. Right: Same as left for the natural logarithm of $\lambda$. 
    }
    \label{fig:LamGam}
\end{figure}

In Fig.~\ref{fig:NNDs} we show examples for the nearest-neighbor spacing distribution and the cumulative one for various values of $\gamma$.
In the left part of~\reffig{fig:NNDGSE} we compare the nearest-neighbor spacing distributions obtained from random-matrix simulations for the gRP Hamiltonian~\eqref{eq:HgRP} (black histograms) for two values of $\gamma$ for the transition from Poisson to WD statistics with $\beta =1$ (left column), $\beta =2$ (middle column) and $\beta =4$ (right column) with the distribution $P_{0\to\beta}(s)$ best fitting them. The agreement is as good as that of the exact nearest-neighbor spacing distributions of random matrices from the WD ensembles with the corresponding Wigner surmise~\cite{Dietz1990,Haake2018}. In the left part of~\reffig{fig:LamGam} we show the values of $\lambda$ resulting from the fit of the analytical curves $P_{0\to\beta}(s)$ given in~\eqref{PSGOE}-\eqref{PSGSE} to those obtained from the RMT simulations employing the gRP Hamiltonian~\eqref{eq:HgRP} as function of $\gamma$. Here, we restricted $\lambda$ to $0.03\leq\lambda\leq 3$. Note that the dimension of the gRP Hamiltonian is sufficiently large to discern the differences between the Wigner surmise, which is derived on the basis of $2\times 2$ matrices, and the exact nearest-neighbor spacing distribution of the associated random-matrix ensemble~\cite{Dietz1990,Dietz1991}. This explains the deviation of $\lambda$ from the largest considered value, $\lambda=3$. Furthermore, below $\gamma\simeq 1.4$, the curves resulting from the random-matrix simulations lie nearly on top of each other. The same holds for the Wigner-surmise $P_{0\to\beta}(s)$ in the corresponding range $2.5\lesssim \lambda\leq 3$, implicating that for $\gamma\lesssim 1.4$ the values of $\lambda$ corresponding to the best fitting analytical curve barely change. Deviations from WD behavior are observed above that value, implicating that the nearest-neighbor spacing distribution becomes sensitive to the modification of the WD Hamiltonian in~\eqref{eq:HgRP} only for $\gamma\gtrsim 1.4$. As visible in the logarithmic plot shown in the right part of ~\reffig{fig:LamGam}, $\lambda\propto N^{-B\gamma}$ for $1.6\leq\gamma\leq 2.1$. A linear regression yields $B\simeq 0.5$, as expected from the definitions of the RP and gRP models; see~\eqref{RPH} and~\eqref{eq:HgRP}.

These features confirm that the nearest-neighbor spacing distribution is not sensitive to small perturbations of the random matrices from the WD ensembles. The mean spacing cannot be used as a measure for the transition from WD behavior to Poisson statistics, since the eigenvalues are rescaled to mean spacing unity. However, we observe that the position of the maximum of the nearest-neighbor spacing distribution undergoes a transition from the value for the corresponding WD ensemble to zero when increasing $\gamma$. Therefore we use it as indicator for the transition from chaotic to regular dynamics. It is plotted as function of $\gamma$ in the right part of~\reffig{fig:NNDGSE} for the WD ensembles. A drastic change of the position is visible for all WD classes for $\gamma\gtrsim 1.45$ up to $\gamma\simeq 2$.        

We also analyzed the distribution of the ratios of consecutive spacings~\cite{Oganesyan2007,Atas2013} between nearest-neighbor eigenvalues, $r_j=\frac{E_{j+1}-E_{j}}{E_{j}-E_{j-1}}$  and of $r_j^{min}={\rm min}\left(r_j,\frac{1}{r_j}\right)$. Wigner-surmise like analytical expressions are available for all three WD ensembles~\cite{Atas2013,Atas2013a}. They are applicable to systems for which the spectral density does not exhibit singularities and have the advantage that no unfolding is required, since the ratios are dimensionless~\cite{Oganesyan2007,Atas2013,Atas2013a}. 

\begin{figure}
    \centering
    \includegraphics[width=0.5\columnwidth]{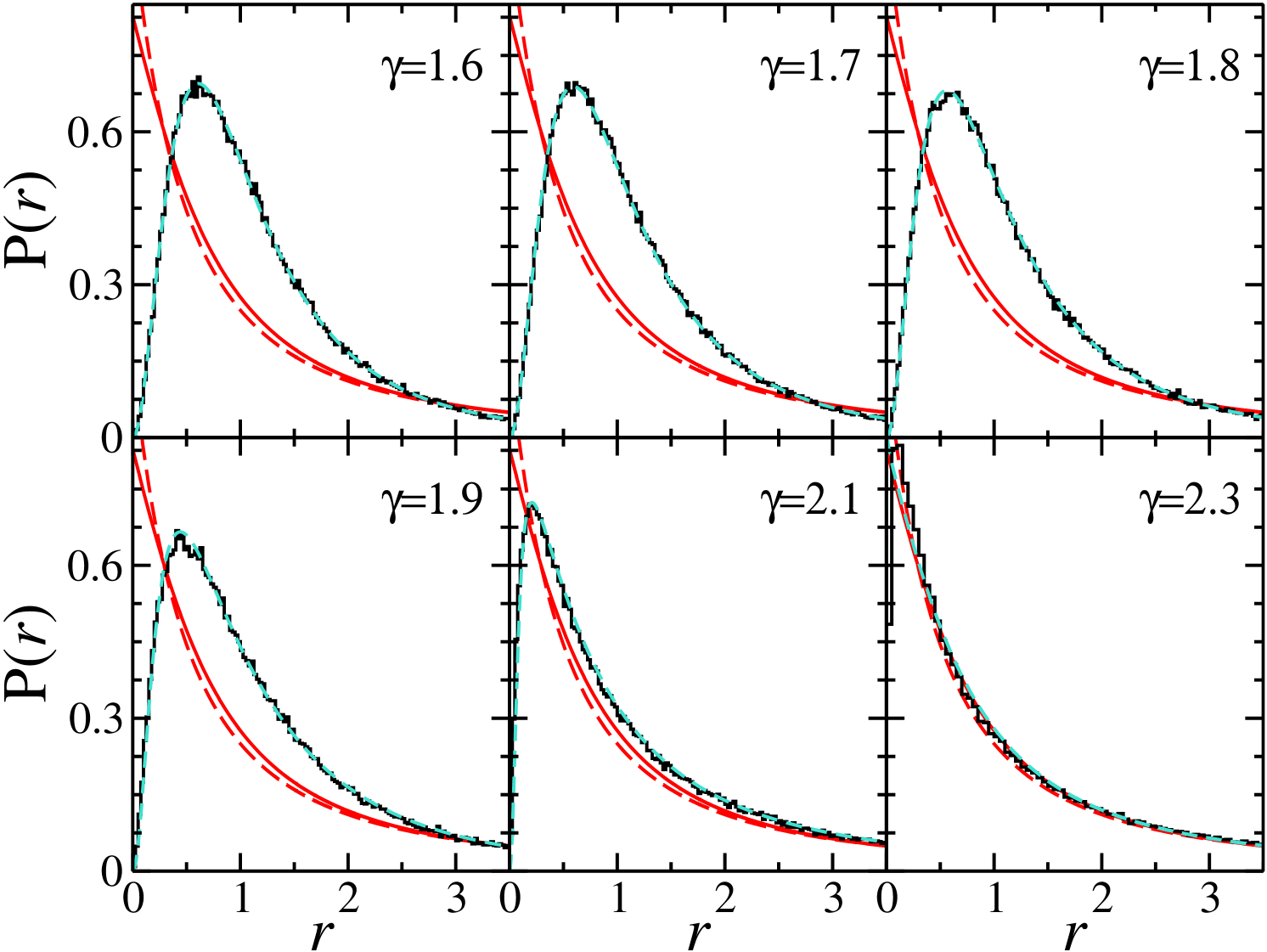}
	\caption{
	    Examples for ratio distributions obtained from random-matrix simulations for the gRP model (black histograms) for the transition from Poisson to GUE. They are compared to the corresponding Wigner-surmise like analytical result obtained from ~\eqref{PRAnalyticGUE0} (turquoise dashed line). 
	    The dashed red lines show the analytical result~\eqref{PRPoi} for Poissonian distributed random numbers~\cite{Atas2013}, the solid red lines exhibit the result for $3\times3$-dimensional diagonal matrices with Gaussian distributed entries given in~\eqref{PRAs1}. For $\gamma=1.6$ the curve is close to the Wigner-surmise like distribution of the GUE~\eqref{PRAs2}. For $\gamma =2.3$ the numerical result is closer to the distribution~\eqref{PRAs1} than to the curve for Poissonian random numbers~\eqref{PRPoi}  above $r\gtrsim 0.3$ whereas, similar to the nearest-neighbor spacing distribution, the ratio distribution of the gRP Hamiltonian deviates from all Wigner-surmise like analytical curves below that value.
    }
   \label{PRRMAnal}
\end{figure}

Based on the joint probability distribution~\eqref{RPHUE} of the eigenvalues of $\hat H^{0\to 2}(\lambda)$ for the transition from Poisson to GUE, we derive in~\appsec{RatioAnalyt} a Wigner-surmise like analytical expression for the ratio distribution, which is given in~\eqref{PRAnalyticGUE},
\ba
    \label{PRAnalyticGUE0}
    P^{0\to 2}(r)=&&\frac{r(r+1)}{R^3}\frac{1}{2\pi}\left\{\sqrt{3}\frac{\alpha^2}{\alpha^2+1}\frac{(3R-2)}{2R}\right.\\
    +&&(2+r)\alpha^2\int_{-\pi}^{\pi}\frac{d\varphi}{\sin\varphi}\left(-X_1+\frac{2}{X_1}+\frac{1}{3X_1^3}\right)\left[1-\frac{2}{\pi}\arctan(X_1)\right]\nonumber\\
    +&&(1+2r)\alpha^2\int_{-\pi}^{\pi}\frac{d\varphi}{\sin\varphi}\left(-X_2+\frac{2}{X_2}+\frac{1}{3X_2^3}\right)\left[1-\frac{2}{\pi}\arctan(X_2)\right]\nonumber\\
    +&&\frac{2}{\pi}\alpha^2\left.\int_{-\pi}^{\pi}\frac{d\varphi}{\sin\varphi}\left[\frac{2+r}{3}\frac{1}{X_1^2(1+X_1^2)}+\frac{1+2r}{3}\frac{1}{X_2^2(1+X_2^2)}\right]\right\}\nonumber
\ea
with $X_1,X_2$ defined in~\eqref{Xi} together with (\ref{Fi}) and (\ref{Ai}), $\alpha=\lambda$ and $R=\frac{2}{3}(1+r+r^2)$. We prove in~\appsec{RatioAnalyt}, that
\be
    P^{0\to 2}(r)\xrightarrow{\alpha\to 0}\frac{3\sqrt{3}}{2\pi(1+r+r^2)},
    \label{PRAs1}
\ee
which is the ratio distribution for the eigenvalues of a $3\times 3$-dimensional diagonal matrix with Gaussian distributed entries, and
\be
    P(r)\xrightarrow{\alpha\to\infty}\frac{81\sqrt{3}}{4\pi}\frac{\left[r(1+r)\right]^2}{(1+r+r^2)^4}\, ,
    \label{PRAs2}
\ee
which is the ratio distribution for the Wigner-surmise like analytical result for the GUE. Examples are shown in~\reffig{RatioAnal}. With increasing $\lambda$ indeed a transition between the limiting cases~\eqref{PRAs1} to~\eqref{PRAs2} takes place. In~\reffig{PRRMAnal} we compare for a few values of $\gamma$ the numerical results to the corresponding analytical ones. Similar to the Wigner-surmise like results for the nearest-neighbor spacing distributions, the numerical evaluation of~\eqref{PRAnalyticGUE0} becomes increasingly cumbersome with increasing $\gamma$, because in the limit $\lambda\to 0$ ($\gamma\to\infty$) the integrand turns into a $\delta$-function as outlined in~\appsec{RatioAnalyt} [see~\eqref{Deltafct}], reflecting the abrupt transition of the ratio distribution occurring when increasing $\lambda$ from zero to any small value in~\eqref{RPH}~\cite{Lenz1991,Lenz1992}.

\begin{figure}
    \centering
    \includegraphics[width=0.4\columnwidth]{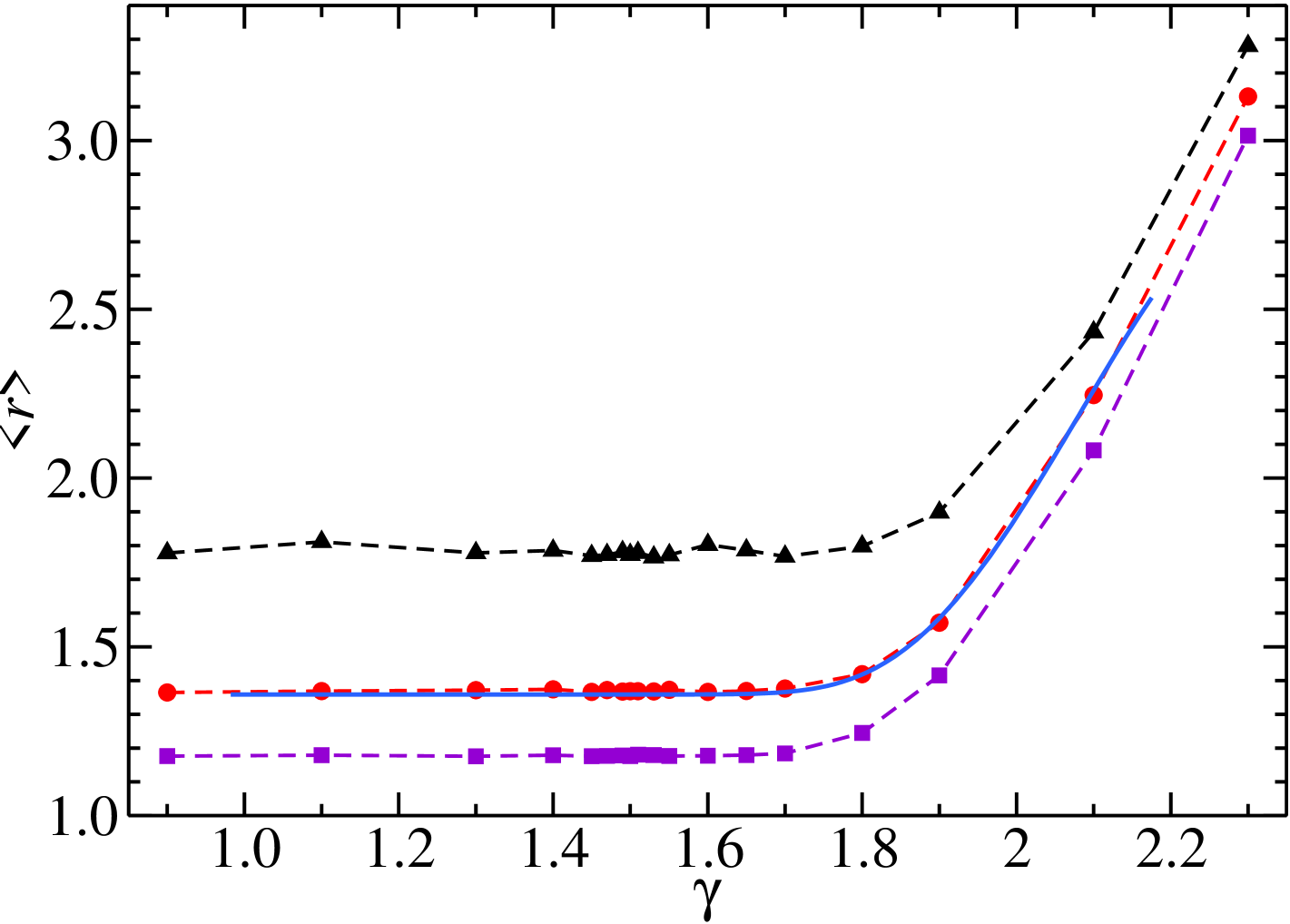}
    \includegraphics[width=0.4\columnwidth]{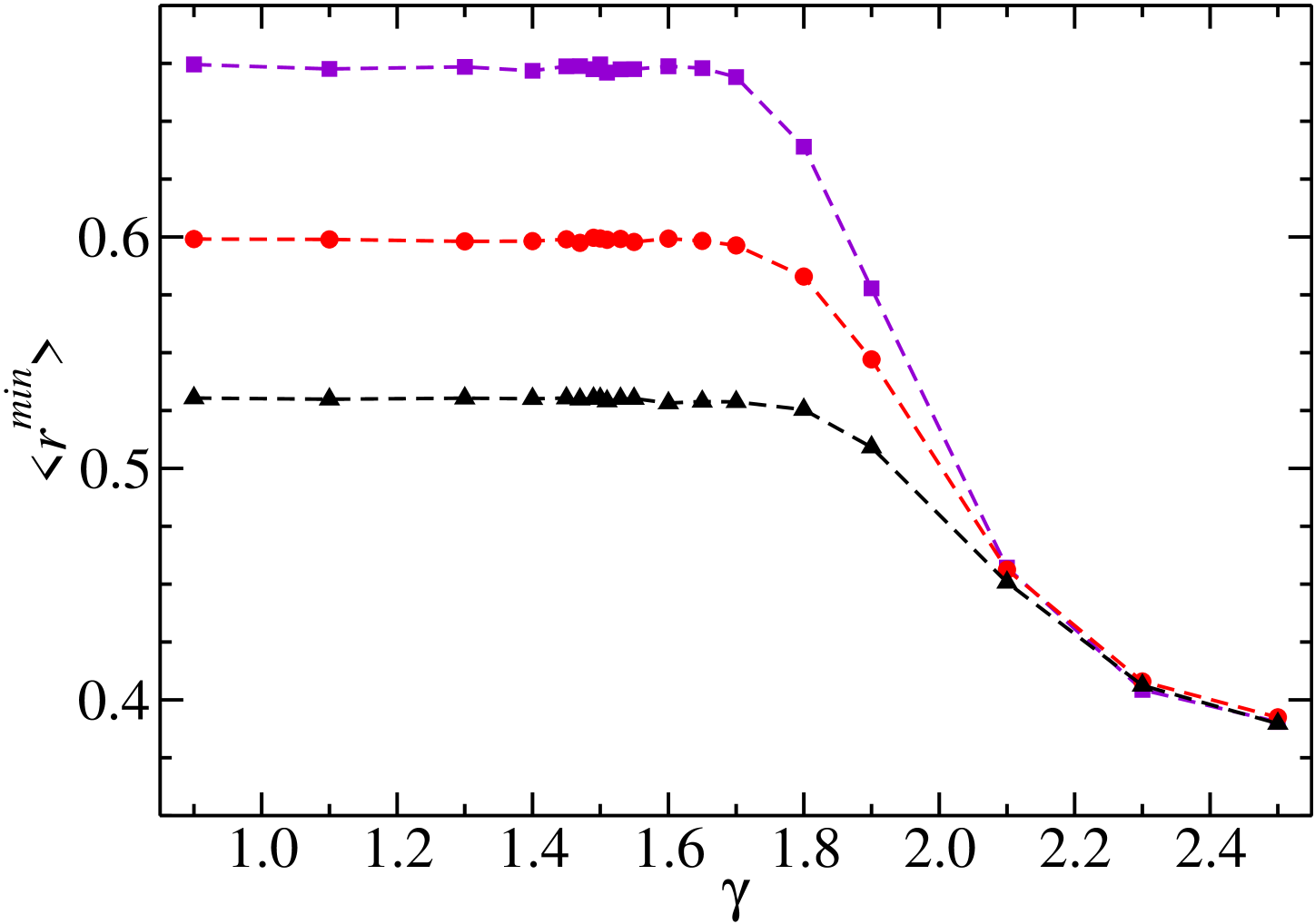}
    \caption{
	    Average values for $\langle r\rangle$ (left) and $\langle r^{min}\rangle$ (right) for the transitions from the WD ensembles to Poisson (black triangles: $\beta =1$, red circles: $\beta =2$, purple squares: $\beta =4$) as function of $\gamma$. 
        The blue solid line exhibits the corresponding analytical result.
        Note, that the ratio distribution~\eqref{PRAnalyticGUE0} becomes indistinguishable from the result~\eqref{PRAs1} (except for $\lambda < r$) for $\gamma\gtrsim 2.1$, so we don't show results above that value.
    }
   \label{fig:Ratios}
\end{figure}

\begin{figure}
    \centering
    \includegraphics[width=0.2\columnwidth]{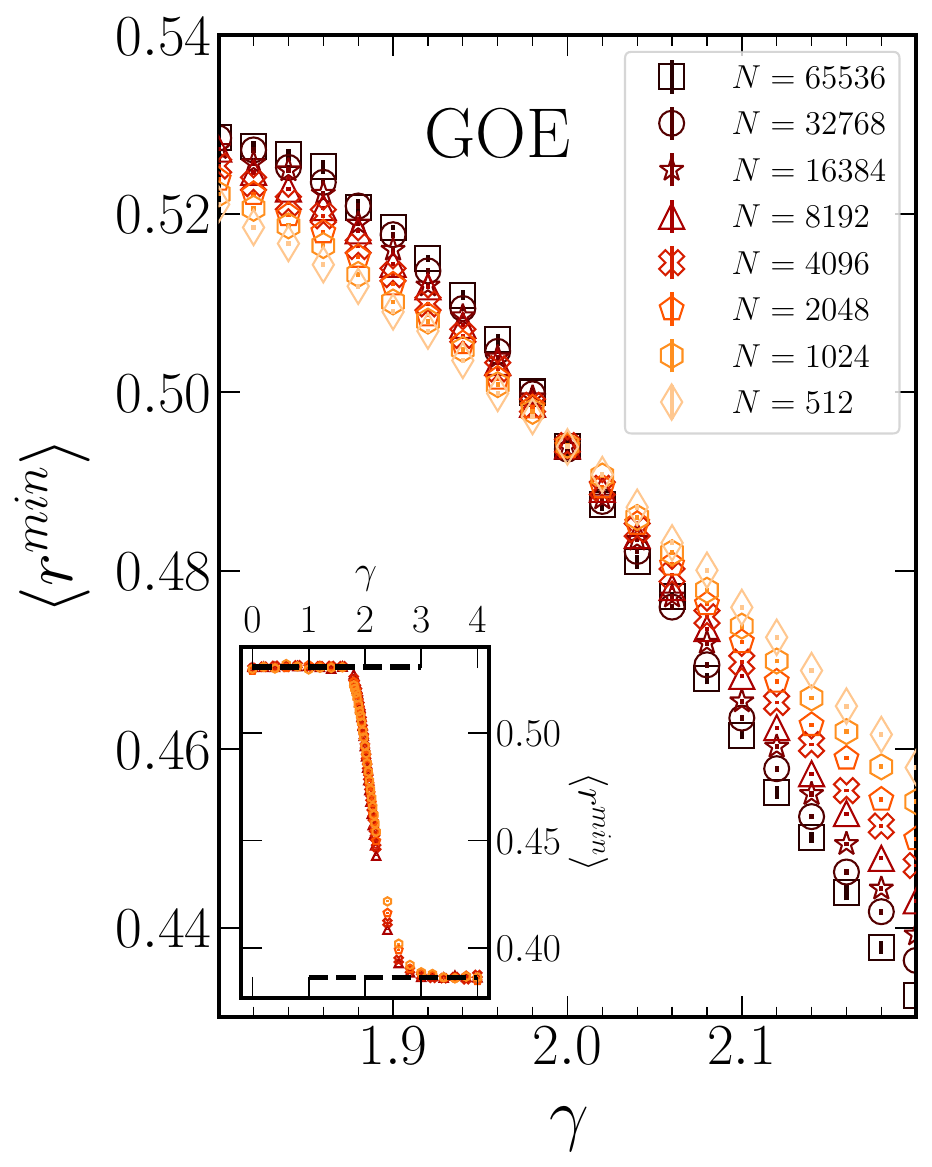}\includegraphics[width=0.2\columnwidth]{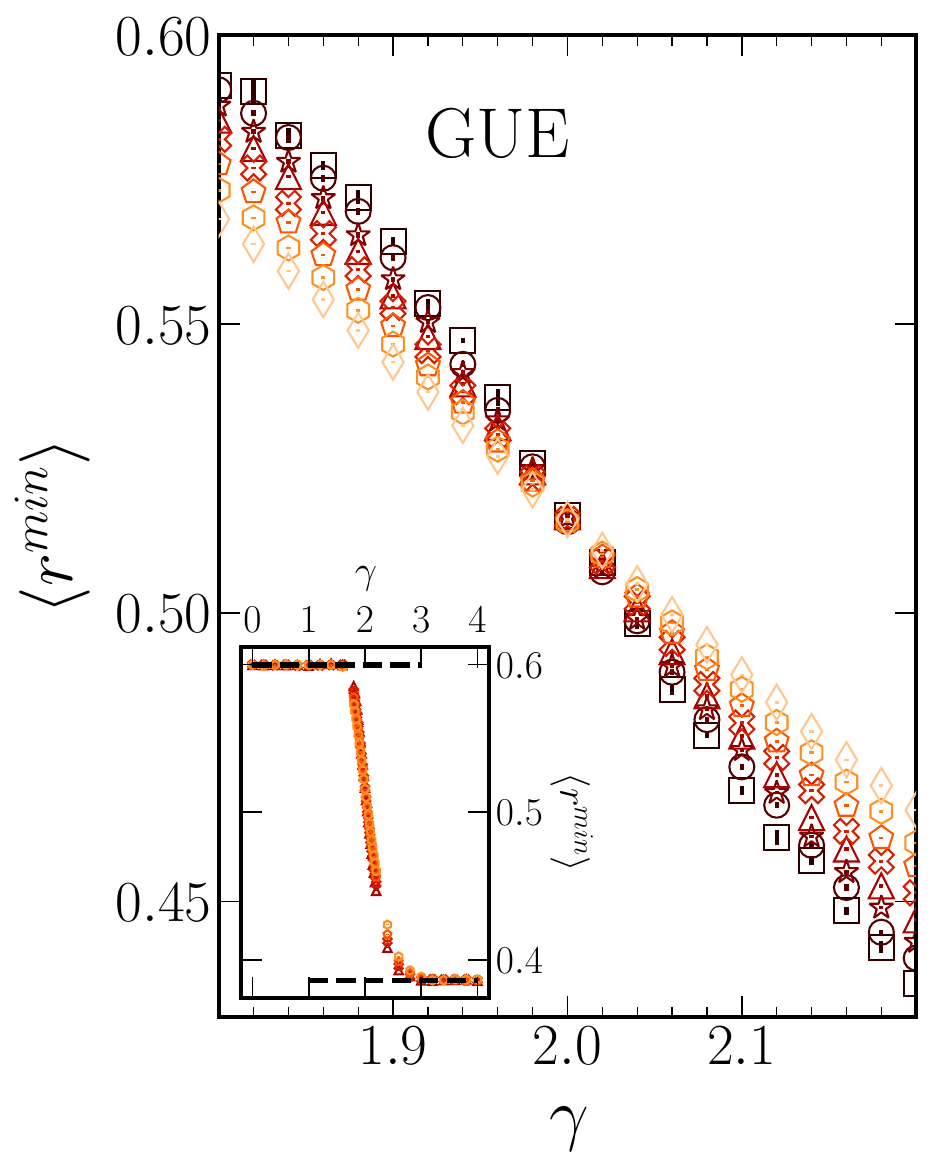}\includegraphics[width=0.2\columnwidth]{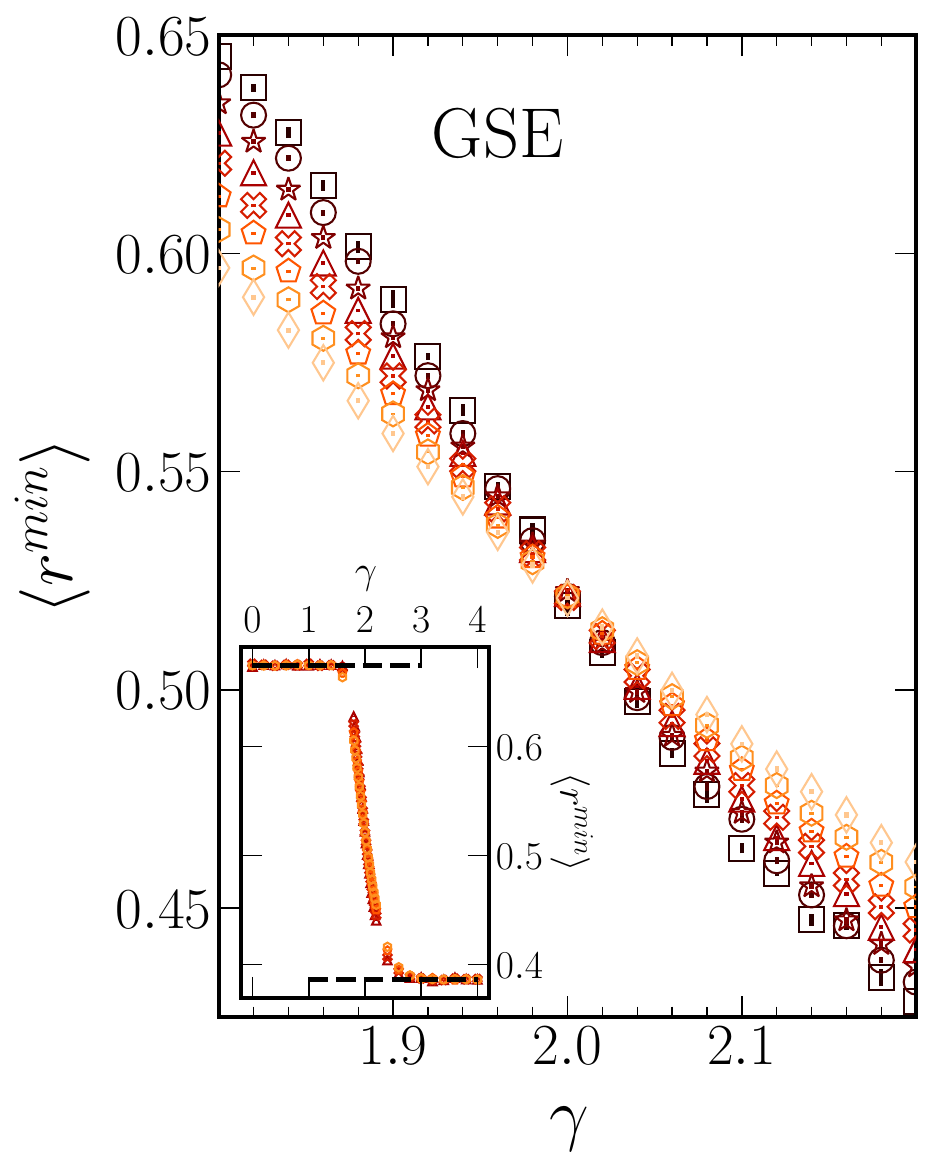}
    \caption{
	    System size dependence of the average values $\langle r^{min}\rangle$ for transitions from the WD ensembles to Poisson as a function of $\gamma$. The number of realizations used was at least 4000, 6000, 3000, 1200, 498, 89, 39, 5 for system sizes from 512 to 65536, respectively and 20$\%$ of the states around the band center was used. The standard errors of mean are smaller than the size of the symbols. 
        The insets show the extended $\gamma$ region for four system sizes, with the ergodic and Poisson values marked by dashed black lines.
    }
   \label{fig:Ratios_transitions}
\end{figure}

\begin{figure}
    \centering
    \includegraphics[width=0.325\columnwidth]{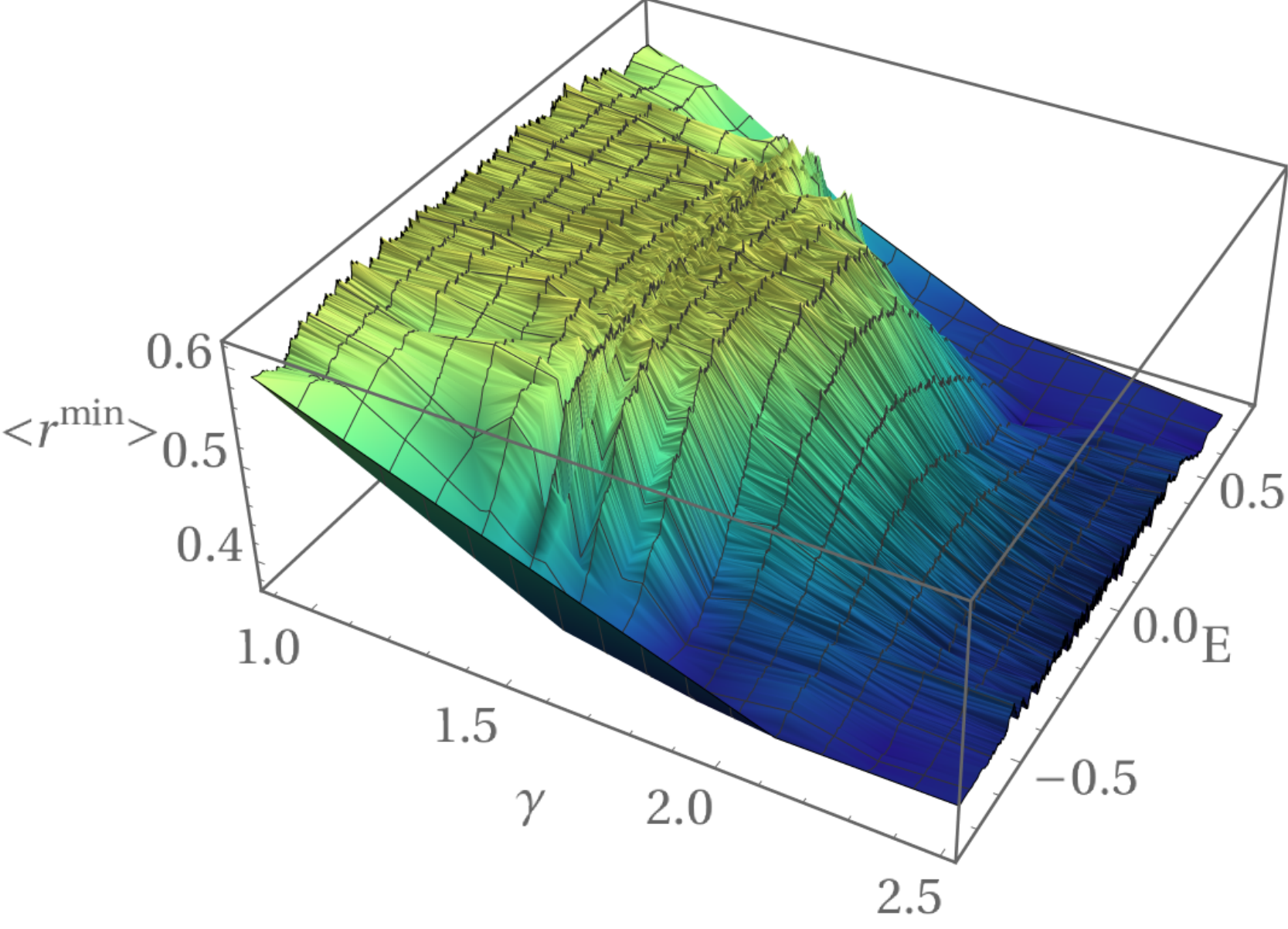}
    \includegraphics[width=0.325\columnwidth]{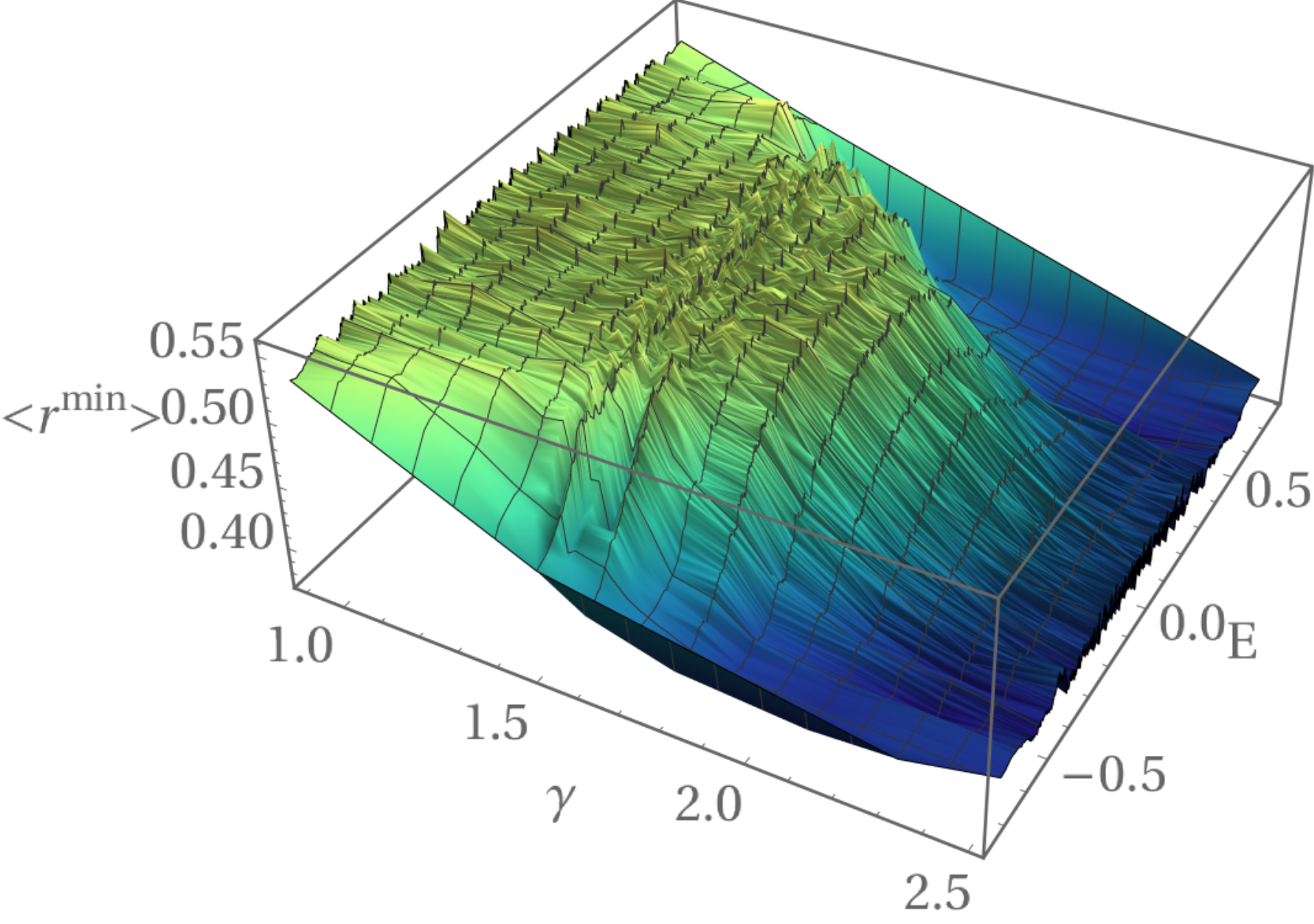}
    \includegraphics[width=0.325\columnwidth]{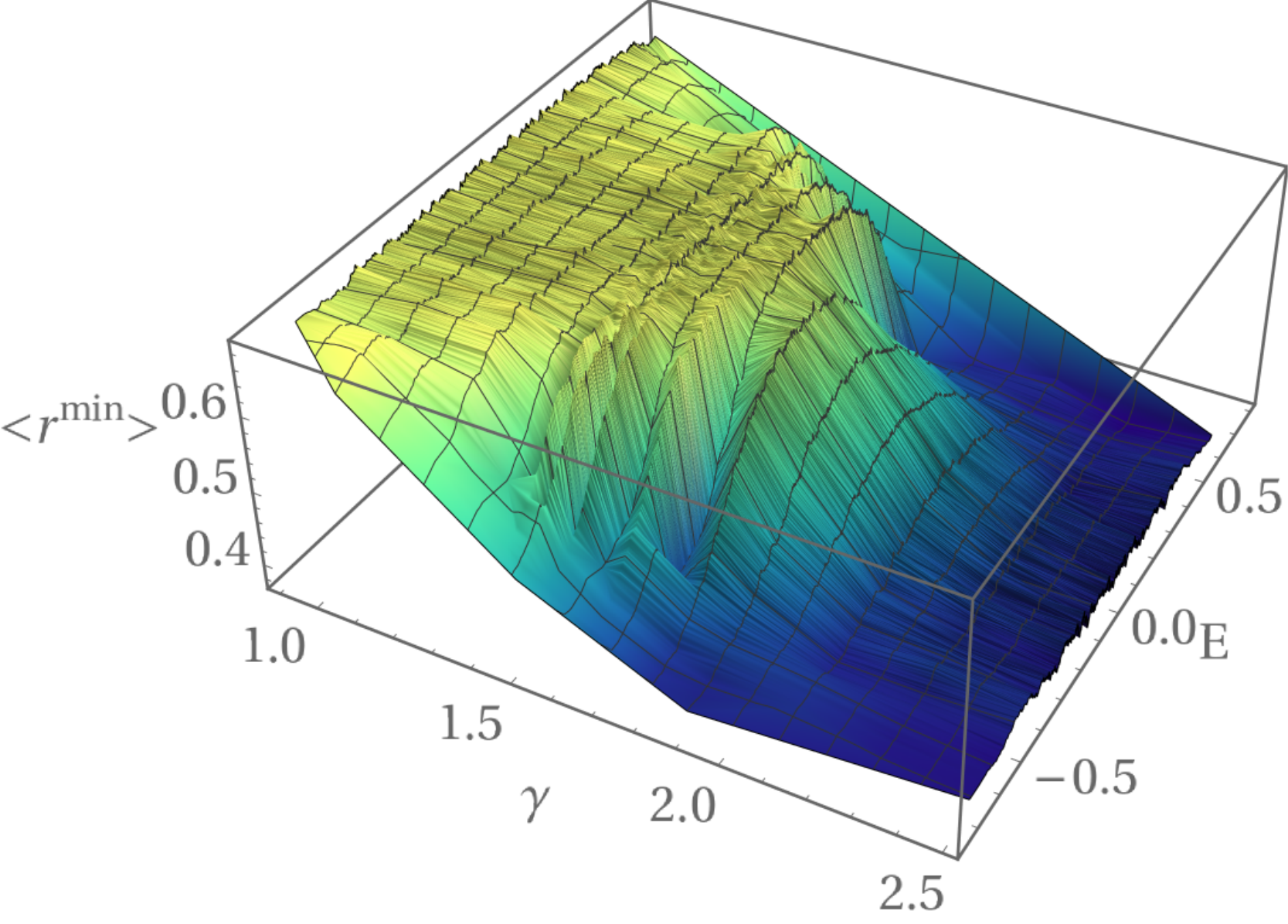}
    \caption{
            Average r-ratios $\langle r^{\rm min}\rangle$ as a function of \(\alpha\) and energy center of a sliding window of 500 levels for N=65534. In the 3D plot Dark blue corresponds to the result for Poisson, yellow to that for GUE (left), GOE (middle) and GSE (right). Here, the eigenvalues were shifted such that the band center is at zero and then divided by their maximum value so that their values range from -1 to +1.
    }
\label{fig:ratios3D}
\end{figure}

For all three WD ensembles analytical results have been obtained for $\langle r\rangle$ and $\langle r^{min}\rangle$~\cite{Atas2013a}, $\langle r\rangle =1.75,1.36.1.17$ and $\langle r^{min}\rangle=0.53,0.6,0.67$ for the GOE, GUE and GSE, respectively, and for Poissonian random numbers they are given by $\langle r\rangle =\infty$ and $\langle r^{min}\rangle=0.39$.
These values are attained for $\beta =1,2,4$ in the limits of small and large $\gamma$, respectively.
This is illustrated in~\reffig{fig:Ratios}, where we also show the analytical result as blue solid line for the transition from Poisson to GUE.
Marginal deviations from the results for the WD ensembles are observed for $\lambda\gtrsim 1.45$, however, clear changes occur only above $\lambda\gtrsim 1.6$, thus implying that the ratio distributions are even less sensitive to small perturbations of the WD matrices~\cite{pino2019from, berkovits2020super} than the nearest-neighbor spacing distribution. Nevertheless, they are commonly used to get information on presence or absence of quantum-chaotic behavior, the reason being that no unfolding of the eigenvalues is required.
In~\reffig{fig:Ratios_transitions} we show for all three WD ensembles the average values $\langle r^{min}\rangle$ for different system sizes $N$.
We observe for all cases ($\beta=1,2,4$) a crossing of the curves at $\gamma =2$, implying that at that value $\langle r^{min}\rangle$ does not depend on $N$. As outlined in Ref.~\cite{pino2019from} at such crossings a discontinuity develops with increasing $N$ leading to a non-analytical point in the thermodynamic limit $N\to\infty$, thus indicating a phase transition from extended to localized at  $\gamma =2$.
Remarkably, the transition takes place at the same value of $\gamma$ for all three universality classes. Figure~\ref{fig:ratios3D} exhibits the energy-resolved average ratios $\langle r^{\rm min}\rangle$ as function of $\gamma$ and of the center of a sliding energy window comprising 500 eigenvalues for all three WD ensembles. The plots clearly show an energy dependent transition from WD to Poisson statistics, thus indicating a mobility edge for the associated eigenvectors.  

\subsection{Long-range correlations}

\begin{figure}
    \centering
    \includegraphics[width=0.4\columnwidth]{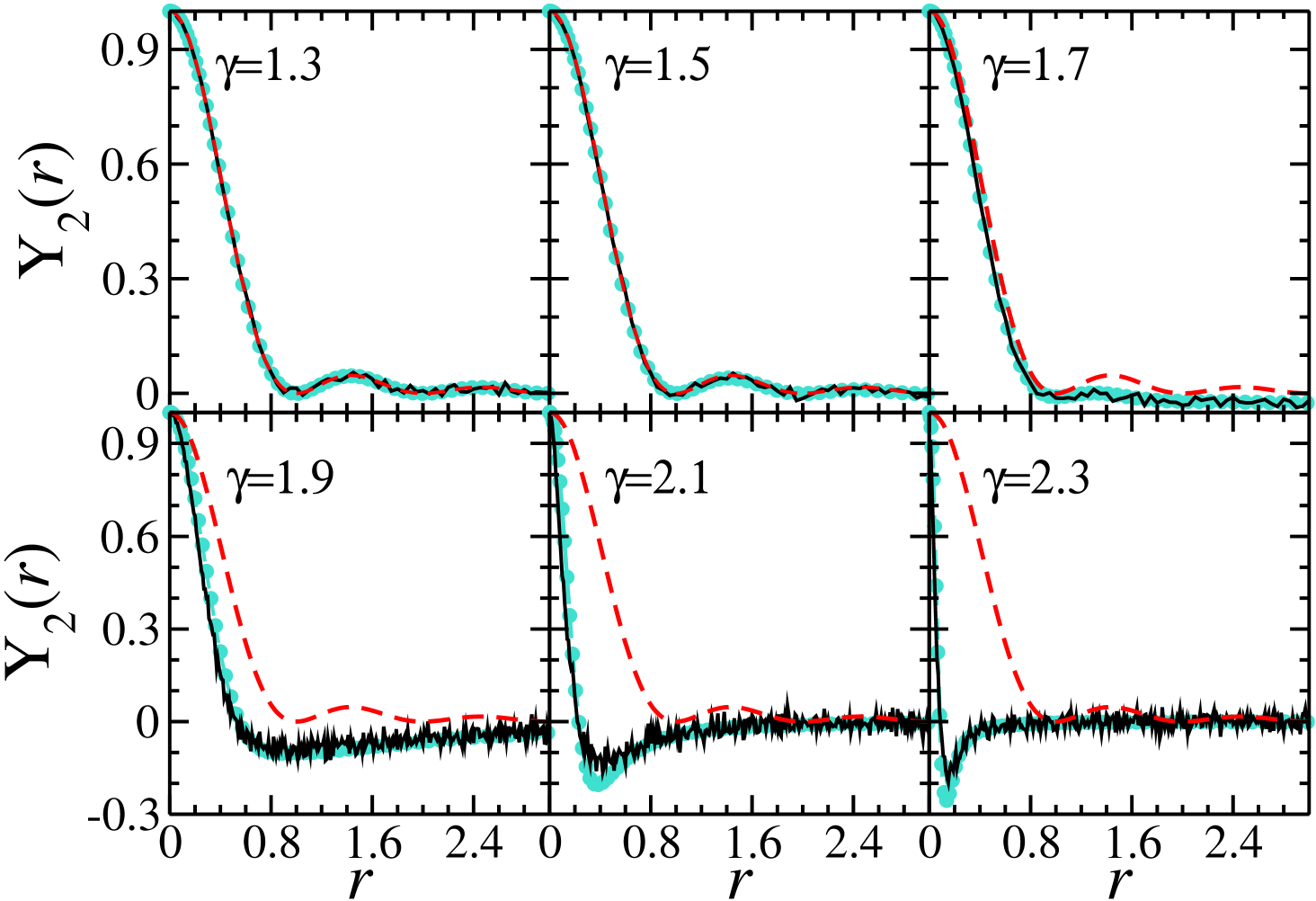}
    \includegraphics[width=0.4\columnwidth]{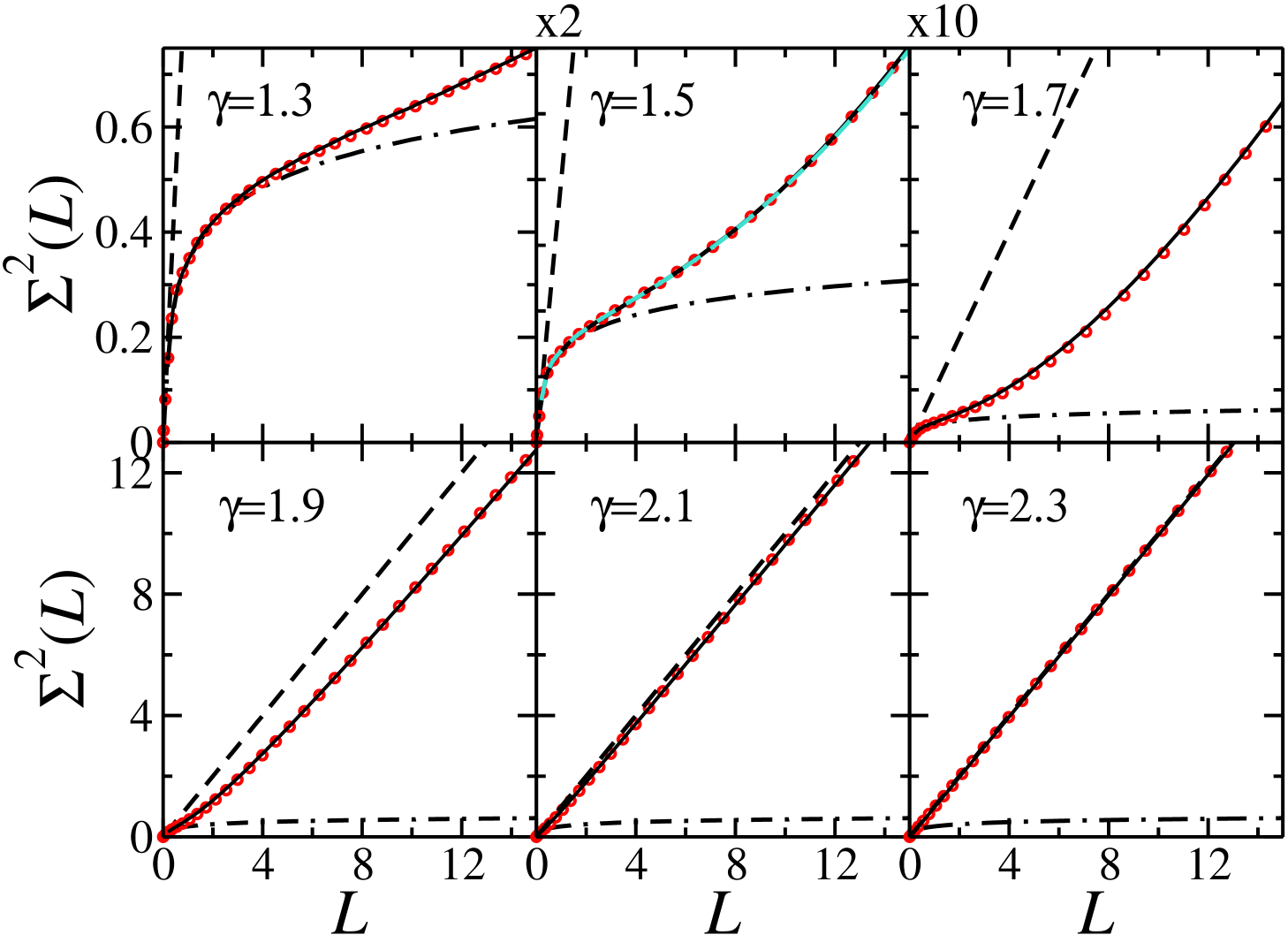}
    \caption{
	    Left: Comparison of the two-point cluster function obtained from the random-matrix simulations for the GUE gRP Hamiltonian~\eqref{eq:HgRP} (black) with the analytical result~\eqref{Y2Anal} (turquoise dots) for various values of $\gamma$ indicated in the panels. Here, $L$ denotes the length of the energy interval in units of mean spacing. The red dashed line exhibits the result for the  WD ensemble with $\beta =2$.
	    Right: Comparison of the number variance obtained from the random-matrix simulations for the GUE gRP Hamiltonian~\eqref{eq:HgRP} (black) with the analytical result~\eqref{Sigma2Anal} (red) for various values of $\gamma$ indicated in the panels. The turquoise line shows one example for random matrices of dimension $N=100000$. It is indistinguishable from the result for $N=2^{16}$. The dashed and dash-dotted black lines exhibit the results for Poisson and GUE statistics, respectively. 
    }
    \label{fig:Sigma2}
\end{figure}

\begin{figure}
    \centering
    \includegraphics[width=0.4\columnwidth]{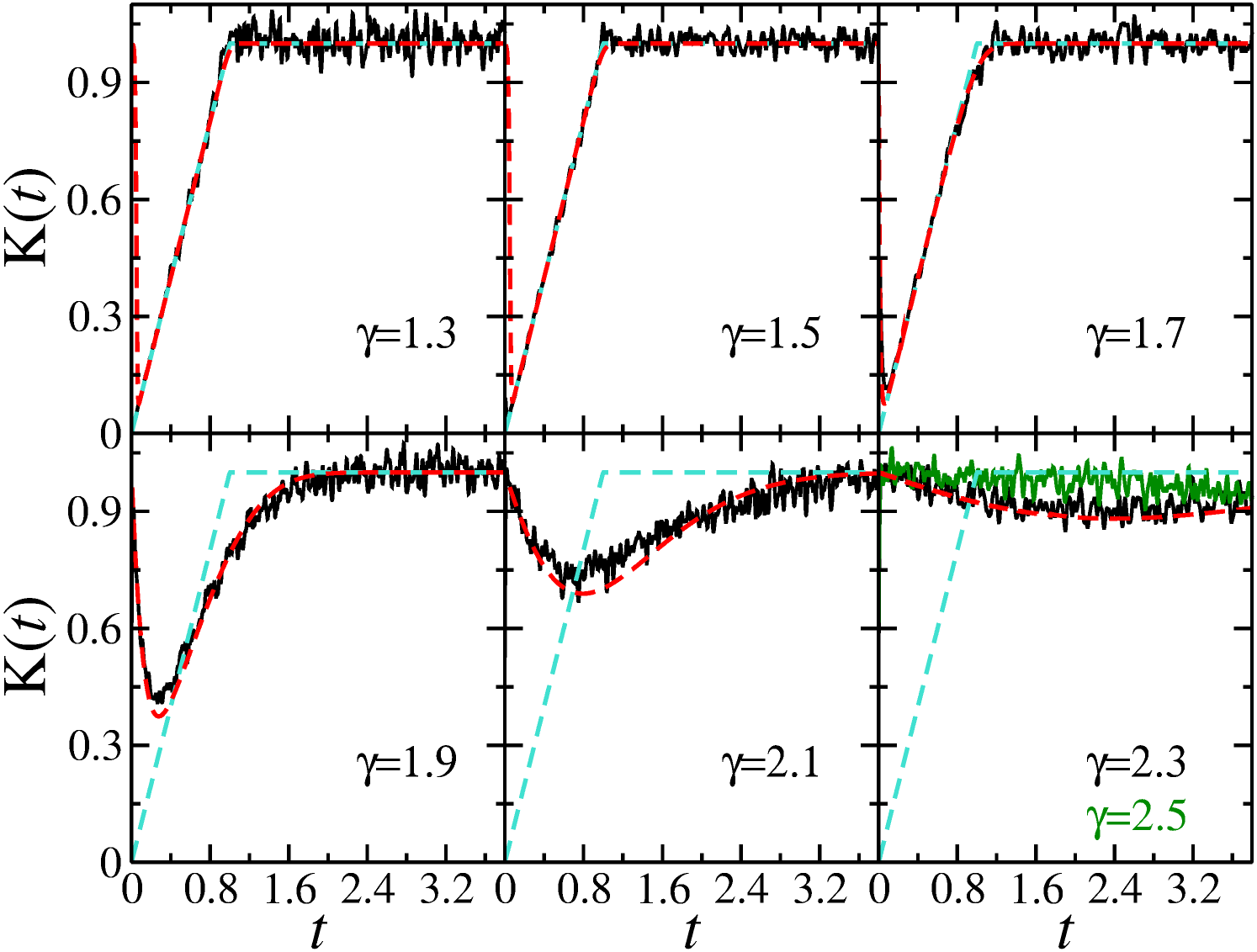}
    \includegraphics[width=0.4\columnwidth]{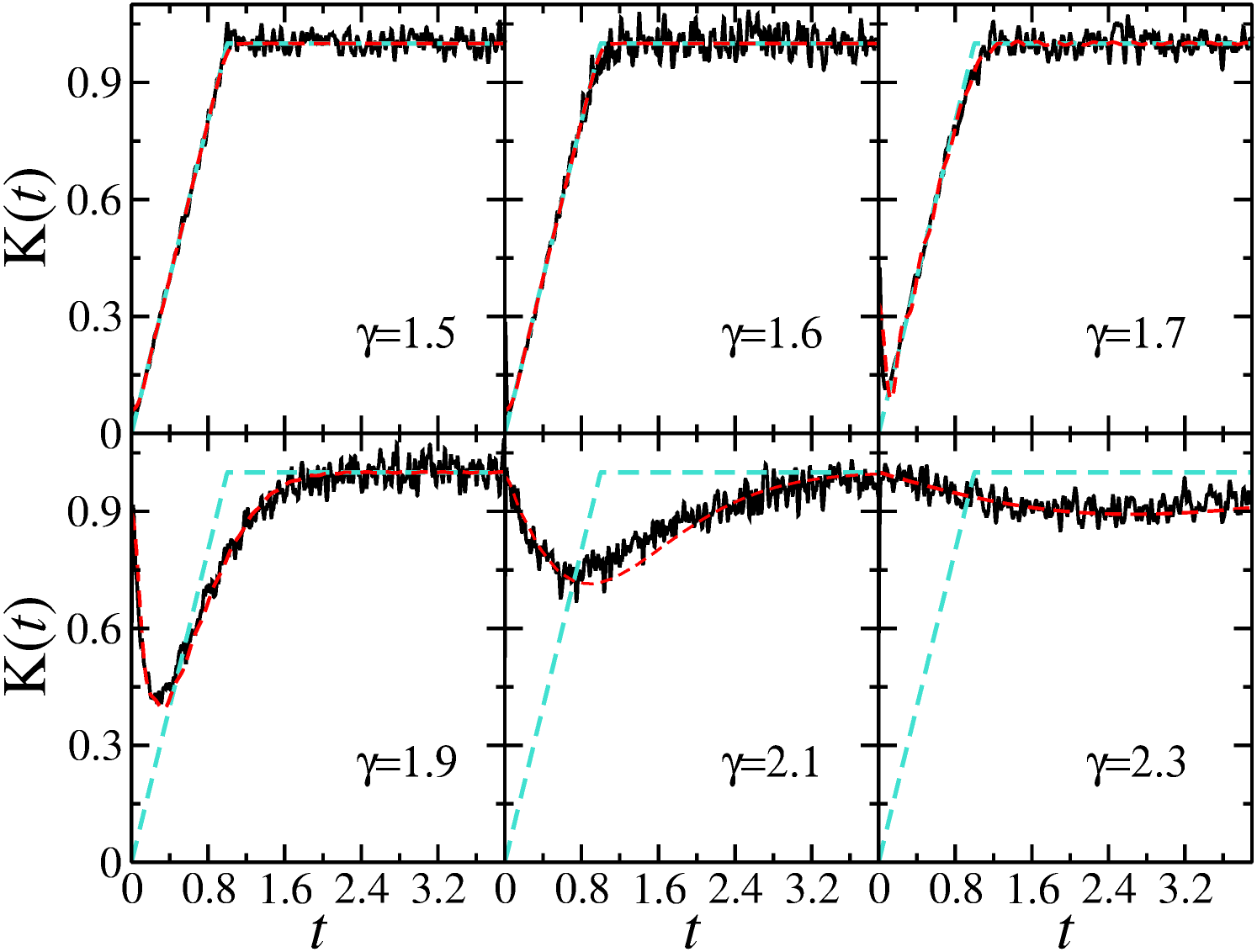}
    \caption{
	    Left: Comparison of the spectral form factor obtained from the random-matrix simulations for the GUE gRP Hamiltonian~\eqref{eq:HgRP} (black) with the analytical result~\eqref{B2Anal} (red) for various values of $\gamma$. The corresponding values of $\lambda$ are obtained by fitting the analytical curves to the numerical ones. We do not show them, since they agree with the values obtained from the nearest-neighbor spacing distributions. The turquoise dashed line shows the results for the WD ensemble with $\beta =2$.
        Right: Same as in the left panel, the only difference being that the analytical curve is obtained from the Fourier transform of the analytical result~\eqref{Y2Anal} for the two-point cluster function (red). Some slight discrepancies are observed for small values of $t$ for the case $\gamma =1.7$, but otherwise agreement with the numerical ones is as good as in the left panel, if not better.}
    \label{fig:K}
\end{figure}
Based on the RP model~\eqref{RPH}, in Ref.~\onlinecite{Lenz1992} approximate analytical results were obtained for the two-point cluster function for the transition from Poisson to GOE and an analytical expression for the transition from Poisson to GUE, which is exact for all values of $\lambda$ and  $N$, however, the computation of the limit $N\to\infty$ starting from that expression was impossible~\cite{Lenz1992}. In~\cite{Venturelli2023} the replica approach was applied to the gRP model for the transition from Poisson to GOE to compute the average spectral density and level compressibility. Yet, exact analytical results for statistical measures of long-range correlations of random matrices from the RP ensemble~\eqref{RPH} are only available for $\beta=2$, see ~\appsec{Analytic}. In Fig.~\ref{fig:Sigma2} we compare the analytical results to random-matrix simulations with the gRP Hamiltonian~\eqref{eq:HgRP} for the two-point cluster function and number variance, and for the spectral form factor in the left panel of~\reffig{fig:K}. Deviations from the WD ensemble with $\beta=2$ are visible for the two-point cluster functions for $\gamma\gtrsim 1.5$, and for the number variance for $\gamma\gtrsim 1$. The values of $\lambda$ obtained from the fit of the analytical result for $\Sigma^2(L)$ to the numerical results as function of $\gamma$ agree with those shown in~\reffig{fig:LamGam}, that were obtained from the fit of $P_{0\to 2}(s)$ given in~\eqref{PSGUE} to the nearest-neighbor spacing distribution of the gRP Hamiltonian~\eqref{eq:HgRP} with $\beta =2$. An anlytical expression for the spectral form factor can also be deduced from the Fourier transform of the analytical result for the two-point cluster function~\cite{Frahm1998} given in~\eqref{Y2Anal}. The result is exhibited in the right panel of~\reffig{fig:K}. Comparison with the analytical result for $K(\tau)$~\cite{Kunz1998}~\eqref{B2Anal} depicted in the left panel of~\reffig{fig:K} shows, that the agreement with the numerical results for $K(\tau)$ is comparable. For the matter of completeness, we would like to mention, that approximations have been derived for $Y_2^{0\to 2}(r)$ for $\lambda \ll 1$ and $\lambda\gg 1$ in Refs.~\onlinecite{French1988,Leyvraz1990,Lenz1992,Pandey1995,Brezin1996,Guhr1996a,Guhr1996,Guhr1997,Altland1997,Kunz1998,Frahm1998}. 

\begin{figure}
    \centering
    \includegraphics[width=0.63\columnwidth]{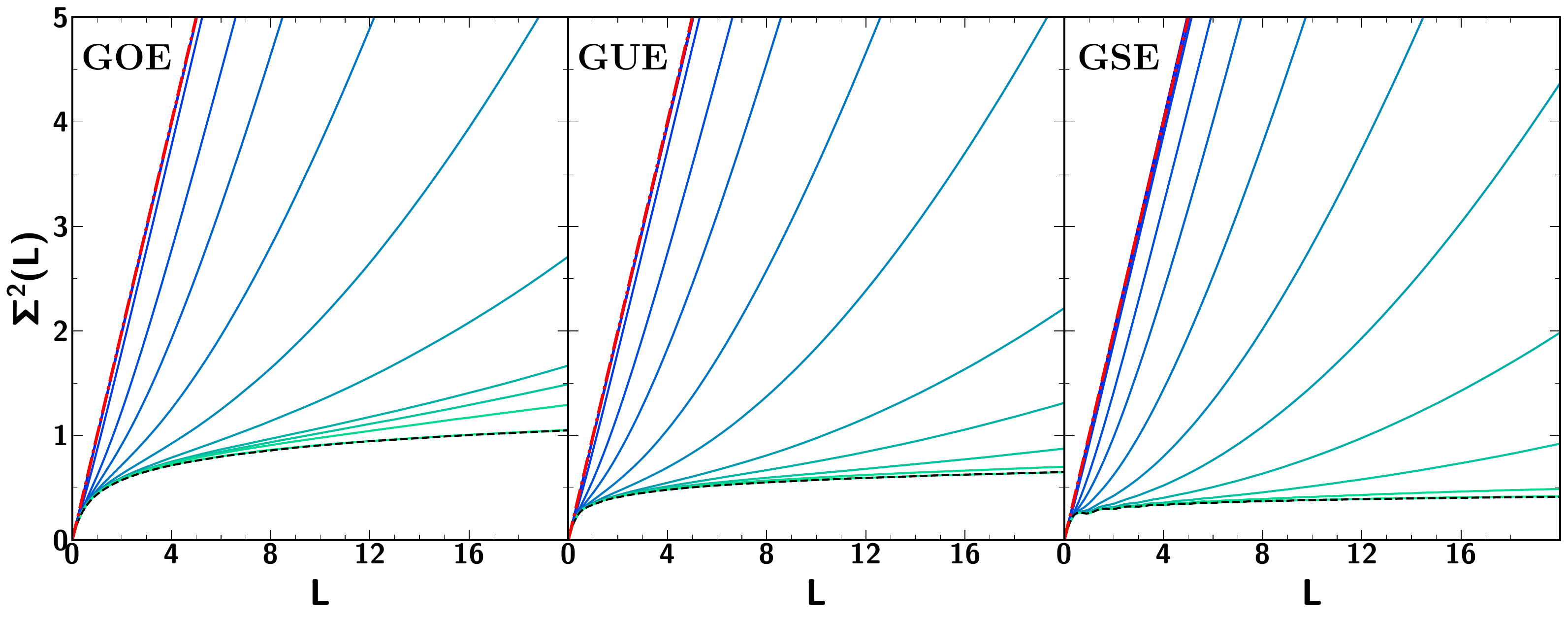}
    \includegraphics[width=0.35\columnwidth]{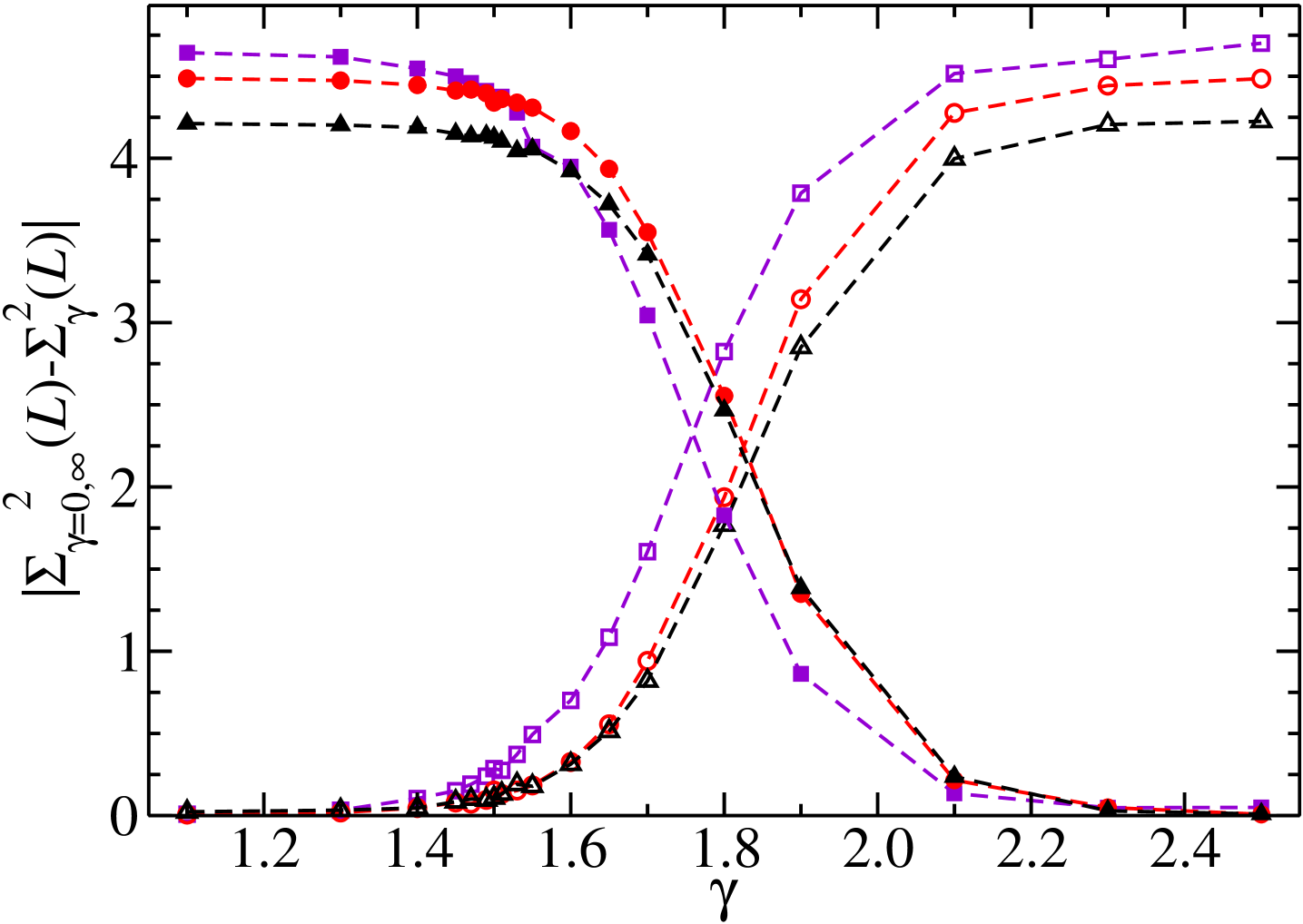}
    \caption{
	    Left: Number variance $\Sigma^2(L)$ obtained from random-matrix simulations for the gRP model for $\beta =2$ (left), $\beta =1$ (middle) and $\beta =4$ (right) for the same values of $\gamma$ and color coding as in~\reffig{fig:NNDs}. The number variance $\Sigma^2(L)$ experiences a transitiopenon from WD behavior to Poisson statistics. Actually, for $\gamma=0.9$ the curves lie on top of the WD result (black dashed line), whereas already for $\gamma=1.1$ deviations are visible and for $\gamma=2.5$ they lie on top of the curve for Poissonian random numbers (red dash-dotted line).
	    Right: Distance between the number variance $\Sigma^2(L)$ at $L=5$ for $\gamma = 0.9$, where it coincides with that for the corresponding WD ensemble (open symbols), respectively, for Poisson statistics (full symbols) and its value for $\gamma > 0.9$ for the transition from Poisson to GOE (black triangles), to GUE (red circles) and to GSE (purple squares).  
    }
   \label{fig:Sigma2_GOE_GUE_GSE}
\end{figure}

\begin{figure}
    \centering
    \includegraphics[width=0.396\columnwidth]{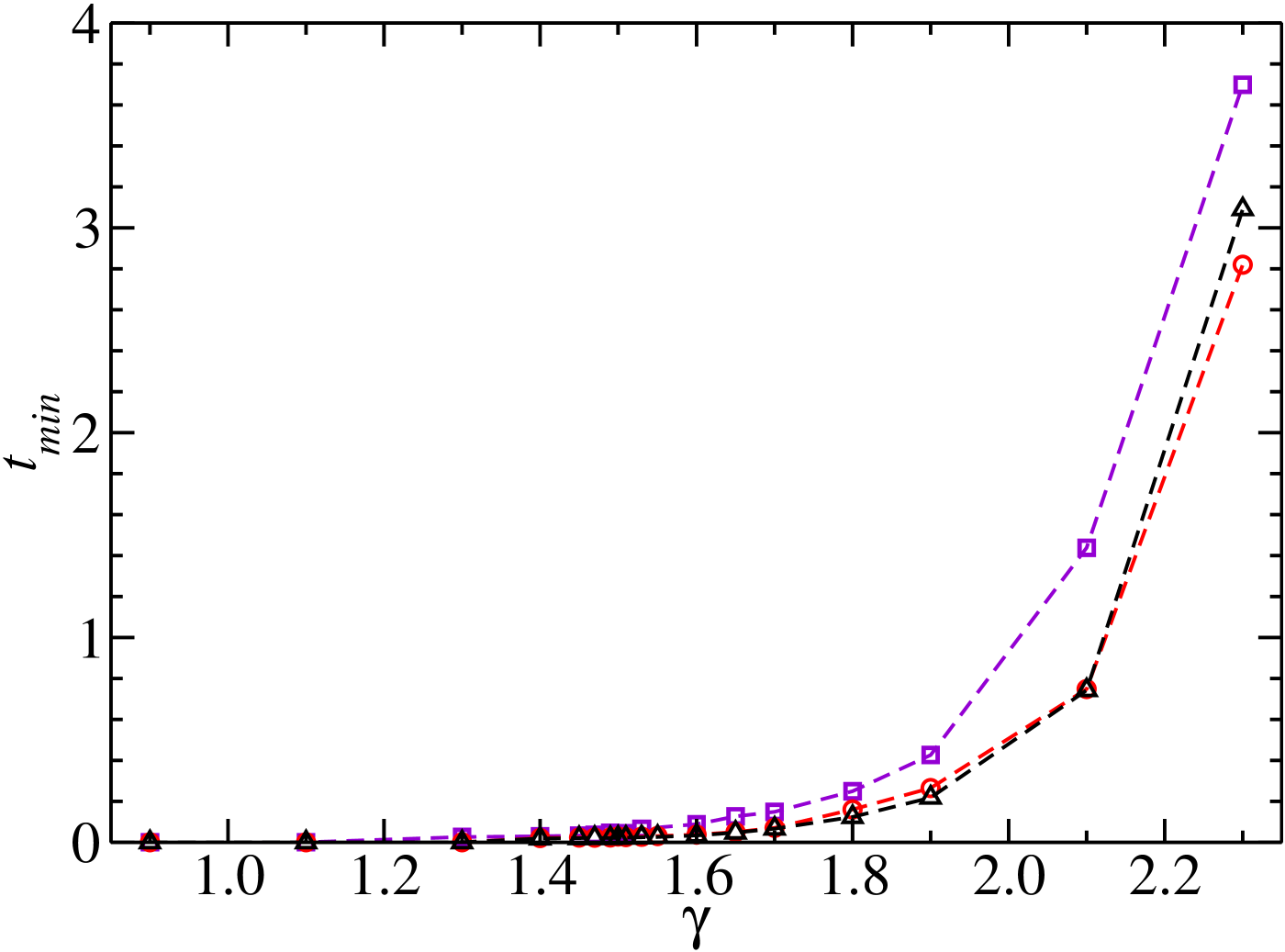}
    \includegraphics[width=0.41\columnwidth]{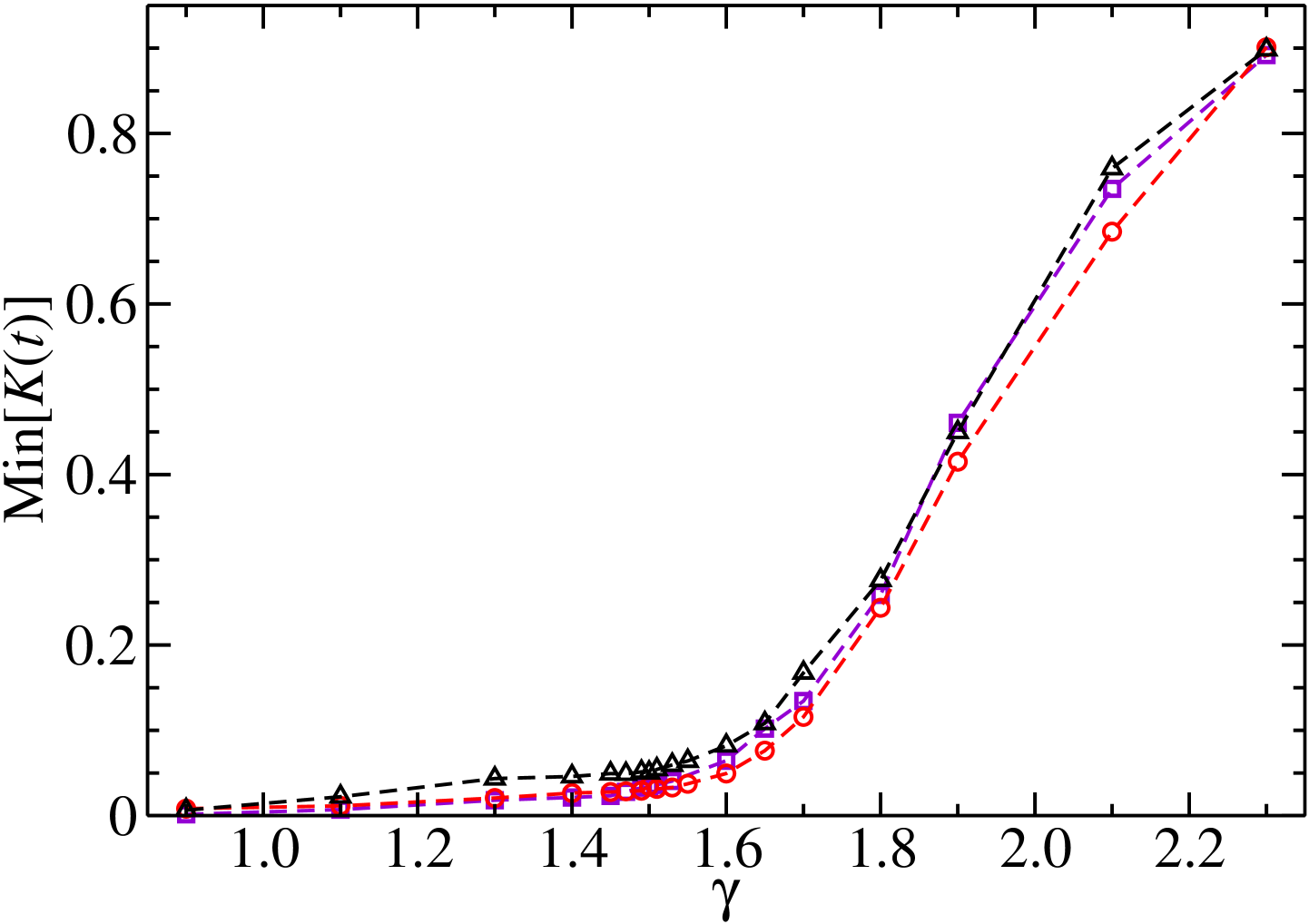}
    \caption{
            Left: Position $t_{min}$ of the minimum of the form factor $K(t)$ for the GOE (black triangles), GUE (red circles) and GSE (purple squares).
            Right: Same as left for the value of the form factor at the position of the minimum, $K(t_{min})$. 
    }
    \label{fig:Kmin}
\end{figure}

In the left part of~\reffig{fig:Sigma2_GOE_GUE_GSE} results are shown for the number variance for the gRP Hamiltonian~\eqref{eq:HgRP} with~\eqref{eq:gRP_variances} for all WD universality classes $\beta =1,2,4$. For $\gamma =0.9$ the curves obtained from the random-matrix simulations lie for all values of $\beta$ on top of the curve for the corresponding WD ensemble, and for $\gamma=2.5$ they lie on top of the curve for Poissonian random numbers.
Clear discrepancies between WD behavior and the numerical simulations are observed in $\Sigma^2(L)$ for all values of $\beta$ for $\gamma\gtrsim 1$. To illustrate this, we plot in the right part of~\reffig{fig:Sigma2_GOE_GUE_GSE} its distance from the curve for the corresponding WD ensemble at $L=5$ (open symbols), and similarly the distance from Poisson statistics (full symbols). For $\gamma\gtrsim 1$ the distances from WD behavior are nonzero but small and they increase rapidly for $\gamma\gtrsim 1.4$ and saturate for $\gamma\gtrsim 2.1$ and are close to Poisson then.

In Fig.~\ref{fig:K_GOE_GSE} we show the numerical results for the spectral form factor $K(t)$ for the gRP models with $\beta =1,4$. The turquoise lines show the results for the corresponding WD ensemble. For all three universality classes the value of $t$ at the minimum, $t_{min}$ and of the minimum, $K(t_{min})$, itself are zero in the ergodic limit, whereas for Poissonian random numbers the spectral form factor is constant, $K(t)=1$, and thus doesn't exhibit a minimum. For $\gamma\gtrsim 1.1$ slight deviations from the corresponding WD ensemble occur around the minimum at $t\simeq 0$, and for $\gamma\gtrsim 1.5$ discrepancies between the gRP and WD ensembles are clearly visible. To illustrate this, we plot in~\reffig{fig:Kmin} the value of $t$ at the minimum of $K(t)$ (left), denoted by $t_{min}$, and the value of the minimum, $K(t_{min})$, itself. We find that $K(t_{min})$ is nonzero but small for $\gamma\gtrsim 1$ and for $2.1\gtrsim\gamma\gtrsim 1.6$ increases drastically and disappears in the considered range of $t\leq 5$ for $\gamma =2.5$.

\begin{figure}
    \centering
    \includegraphics[width=0.64\columnwidth]{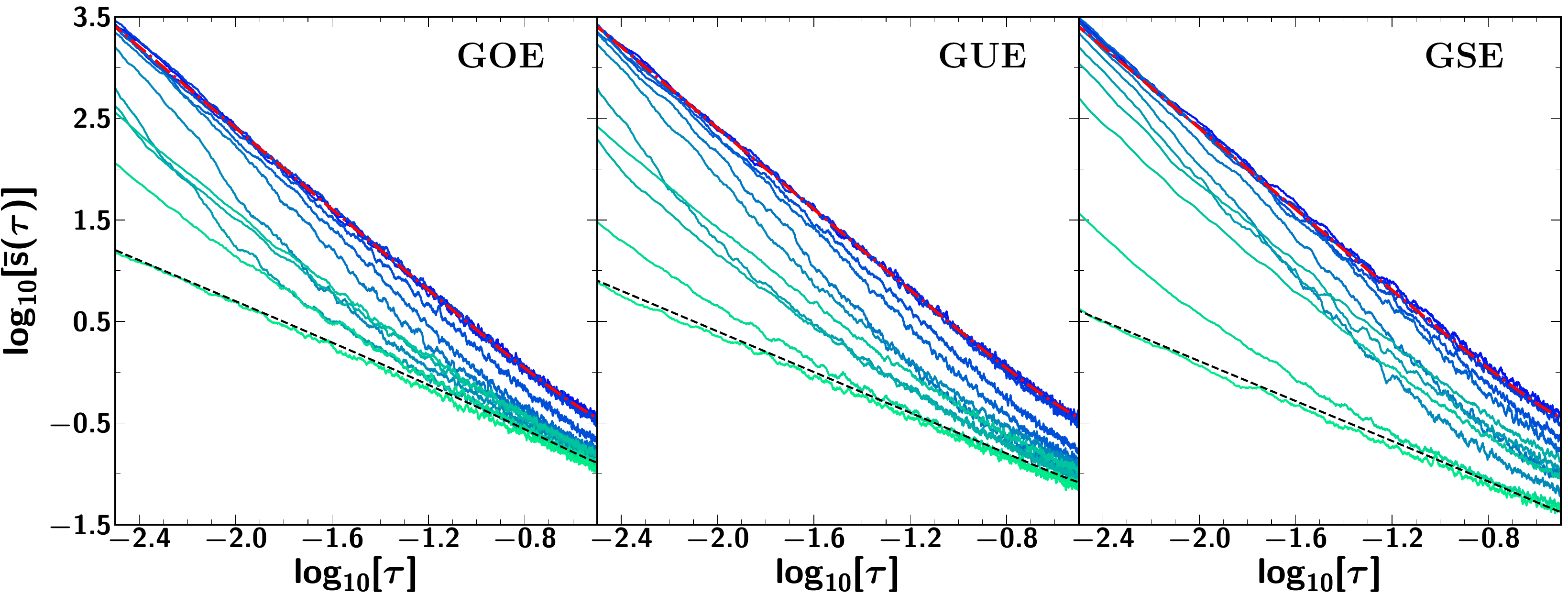}
    \includegraphics[width=0.34\columnwidth]{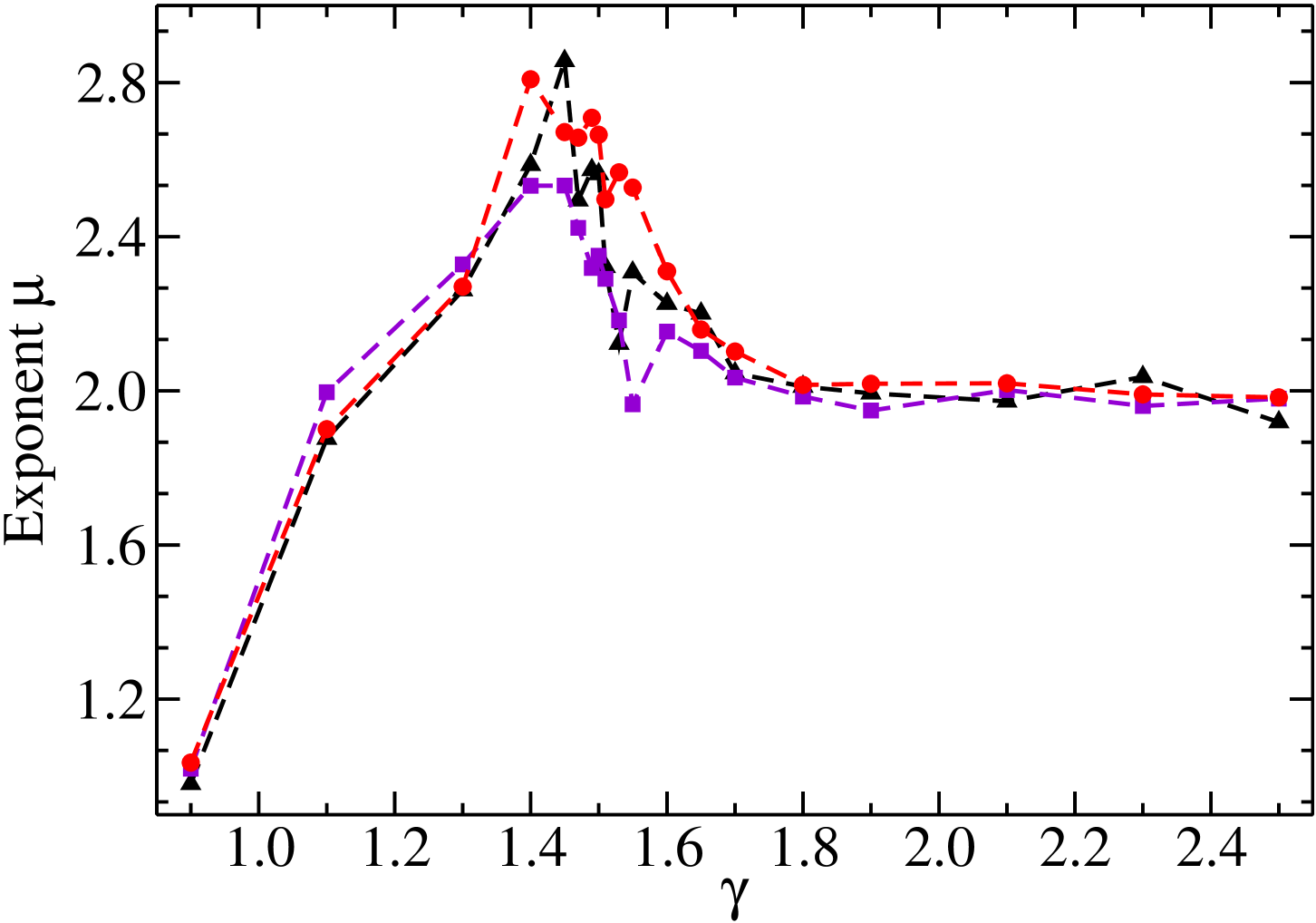}
    \caption{
            Left: Power spectrum obtained from random-matrix simulations for the gRP model for $\beta =1$ (left), $\beta =2$ (middle) and $\beta =4$ (right) for the same values of $\gamma$ and color coding as in~\reffig{fig:NNDs}. For $\gamma=0.9$  the curve is close to an approximate result for the corresponding WD ensemble (turquoise dashed line), and for $\gamma=2.5$ it lies on top of the approximate result for Poissonian random numbers (turquoise dash-dot line) (see main text).
            Right: Same as left for the exponent $\mu$ of the power law $\langle s(\tau\ll 1)\rangle\propto \tau^{-\mu}$ expected for the limiting cases, i.e., for Poisson statistics and the WD ensembles. It is obtained by fitting in the asymptotic region $\tau\lesssim 10^{-2}$ a straight line $y=-\mu\log_{10}[\tau]+const.$ to the logarithm of the power spectra, $\log_{10}[s(\tau)]$ as function of $\log_{10}[\tau]$ shown in the left part.
    }
   \label{fig:PowerSp}
\end{figure}
Another measure for long-range correlations is the power spectrum defined in~\eqref{eq:power}. Exact analytical results were obtained for the power spectra in Refs.~\onlinecite{Riser2017,Riser2023} for fully chaotic quantum systems with violated \Ti-invariance, however, we are not aware of any analytical results for the RP model. The power spectrum exhibits for $\tau\ll 1$ a power law behavior $\langle s(\tau)\rangle\propto \tau^{-\mu}$, where $\mu =2$ for Poisson distributed random numbers and $\mu =1$ independently of the universality class for the WD ensembles. In the left part of~\reffig{fig:PowerSp} we show logarithmic plots of the power spectra versus $\log_{10}[\tau]$ obtained from the gRP model for all three universality classes. These figures illustrate that the power spectra indeed increase linearly with decreasing $\log_{10}[\tau]$ below $\log_{10}[\tau]\lesssim -1.0$ close to $\gamma\lesssim 1.1$ and for $\gamma\gtrsim 1.8$. For the other cases they increase linearly below $\log_{10}[\tau]\lesssim -1.6$ for $\beta =4$ and below $\log_{10}[\tau]\lesssim -2.0$ for $\beta =1,2$. For $1.8\gtrsim \gamma\gtrsim 1.1$ their slopes change drastically with increasing $\gamma$. The power spectra are compared to theoretical approximations in terms of the spectral form factor derived in Ref.~\onlinecite{Molina2007} for the WD ensembles and Poisson statistics, that have been tested experimentally for all WD ensembles~\cite{Bialous2016,Che2021} for spectra consisting of several hundreds of eigenvalues. For $\gamma =0.9$ the power spectrum agrees with that of the corresponding WD ensemble and it approaches the result for Poisson statistics for $\gamma\gtrsim 2$. Accordingly, we may use the slope of the straight line best fitting $\log_{10}\left[s\left(\tau\right)\right]$ for $\log_{10}[\tau]\lesssim -2$ as indicator for the onset of deviations from WD statistics and agreement with Poisson. We show the values of $\mu$ obtained from linear regression of the logarithm of the power spectra as function of the logarithm of $\tau$ for $\tau\lesssim 10^{-2}$ for the three WD ensembles in the right part of~\reffig{fig:PowerSp}. Deviations from WD statistics are visible for $\gamma >1$ and at $\gamma\simeq 1.45$ an abrupt change is observed. There, actually, the range of $\tau$ values available for the linear fit was smaller than for the other cases (about 300 levels), however the accuracy suffices to get information on the qualitative behavior, which clearly deviates from that expected in the limiting cases. 

We wondered whether the approximation~\cite{Molina2007} of the power spectrum in terms of the spectral form factor also applies to the intermediate case between WD behavior and Poisson statistics. Accordingly we compared for the transition from Poisson to GUE the curves obtained by replacing the spectral form factor in this approximation by the analytical result~\eqref{B2Anal} to the power spectra obtained from the random-matrix simulations. We found deviations especially for $1.4\lesssim\gamma\lesssim 1.7$. Even for $\gamma =0.9$, which is close to the GUE curve, slight differences are visible. We attribute this to the high dimensions of the matrices used, that are large enough to reveal deviations from the approximations. A few examples are shown in~\reffig{fig:Power_GUE_Approx}.

Summarizing the results of~\refsec{RMTSpectr}, we observe in the short-range correlations changes in the position of the maximum of the nearest-neighbor spacing distributions and the average ratio distributions above $\gamma\approx 1.45-1.6$ and saturation above $\gamma\simeq 2$, whereas in the long-range correlations changes are visible for $\gamma\gtrsim 1$. Yet, for $\gamma\gtrsim 2$ deviations of $\Sigma^2(L)$ from the corresponding WD and Poisson curves, the power $\mu$ of the asymptotic algebraic decay of the power spectrum and the position of the minimum of the form factor saturate at values close to those of Poissonian random numbers.

\section{Properties of the eigenvectors of the gRP model}
\label{RMTWFs}

Long-range correlations like the number variance and the spectral form factor clearly indicate a transition from WD behavior to Poisson statistics at $\gamma\gtrsim 1$, which indicates a change from ergodic to localized states however for the unambiguous determination of the fractal phase and the Anderson transition, the analysis of properties of the eigenvectors of the gRP Hamiltonian $\hat H^{\rm gRP}(\gamma)$ in~\eqref{eq:HgRP} is needed.
In this section we investigate them in terms of fractal dimensions~\cite{kravtsov2015random}, participation ratios~\cite{bogomolny2018eigenfunction} and participation entropy~\cite{pino2019from}, Kullback-Leibler divergences~\cite{kullback1951on, pino2019from,khaymovich2020fragile} and fidelity susceptibility~\cite{sels2021dynamical, skvortsov2022sensitivity}.
Due to Kramer's degeneracies for the gRP model with symplectic universality class, we only consider one half of the eigenvectors, namely those with an odd index, for that case.

\subsection{Fractal dimensions}

\begin{figure}
    \centering
    \includegraphics[width=0.75\columnwidth]{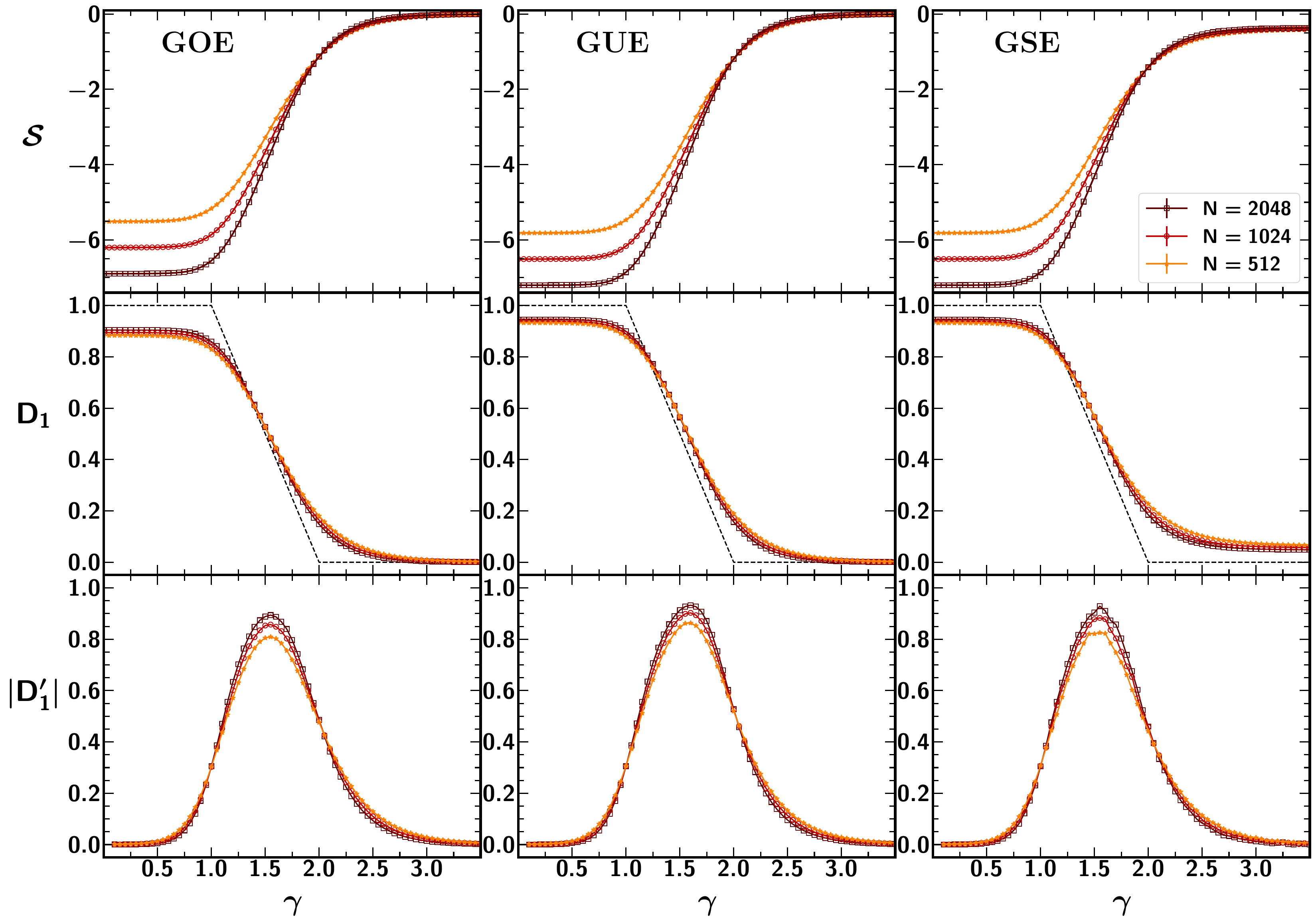}
    \caption{
	    Participation entropy ${\cal{S}}$, fractal dimension $D_1$ and the derivative of the fractal dimension $|D_1^{\prime}|$ with respect to $\gamma$ for the GOE, GUE and GSE gRP model for system sizes $N = 2^n$, with $n = 9, 10, 11$ (yellow to brown).
            The dotted line in the fractal dimension is the analytical result~\cite{kravtsov2015random}.
    }
    \label{fig:fractal_dimensions}
\end{figure}

\begin{figure}
    \centering
    \includegraphics[width=0.75\columnwidth]{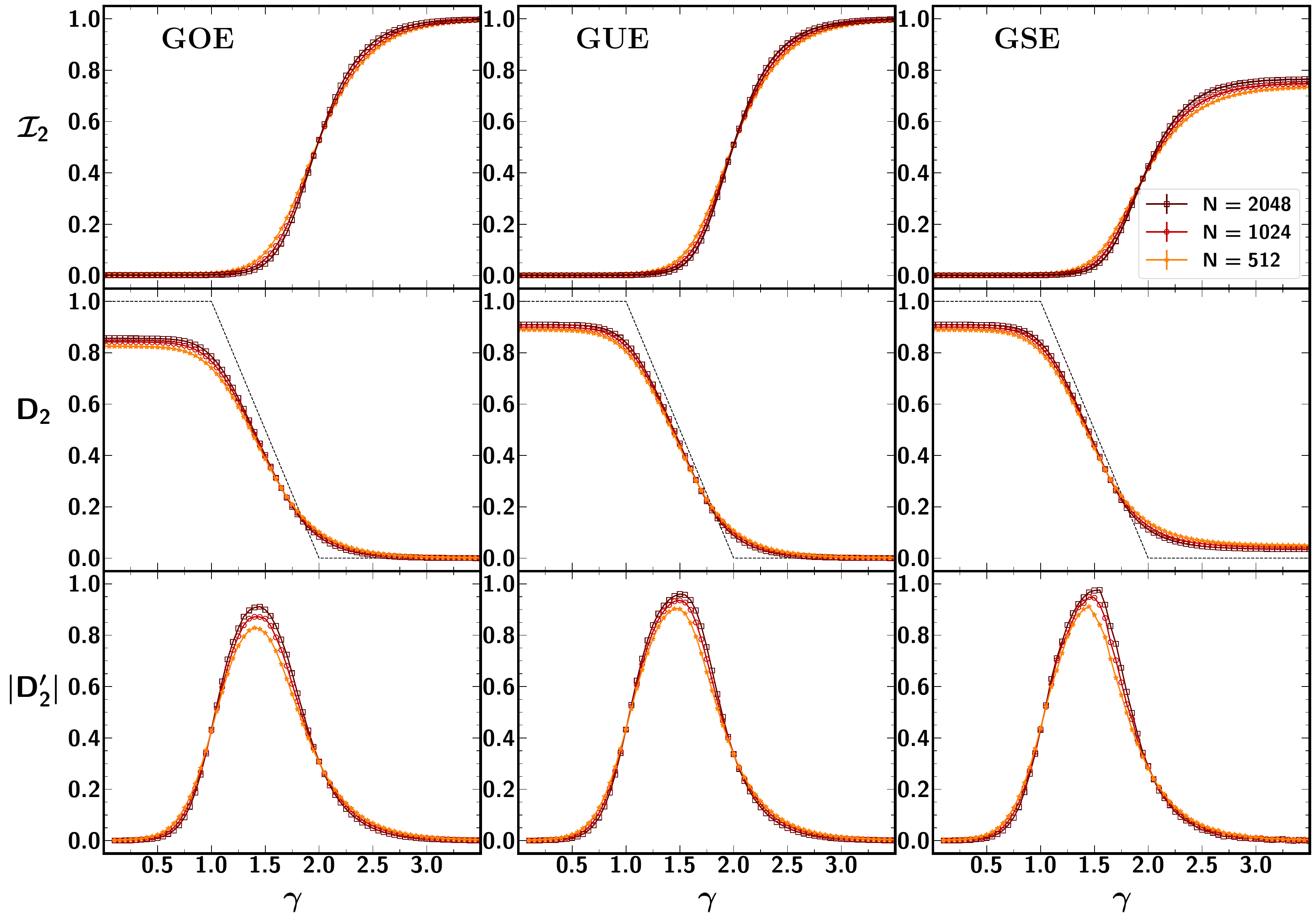}
    \caption{
            Participation number ${\cal{I}}_2$, fractal dimension $D_2$ and the derivative of the fractal dimension $|D_2^{\prime}|$ with respect to $\gamma$ for the GOE, GUE and GSE gRP model for system sizes $N = 2^n$, with $n = 9, 10, 11$ (yellow to brown).
            The dotted line in the fractal dimension is the analytical result~\cite{kravtsov2015random}.
    }
    \label{fig:PNs}
\end{figure}

We analyzed several measures to obtain information on the properties of the eigenvectors, one being the generalized participation numbers (PN),
\begin{gather}
    {\cal{I}}_{q} = \langle \sum_i |\psi_{\mu}(i) |^{2 q} \rangle,
\end{gather}
where the normalized \(\mu-\)th eigenstate of the Hamiltonian, \(H |\psi_{\mu}\rangle = E_{\mu} |\psi_{\mu} \rangle\), is written in the computational basis \( |\psi_{\mu}\rangle = \sum_i \psi_{\mu}(i) | i \rangle\), and the average is taken over a chosen energy window around the band center as well as over multiple disorder realizations. The participation entropy is defined as
\begin{gather}
        {\cal{S}} = \langle \sum_i |\psi_{\mu}(i) |^{2} \log(|\psi_{\mu}(i) |^{2})\rangle ,
\end{gather} 
 and the fractal dimension $D_q$ was introduced in the context of chaotic dynamics in Refs.~\cite{Grassberger1983,Hentschel1983}. For ${\cal{I}}_{q}$ it is defined as
\begin{gather}
	D_q = \lim_{N\to\infty}\log_N({\cal{I}}_{q} )/(1-q).
\end{gather}
The fractal dimension for $q=1$ is obtained with help of l'H\^opital's rule, 
\be
D_1 = \lim_{N\to\infty}\log_N(\sum_i |\psi_{\mu}(i) |^{2} \log(|\psi_{\mu}(i) |^{2}).  
\ee
The values of the fractal dimensions are for the localized, fractal and extended phases \( D_q = 0\), \(0 < D_q < 1\), and \(D_q = 1\), respectively. For sufficiently large $N$ the PN is well approximated by ${\cal{I}}_{q}=\mathcal{C}N^{(q-1)D_q}$ with $\mathcal{C}=o(1)$. Accordingly, as commonly done~\cite{Ott2002,pino2019from}, we consider only $N\gg 1$, drop the limit operation in the definition of $D_q$ and define
\begin{gather}
        D_q = \log_N({\cal{I}}_{q} )/(1-q),
\end{gather}
where we disregard the constant $\mathcal{C}$. Note, that for sufficiently large $N$ and the fractal dimension for \(q=1\) is related to the participation entropy,
\begin{gather}
	{\cal{S}} \simeq - D_1 \log(N)+\log_N{\mathcal{C}} 
\end{gather}
In Figs.~\ref{fig:fractal_dimensions} and~\ref{fig:PNs} we show for all three WD ensembles the participation entropy and participation numbers for dimensions $N=2^n,\ n=9,10,11$ together with $D_1$ and $D_2$, respectively. 
For $D_1$ and $D_2$ we also show the analytical result $D_q= 1, 2-\gamma, 0$ for $\gamma < 1$, $1 < \gamma < 2$ and $\gamma > 2$, respectively, derived in Ref.~\onlinecite{kravtsov2015random} for the transitions from Poisson to GOE and GUE.
We also plot it for the transition to GSE. Deviations between the analytical curve and numerical results are of the same size for the unitary and symplectic universality classes for $1\leq\gamma\lesssim 2$.
However, for $\gamma\gtrsim 2$ $D_1$, $D_2$ and ${\cal{S}}$ approach a non-zero value and, accordingly, ${\cal{I}}_2$ is less than unity for the symplectic case.
Note that, due to Kramer's degeneracy the dimension is effectively one half of that of the other two cases.
More importantly, any linear combination of the associated eigenvectors are eigenvectors of $\hat H^{\rm gRP}$ in~\eqref{eq:HgRP}, so that the occupation probabilities are spread over two eigenstates.
This explains, why for these measures the values expected for complete localization are not yet attained for the highest considered dimension and $q$ value. 

Similar to the observations made for the ratios of adjacent spacings in~\reffig{fig:Ratios_transitions}, the ergodic and Anderson transitions are identified in the corresponding derivatives with respect to $\gamma$, $|D_1^{\prime}|$ and $|D_2^{\prime}|$, as the values of $\gamma$, where the curves for different $N$ cross. These values agree well with the predicted values $\gamma_E=1$ and $\gamma_A=2$, respectively~\cite{pino2019from}. Note, that the positions of the maxima of $|D_1^{\prime}|$ and $|D_2^{\prime}|$, that is of the inflection points of $D_1$ and $D_2$, as function of $\gamma$ are at the value $\gamma\simeq 1.5$, where deviations from the WD ensembles and drastic changes set in for the short-range correlations (see Figs.~\ref{fig:NNDGSE} and~\ref{fig:Ratios}) and long-range correlations (see Figs.~\ref{fig:PowerSp} and~\ref{fig:Kmin}), respectively.

\subsection{Kullback-Leibler divergence}

The Kullback-Leibler (KL) divergence~\cite{kullback1951on} or relative entropy is commonly used as a measure to compare two probability densities, exhibiting nonzero values when they differ, and values close to zero when they are similar.
In our case the probability density of interest is that of the eigenstate occupations $|\psi_{\mu}(i)|^{2}$.
The KL divergences that we study are defined as~\cite{pino2019from, khaymovich2020fragile}
\begin{gather}
    {\cal{K}}_A = \langle \sum_i |\psi_{\mu}(i) |^{2} \log(\frac{|\psi_{\mu}(i)|^{2}}{|\psi_{\mu+1}(i)|^{2}}) \rangle
	\label{KL1}\\
    {\cal{K}}_E = \langle \sum_i |\psi_{\mu}(i) |^{2} \log(\frac{|\psi_{\mu}(i)|^2}{|\tilde{\psi}_{\mu^{\prime}}(i)|^{2}}) \rangle.
	\label{KL2}
\end{gather}
Here, \({\cal{K}}_A\) compares the occupation probability density of two eigenstates corresponding to nearest-neighbor eigenvalues within the same disorder realization and, accordingly, provides an appropriate measure to determine the Anderson localization phase transition, whereas \({\cal{K}}_E \) compares the distributions of two eigenstates from different realizations and yields a suitable indicator of the ergodic phase transition. We would like to stress, that especially for the Anderson transition in the GSE gRP model, it was crucial to use for the analysis of KL divergences only the eigenvectors of one of the pairs of degenerate eigenvalues, e.g., only those with an odd index $\mu$ as we did. Considering all eigenvectors, or linear combinations of those corresponding to the degenerate eigenvalues, doesn't yield meaningful results, as may be expected from their definitions~\eqref{KL1} and~\eqref{KL2}.   

\begin{figure}
    \centering
    \includegraphics[width=0.7\columnwidth]{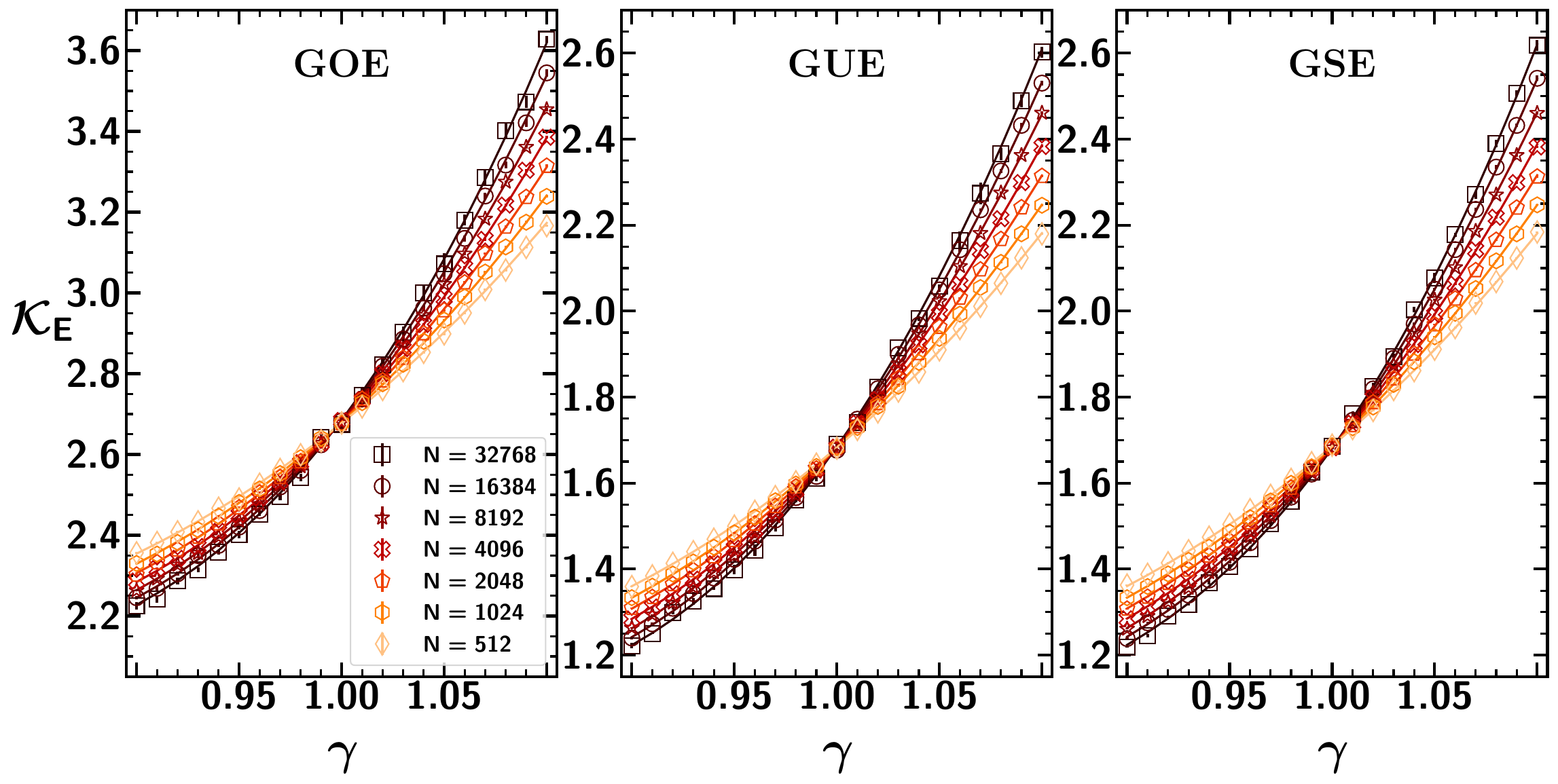}
    \caption{
        The KL divergence ${\cal{K}}_E$ near the ergodic transition for all three WD symmetry classes. The standard errors of mean are shown and are smaller than the symbol size.
        The lines are the best fits obtained using the minimization of the \(\chi^2\) statistics with $n_1 = 3, m_1 = 2, n_2 = m_2 = 0$, see also Table~\ref{table:ergodic}.}
    \label{fig:KL_divergences_ergodic}
\end{figure}

\begin{figure}
    \centering
    \includegraphics[width=0.7\columnwidth]{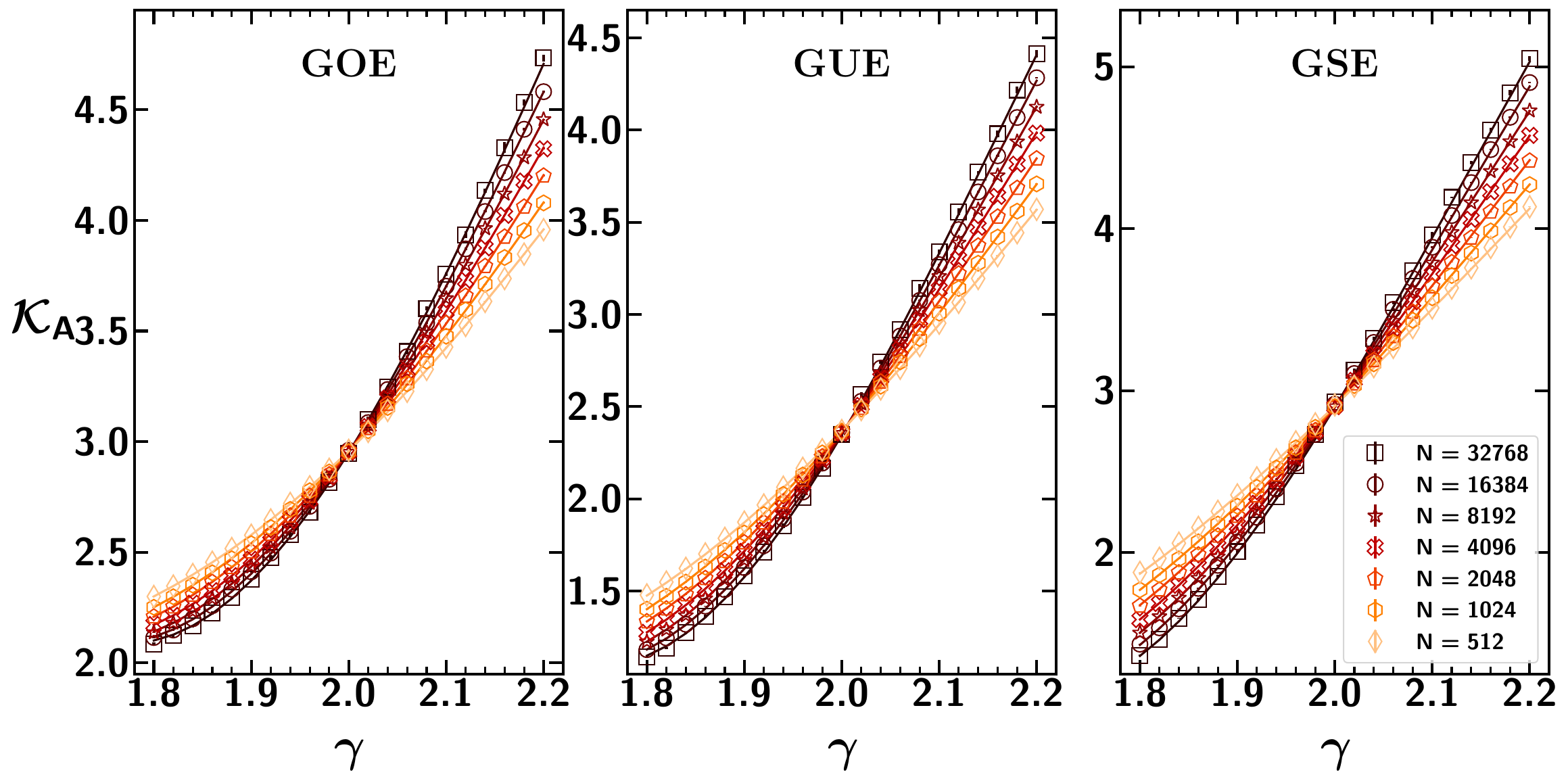}
    \caption{
        The KL divergence ${\cal{K}}_A$ near the Anderson transition for all three WD symmetry classes. 
	The standard errors of mean are shown and are smaller than the symbol size.
        The lines are the best fits obtained using the minimization of the \(\chi^2\) statistics with $n_1 = 3, m_1 = 2, n_2 = m_2 = 0$, see also Table~\ref{table:Anderson}.
    }
    \label{fig:KL_divergences_Anderson}
\end{figure}

To determine the two transition points and the corresponding critical exponents we use the \emph{finite size scaling} (FSS) analysis as described in Ref.~\onlinecite{slevin2014critical}.
The KL divergences are assumed to be given by a scaling law
\begin{gather}
    \label{eq:scaling}
    {\cal{K}}_l = F(\Phi_1, \Phi_2),
\end{gather}
with the scaling variables \(\Phi_1, \Phi_2\), given as
\begin{gather}
    \Phi_j = u_j(w) \bigl[\log(N)\bigr]^{\alpha_j},
\end{gather}
where \(w = (\gamma-\gamma_c)/\gamma_c\) is the reduced parameter of the gRP model and \(\gamma_c\) is the transition point to be determined.
The logarithmic system size dependence in the scaling variables was first used in Ref.~\onlinecite{pino2019from} and further justified in Ref.~\onlinecite{khaymovich2020fragile} and the corresponding scaling exponent are \(\alpha_1\) and \(\alpha_2\).
For the \emph{relevant} variables , \(\Phi_1\), and the \emph{irrelevant} ones, \(\Phi_2\), the scaling exponents are given in terms of $\nu$ and $y$, respectively, with \(\alpha_1 = 1/\nu\) and \(\alpha_2 = y\).
In the vicinity of the transition the functions \(u_i\) are Taylor expanded 
\begin{gather}
    \label{eq:u_i}
    u_i(w) = \sum_{j=0}^{m_j} b_{i,j} w^j,
\end{gather}
where the cutoff integer \(m_j\) is a parameter of the FSS and \(b_{i,j}\) are additional fitting coefficients. Similarly the scaling function \(F\) is Taylor expanded in powers of the scaling variables
\begin{gather}
    \label{eq:scaling_Taylor}
    F(\Phi_1, \Phi_2) = \sum_{j_1=0}^{n_1} \sum_{j_2=0}^{n_2} a_{j_1,j_2} \Phi_1^{j_1} \Phi_2^{j_2},
\end{gather}
with fitting coefficients \(a_{j_1,j_2}\). 
To avoid disambiguity we set \(a_{1,0} = a_{0,1} = 1\) and \(b_{1,0} = 0\). 
Then the total number of free parameters is \(N_P = 2 + m_1 + m_2 + (n_1 + 1) (n_2 + 1)\) and in order to determine them we minimize the \(\chi^2\) statistics, given as
\begin{gather}
    \label{eq:chi_square}
    \chi^2 = \sum_{l=1}^{N_D} \frac{(F_l - {\cal{K}}_l)^2}{\sigma_l^2}.
\end{gather}
In the numerical analysis, the \({\cal{K}}_l\) are obtained for matrix sizes \(N = 512 - 32768\):
(i) for \({\cal{K}}_E\) by extracting a single state closest to the energy 0 and averaging over multiple realization of the gRP matrices;
(ii) for \({\cal{K}}_A\) by averaging KL divergence values at different parameters \(\gamma\) over $\pm 10\%$ of the $N$ eigenstates around the band center, which is at energy zero, and then averaging the resulting mean values over multiple realizations of gRP matrices.
We find that the values \({\cal{K}}_E\) for nearby states within the same random matrix realization are highly correlated.
Thus we take a single state closest to the band center from each realization.
The associated standard errors of the mean yield \(\sigma_l\) entering~\eqref{eq:chi_square}.
The total number of data points is \(N_D\).
The results are shown in Figs.~\ref{fig:KL_divergences_ergodic} and~\ref{fig:KL_divergences_Anderson} (symbols) together with the curves best fitting them (solid lines of corresponding color).
For the minimization of Eq.~\eqref{eq:chi_square} we use the Levenberg–Marquardt (LM) algorithm as implemented in the LMFIT package in Python.
We have also performed Monte-Carlo (MC) simulations of the synthetic data sets (as described in Ref.~\onlinecite{press2002numerical}, Chapter 15.6) using \(300\) to \(1000\) sets. 

The results of the fitting procedure are summarized in Tables~\ref{table:ergodic} and~\ref{table:Anderson} for the ergodic and Anderson transitions, respectively.
For the ergodic transition the fitting without the irrelevant scaling variable gives consistent results with very high precision.
For the GUE and GSE the irrelevant scaling variable is needed at the Anderson transition, as indicated by high values of \(\chi^2\) if it is not used.
We find that the stability of fitting is better for the ergodic transition.
We suspect that this is due to the underestimation of \(\sigma_l\) in Eq.~\eqref{eq:chi_square}, which might originate from the correlations between the fractal states within the same disorder realization as reported in Ref.~\onlinecite{kravtsov2015random} and is known to occur for the Anderson transition in the 3D Anderson model~\cite{rodriguez2010critical,rodriguez2011multifractal}.
Errors are largest for the GSE case, which we again attribute to Kramer's degeneracy, implicating halving of the dimension and the superposition of the associated pairs of eigenmodes.
Nevertheless the fitting results as given in Tables~\ref{table:ergodic} and~\ref{table:Anderson} agree very well for the different settings and clearly confirm  up to the errors the prediction $\nu_A=\nu_E=1$, thus showing the superuniversality of the transitions in the gRP models.

\begin{table}
    \begin{center}
        \begin{tabular}{c|c|c|c|c|c|c} 
             class & $n_1, m_1, n_2, m_2$ &  $\gamma_E$ & $\nu_E$ & $\chi^2$ & $N_{D}$ & MC sets\\
             \hline % GOE 
             & $3, 2, 0, 0$ & $0.9971 \pm 0.0006$ & $1.0243 \pm 0.0091$ & 147.4 & 147 & \\
             GOE & $3, 2, 0, 0$ & $0.9971 \pm 0.0007$ & $1.0244 \pm 0.0090$ &  & & 1000 \\
             %GOE & $4, 2, 0, 0$ & $0.9972 \pm 0.0006$ & $1.0245 \pm 0.0092$ & 146.7 & 147 & \\ 
             %& $4, 2, 0, 0$ & $0.9972 \pm 0.0006$ & $1.0246 \pm 0.0088$ & & & 1000 \\ 
              & $2, 4, 0, 0$ & $0.9965 \pm 0.0006$ & $1.0037 \pm 0.0078$ & 154.0 & 147 & \\
             & $2, 4, 0, 0$ & $0.9965 \pm 0.0005$ & $1.0035 \pm 0.0076$ & & & 1000 \\

             \hline % GUE 
             & $3, 2, 0, 0$ & $0.99937 \pm 0.00036$ & $1.0032 \pm 0.0051$ & 189.8 & 147 & \\
             GUE & $3, 2, 0, 0$ & $0.99937 \pm 0.00031$ & $1.0032 \pm 0.0044$ &  & & 1000 \\
             %GUE & $4, 2, 0, 0$ & $0.99965 \pm 0.00038$ & $1.0041 \pm 0.0052$ & 185.2 & 147 & \\ 
             %& $4, 2, 0, 0$ & $0.99967 \pm 0.00034$ & $1.0042 \pm 0.0044$ & & & 1000 \\ 
             & $2, 4, 0, 0$ & $0.99924 \pm 0.00034$ & $0.9997 \pm 0.0053$ & 189.4 & 147 & \\
             & $2, 4, 0, 0$ & $0.99924 \pm 0.00028$ & $0.9998 \pm 0.0045$ & & & 1000 \\

             \hline % GSE 
             & $3, 2, 0, 0$ & $1.00192 \pm 0.00040$ & $1.0046 \pm 0.0053$ & 203.6 & 147 & \\
             GSE & $3, 2, 0, 0$ & $1.00192 \pm 0.00045$ & $1.0047 \pm 0.0044$ &  & & 1000 \\
             %GSE & $4, 2, 0, 0$ & $1.00193 \pm 0.00038$ & $1.0031 \pm 0.0051$ & 189.2 & 147 & \\ 
             %& $4, 2, 0, 0$ & $1.00193 \pm 0.00033$ & $1.0031 \pm 0.0046$ &  & & 1000 \\
             %\hline
             & $2, 4, 0, 0$ & $1.00091 \pm 0.00035$ & $0.9877 \pm 0.0054$ & 213.8 & 147 & \\
             & $2, 4, 0, 0$ & $1.00092 \pm 0.00028$ & $0.9876 \pm 0.0043$ & & & 1000 \\
        \end{tabular}
    \end{center}
    \caption{
        The FSS analysis for the ergodic transition in the gRP models for all three WD symmetry classes.
        The parameter range \(\gamma \in [0.9, 1.1]\) was used.
        The rows where \(\chi^2\) and the number of MC sets are given show the results from the LM algorithm and the MC simulations, respectively.
        A single state closest to the band center was used per realization.
    }
    \label{table:ergodic}
\end{table}

\begin{table}
    \begin{center}
        \begin{tabular}{c|c|c|c|c|c|c|c}
             class & $n_1, m_1, n_2, m_2$ &  $\gamma_A$ & $\nu_A$ & y & $\chi^2$ & $N_{D}$ & MC sets\\
             \hline % GOE 
             %& $3, 2, 0, 0$ & $2.0025 \pm 0.0004$ & $0.9964 \pm 0.0032$ & & 211.6 & 147 & \\
             %GOE & $3, 2, 0, 0$ & $2.0025 \pm 0.0003$ & $0.9966 \pm 0.0026$ & & & & 1000 \\
             GOE & $5, 2, 0, 0$ & $2.0036 \pm 0.0004$ & $0.9984 \pm 0.0029$ & & 159.0 & 147 & \\
             & $5, 2, 0, 0$ & $2.0035 \pm 0.0004$ & $0.9984 \pm 0.0025$ & & & & 1000 \\

             \hline % GUE 
             %& $3, 2, 0, 0$ & $2.0068 \pm 0.0005$ & $0.9930 \pm 0.0033$ & & 404 & 147 & \\
             %GUE & $3, 2, 0, 0$ & $2.0068 \pm 0.0003$ & $0.9930 \pm 0.0020$ & & & & 1000 \\
             GUE & $3, 2, 1, 1$ & $1.9989 \pm 0.0021$ & $1.0061 \pm 0.0068$ & $-5.7 \pm 2.0$ & $158.4$ & 147 & \\
             & $3, 2, 1, 1$ & $1.9986 \pm 0.0027$ & $1.0062 \pm 0.0068$ & $-6.4 \pm 2.6$ & & & 300 \\

             \hline % GSE 
             %& $3, 2, 0, 0$ & $2.0020 \pm 0.0007$ & $0.9461 \pm 0.0047$ & & 1414 & 147 & \\
             %GSE & $3, 2, 0, 0$ & $2.0020 \pm 0.0002$ & $0.9461 \pm 0.0014$ & & & & 1000 \\
             GSE & $3, 2, 1, 1$ & $1.965 \pm 0.011$ & $0.972 \pm 0.035$ & $-1.3 \pm 0.7$ & $203.1$ & 147 & \\
             & $3, 2, 1, 1$ & $1.965 \pm 0.008$ & $0.970 \pm 0.020$ & $-1.4 \pm 0.6$ & & & 300 \\
        \end{tabular}
    \end{center}
    \caption{
        The FSS analysis for the Anderson transition in the gRP models for all three WD symmetry classes.
        The parameter range \(\gamma \in [1.8, 2.2]\) was used.
        The rows where \(\chi^2\) and the number of MC sets are given show the results from the LM algorithm and the MC simulations, respectively.
        In total twenty percent of the states around the band center were used for each realization.
        Note that for $n_2=m_2=0$ no irrelevant scaling variables are used and thus $y$ equals zero.
    }
    \label{table:Anderson}
\end{table}

\subsection{Fidelity susceptibility}

In Ref.~\onlinecite{Pandey2020} a new measure has been introduced which depends on the eigenvalues \emph{and} eigenvectors of a parameter-dependent Hamiltonian, and probes ergodicity in terms of the adiabatic deformation of these eigenstates.
The Hamiltonian is obtained by perturbing the gRP Hamiltonian $\hat H^{\rm gRP}(\gamma)$ in~\eqref{eq:HgRP} with a parameter-dependent perturbation, $\hat H(\epsilon)=\hat H+\epsilon\hat V$.
The adiabatic gauge potential (AGP) which generates the adiabatic deformation of the eigenstates $\{E_l(\epsilon),\vert l(\epsilon\rangle\}$ of $\hat H(\epsilon)$, obtained from the eigenvalue equation $\hat H(\epsilon)\vert l(\epsilon)\rangle =E_l(\epsilon)\vert l(\epsilon)\rangle$, is defined as
\be
    \mathcal{A}_\epsilon\vert l(\epsilon)\rangle =i\partial_\epsilon\vert l(\epsilon)\rangle.
\ee
Differentiation of the eigenvalue equation with respect to $\epsilon$ yields for $\epsilon\to 0$~\cite{Pechukas1983,Haake1991}
\be
\langle m\vert\mathcal{A}_\epsilon\vert l\rangle\vert_{\epsilon =0}= -i\frac{\langle m\vert\partial_\epsilon \hat H(\epsilon)\vert_{\epsilon =0}\vert l\rangle}{E_m-E_l},
\ee
where we introduced the notation $\vert l\rangle \equiv \vert l(\epsilon =0)\rangle$ and $E_l\equiv E_l(\epsilon =0)$.
The Hilbert-Schmidt norm of $\mathcal{A}_\epsilon$ yields the fidelity susceptibility~\cite{Sierant2019,skvortsov2022sensitivity},
\be
    \chi_l =\sum_{m\ne l}\frac{\vert\langle m\vert\partial_\epsilon\hat H(\epsilon)\vert_{\epsilon =0}\vert l\rangle\vert^2}{(E_m-E_l)^2},
\ee
which has been shown~\cite{Sierant2019,Pandey2020,nandy2022delayed,Orlov2023} to be a particularly sensitive measure for ergodicity. This can be expected from its structure. Namely, in the ergodic phase eigenfunctions are fully extended and the eigenvalues repel each other, whereas in the fractal phase eigenfunctions are partially localized~\cite{bogomolny2018eigenfunction} and part of the eigenvalues are nearly degenerate.   

We consider an $\epsilon$-independent potential $\hat V$ with box-distributed random entries, which is diagonal in the representation of the unperturbed gRP Hamiltonian $\hat H^{\rm gRP}(\gamma)$ given in~\eqref{eq:HgRP}, and compute for given $\hat V$ and matrix elements $H_{nm}$ in~\eqref{eq:gRP_variances} the fidelity susceptibility $\chi$ as function of the gRP parameter $\gamma$. Here, we take into account $\pm 10\%$ of the eigenstates $\vert l\rangle$ around the band center for various dimensions $N$. Furthermore, we analyze the logarithm of $\chi_l$, $\zeta=\langle\langle\log(\chi_l)\rangle_l\rangle_{\rm ens}$ to mitigate the accidental small denominators in the definition of \(\chi_l\)~\cite{sels2021dynamical}, with $\langle\cdot\rangle_l$ and $\langle\cdot\rangle_{\rm ens}$ denoting the arithmetic mean over $l$ and ensemble average over numerous random-matrix realizations of $\hat H^{\rm gRP}(\gamma)$ defined in equations~\eqref{eq:HgRP} and~\eqref{eq:gRP_variances}, respectively.

\begin{figure}
    \centering
    \includegraphics[width=0.49\columnwidth]{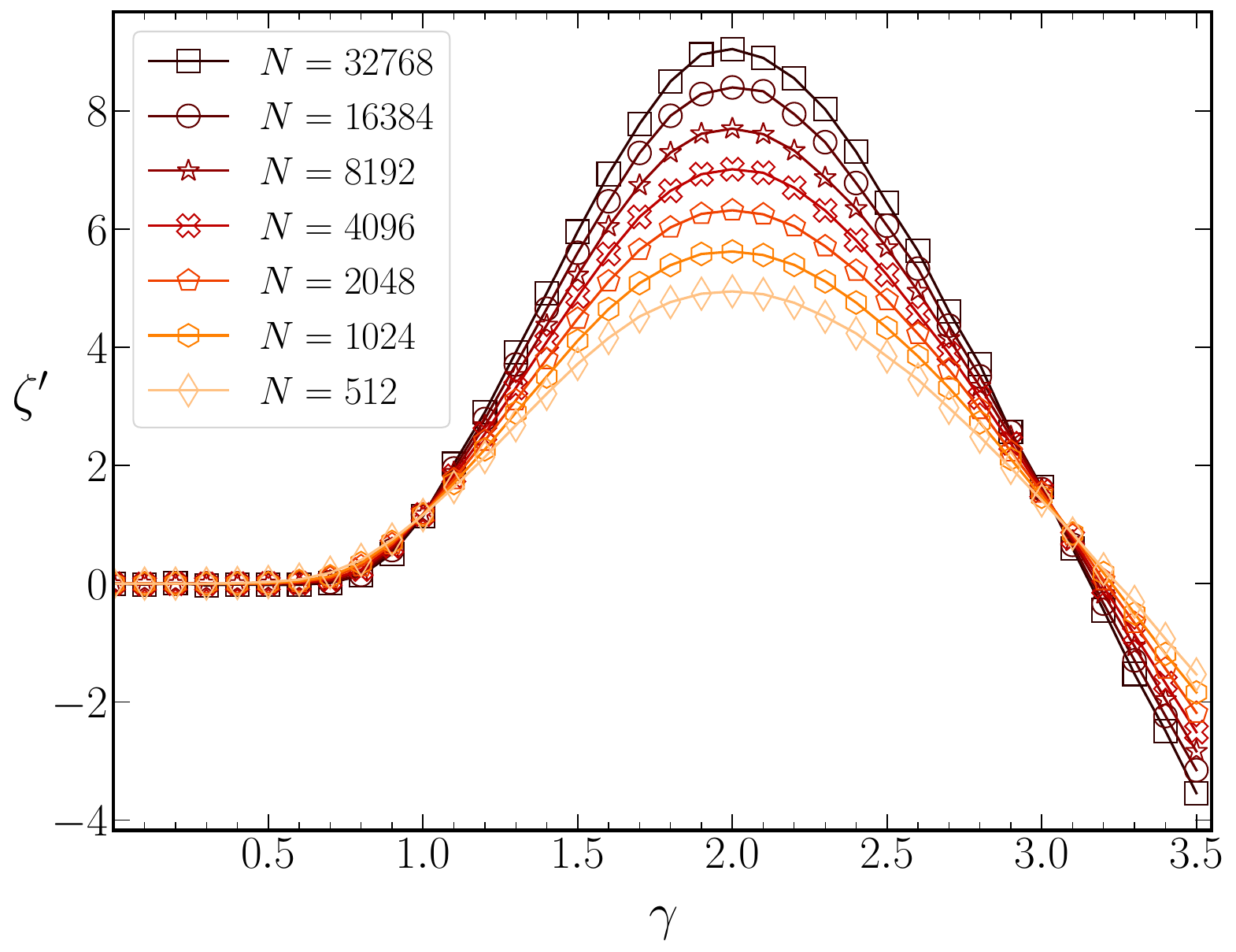}
	\caption{
	    Shifted fidelity susceptibility $\zeta^\prime = \zeta(\gamma) - \zeta(\gamma = 0)$ for the GOE gRP model subject to an $\epsilon$-independent potential $\hat V$ (see main text).
    }
    \label{fig:Fidelity}
\end{figure}

\begin{figure}
    \centering
    \includegraphics[width=0.9\columnwidth]{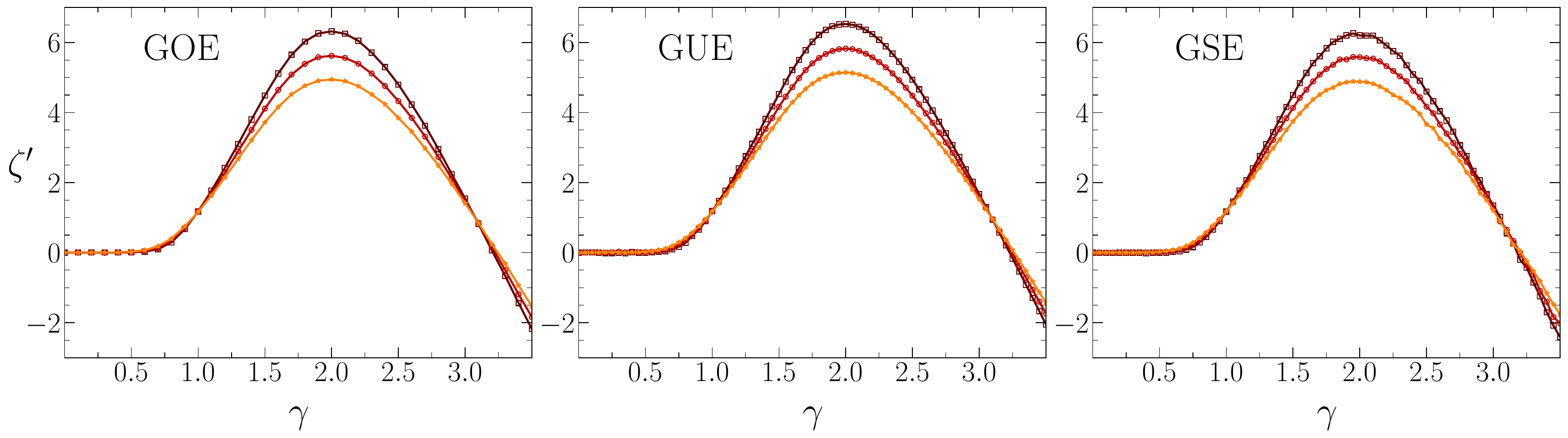}
	\caption{
     Shifted fidelity susceptibility $\zeta^\prime = \zeta(\gamma) - \zeta(\gamma = 0)$ for the three WD gRP models, for system sizes \(N = 512,1024, 2048\).
    }
    \label{fig:Fidelity1}
\end{figure}

\begin{figure}
    \centering
    \includegraphics[width=0.32\columnwidth]{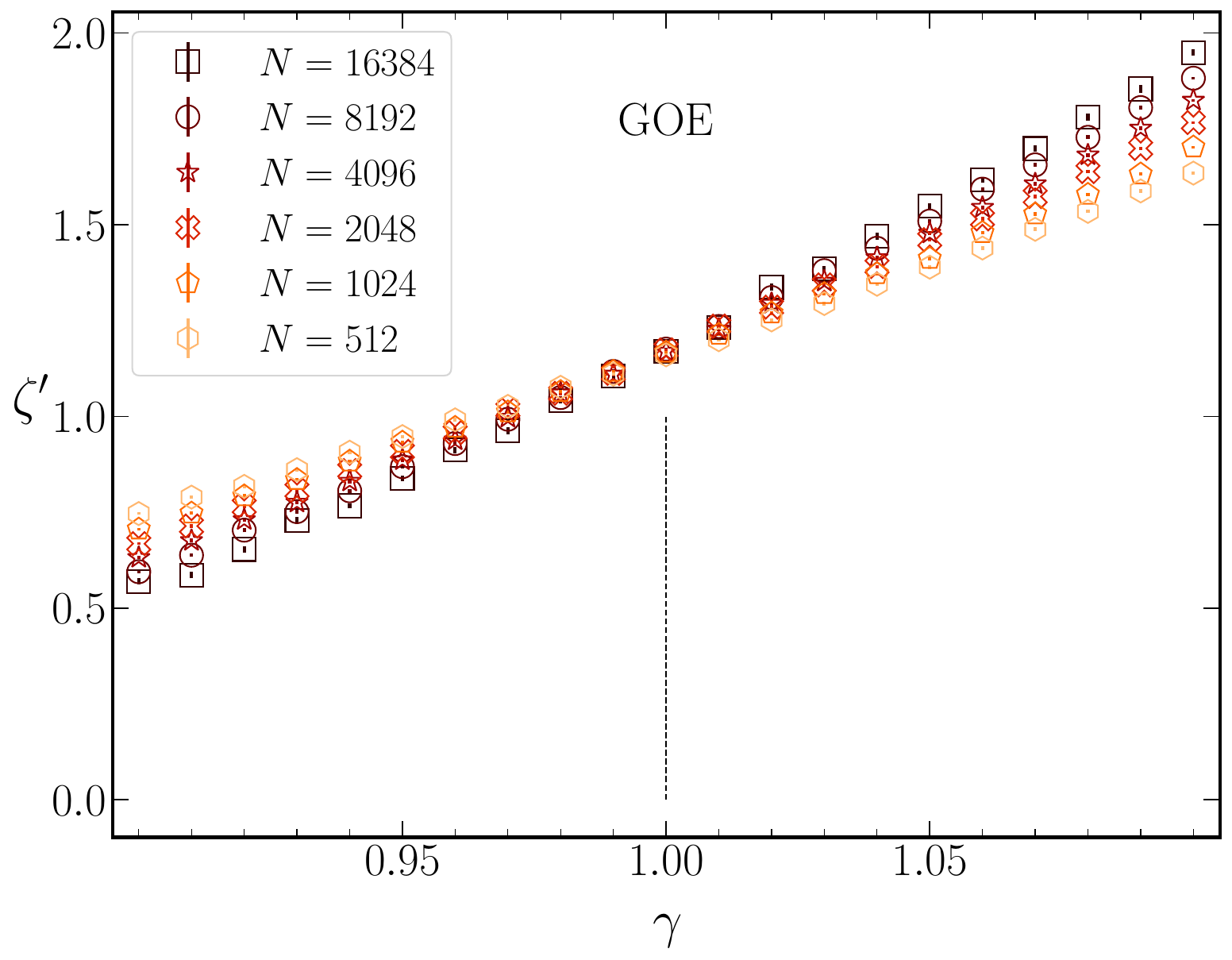}
    \includegraphics[width=0.32\columnwidth]{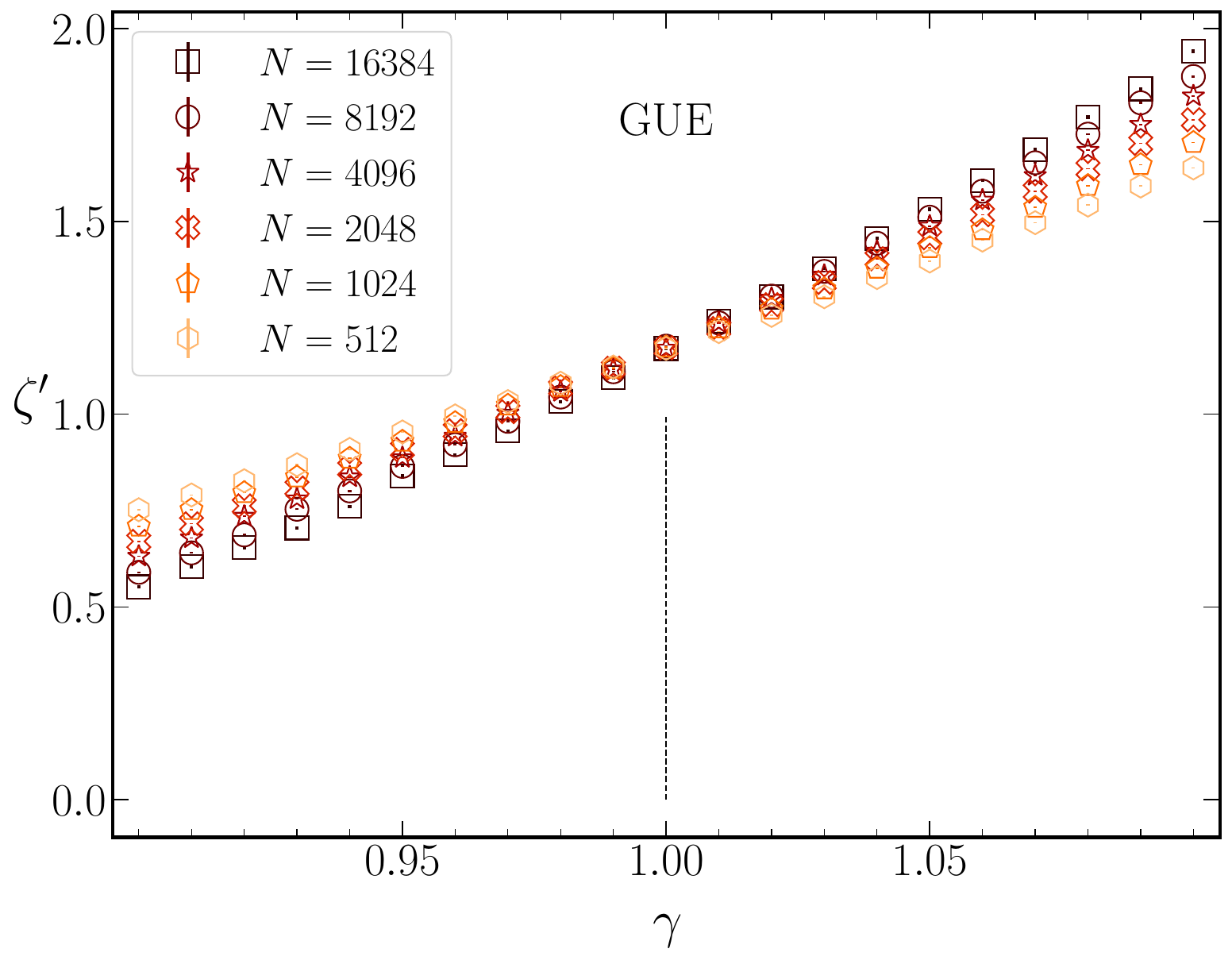}
    \includegraphics[width=0.32\columnwidth]{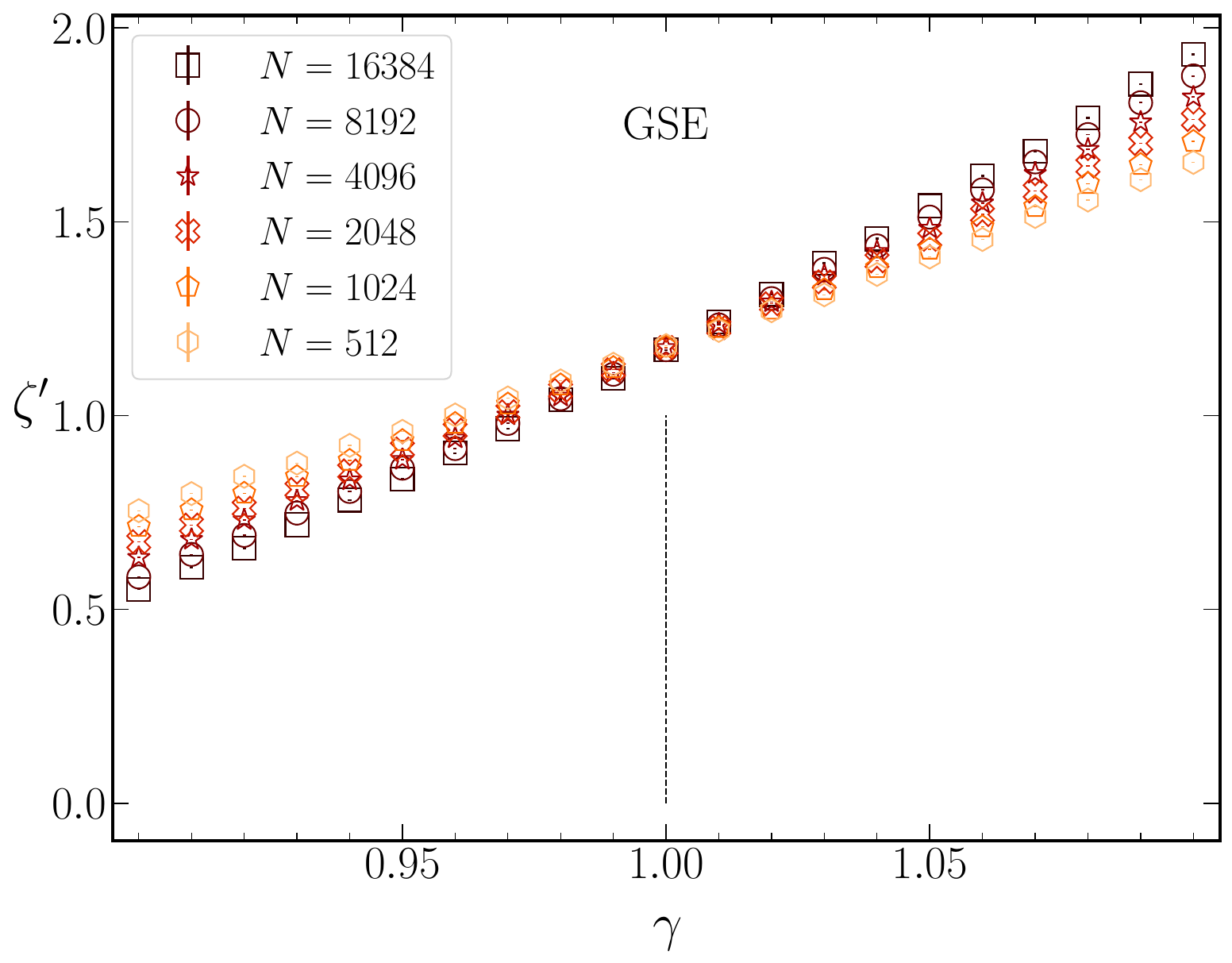}
	\caption{
	    Shifted fidelity susceptibility $\zeta^\prime = \zeta(\gamma) - \zeta(\gamma = 0)$ in the vicinity of the ergodic transition for the three WD gRP models subject to an $\epsilon$-independent potential $\hat V$ (see main text). The standard errors of mean are shown and are smaller than the symbol size.
    }
    \label{fig:Fidelity2}
\end{figure}

\begin{figure}
    \centering
    \includegraphics[width=0.7\columnwidth]{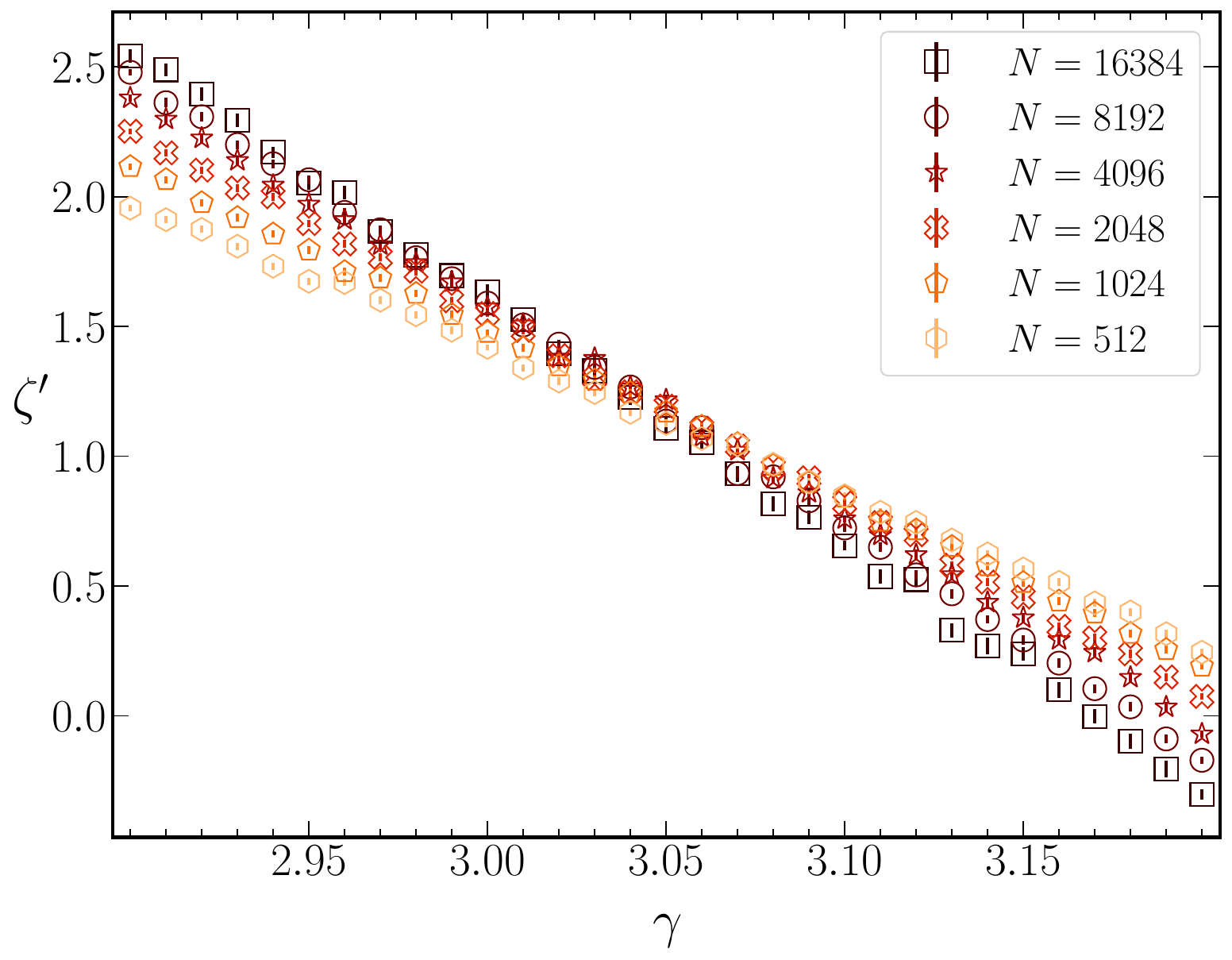}
    \caption{
	    Shifted fidelity susceptibility $\zeta^\prime = \zeta(\gamma) - \zeta(\gamma = 0)$ in the vicinity of $\gamma = 3$ for the GOE gRP model subject to an $\epsilon$-independent potential $\hat V$ (see main text). The standard errors of mean are shown and are smaller than the symbol size.
    }
    \label{fig:Fidelity3}
\end{figure}

In~\reffig{fig:Fidelity} we show the shifted fidelity susceptibility for the GOE gRP for 7 different dimensions $N=2^n$ with $n=9-15$. The number of realizations used was at least 2000, 1000, 500, 200, 100, 20, 3 for system sizes from 512 to 32768, respectively. In~\reffig{fig:Fidelity1} we compare these results with those of the other two WD ensembles for $n=9-11$, where less realizations were used for the GUE (500, 250, 50) and GSE (1000, 500, 100) classes. The potential $\hat V$ changes with the system size and universality class so that we rather compare $\zeta^\prime = \zeta(\gamma) - \zeta(\gamma = 0)$. For all cases the curves for different dimensions $N$ cross each other at $\gamma =1$, indicating that there the transition from the ergodic to the fractal phase takes place. This is illustrated in~\reffig{fig:Fidelity2} showing a zoom into the region around $\gamma=1$. For $\gamma >1$ the curves increase until they reach a maximum at $\gamma =2$, that is, at the value of $\gamma$ where the Anderson transition takes place. Beyond that value $\chi_l$ decreases to zero with increasing $\gamma$ for all WDs, indicating localization. Note, that the curves cross each other again at $\gamma\simeq 3$, however, in distinction to that at $\gamma =1$, there the crossings are spread over a nonzero range of $\gamma$ values and the curves do not change their behavior, that is, continue to decrease with the same slope. We show as example a zoom around $\gamma=3$ for the GOE gRP model in ~\reffig{fig:Fidelity3}. Assuming scaling of $\zeta^\prime$, similarly to the FSS analysis of KL divergences, we could not find an acceptable fit either with or without irrelevant variable for the data around $\gamma\simeq 3$. In contrast, for the ergodic transition, e.g., for the GOE gRP model in the vicinity of $\gamma =1$, as shown in left panel of~\reffig{fig:Fidelity} we obtain $\gamma_E = 0.9933 \pm 0.0009$, $\nu_E = 0.994 \pm 0.011$, with $N_D = 126$ and $\chi^2 = 124.8$ for $n_1, m_1, n_2, m_2 = 3,2,0,0$, which is in good agreement with the results obtained via FSS of the KL divergence, as given in Table~\ref{table:ergodic}.

\section*{Conclusions} 

We analyzed spectral properties and properties of the eigenvectors of random matrices from the gRP model for all the WD ensembles.
We extend the known results for the transition from Poisson to GOE and GUE to the symplectic universality class, i.e., the GSE.
Furthermore, employing high-dimensional random matrices (N=65536), we validate for the transition from Poisson statistics to GUE the existing analytical results for the long-range correlations~\cite{Kunz1998,Frahm1998} and an analytical expression for the ratio distribution derived in~\appsec{RatioAnalyt}.
We also compare for all three WD ensembles the numerically obtained nearest-neighbor spacing distributions to Wigner-surmise like results~\cite{Lenz1991,Schierenberg2012}.
We analyze the transition from chaoticity to integrability in terms of the position of the maximum of the nearest-neighbor spacing distribution and the average ratios for short-range correlations. For long-range correlations, we employ the distance of the number variance from WD statistics, the asymptotic power-law behavior of the power spectrum and the position of the minimum of the spectral form factor to identify the transitions. We find that deviations from WD statistics are observed in the short-range correlations only above $\gamma\simeq 1.5$, whereas they set in immediately beyond $\gamma\simeq 1$ for the long-range correlations, implying that correlations in the eigenvalue spectra need to be probed over several mean spacings to observe changes in the spectral properties when introducing a small perturbation that induces regular behavior or partial localization into a fully chaotic Hamiltonian~\cite{Zhang2023b}. Both the measures for short- and long-range correlations approach the corresponding result for Poissonian random numbers above $\gamma\simeq 2$ for all three WD ensembles.     

To obtain information on the properties of the eigenvectors of the gRP Hamiltonian~\eqref{eq:HgRP} and accurately determine the ergodic and Anderson transition and identify the fractal phase, we analyzed fractal dimensions, including the participation entropy and participation number, and KL divergences.
For the symplectic case, due to Kramer's degeneracy which implies that the occupation probability spreads over the pairs of eigenmodes associated with the degenerate eigenvalues, we find for the localized phase small deviations from the predicted values.
The Anderson transition is seen in the generalized participation numbers, in addition the ergodic one in the derivative of the fractal dimensions $D_1$ and $D_2$.
Both transitions are detected using KL divergences of eigenfunction occupations.
A finite-size scaling analysis shows that all these measures show superuniversality of the transitions in the sense that the values of $\gamma_E$ and $\gamma_A$ are identical for all three WD ensembles, with the critical exponent being consistent with the value $\nu = 1$.
Similarly, the fidelity susceptibility detects the ergodic transition and exhibits a maximum at the Anderson transition.
When looking at the curves for different dimensions $N$, shown in Figs.~\ref{fig:Fidelity} and~\ref{fig:Fidelity1}, one might conclude that there is a further transition at $\gamma=3$.
However, there the curves do not cross at a single point and the curves proceed through these crossing points without changing their behavior, that is, continue to decrease with the same slope implying that they do not identify a genuine third transition. Additionally, an attempt of FSS for, e.g., the data resulting from the GOE gRP model, as shown in~\reffig{fig:Fidelity3}, gives an unacceptably high $\chi^2$.

One interesting question for future research is to confirm the ability of fidelity susceptibility to detect non-ergodic transitions by benchmarking it in other models with non-ergodic phases, and to investigate if (and how) it can discriminate between single fractal and multifractal regions in the parameter space. Another one is to find spectral measures for long-range correlations in addition to those considered in the present work that might provide useful indicators to detect transitions, e.g. a non-ergodic one.

\section*{Acknowledgments} 
\label{sec:acknowledgments}

We acknowledge financial support from the Institute for Basic Science (IBS) in the Republic of Korea through the project IBS-R024-D1.
D.R. thanks FAPESP, for the ICTP-SAIFR grant 2021/14335-0 and the Young Investigator grant 2023/11832-9, and the Simons Foundation for the Targeted Grant to ICTP-SAIFR.
We are indebted to Boris Altshuler, Henning Schomerus, Tomi Ohtsuki, and Keith Slevin for fruitful discussions.

\bibliography{mbl,general,local,BD}

%\printbibliography
\begin{appendix}

\renewcommand{\theequation}{A\arabic{equation}}
\renewcommand{\thefigure}{A\arabic{figure}}

\setcounter{figure}{0}

\section{Analytical Results for the transition $\beta =0\to\beta =2$ from Poisson to GUE}
\label{Anal}

\subsection{The joint-probablity density of the eigenvalues for the transition from Poisson to GUE}
\label{RatAn}

The derivation of the joint-probability distribution of the eigenvalues $\{e_i\}$, $P(\{e_i\};\gamma)$ of random matrices from the RP model
\be
    \label{RPHUE}
    \hat H^{0\to 2}(\Lambda) =\hat H_{0}+\Lambda\hat H^{(\beta =2)},\, \Lambda =\frac{\lambda}{\sqrt{1+\lambda^2}},
\ee
where $\hat H^{(0)}$ denotes a random diagonal matrix and $\hat H^{\beta}$ a random matrix from the GUE with $\beta =2$,
with Gaussian distributed matrix elements with variances
\be
    \label{HRP_variances}
    \left\langle\left(H^{2}_{nm}\right)^2\right\rangle = \sigma^2 = 1.
\ee
involves an integral over the unitary matrices diagonalizing it, which is the Harish-Chandra Itzykson-Zuber integral~\cite{Chandra1958,Pandey1983,Lenz1992}, yielding
\be
    \label{PDF}
    P(\{e_i\})=\left(\frac{1}{\sqrt{2\pi\Lambda^2}}\right)^N\int d\left[\boldsymbol{E}\right]P^{(0)}(\boldsymbol{E})\exp\left[-\frac{1}{2\Lambda^2}\sum_i(e_i-E_i)^2\right]\frac{\prod_{n<m}(e_n-e_m)^2}{\prod_{n<m}(E_n-E_m)^2}.
\ee
The probablity density $P^{(0}(\boldsymbol{E})$ of the matrix elements $E_i$ of $\hat H^{0}$ is arbitrary, however for the numerical simulations we chose them Gaussian distributed with variances
\be
    \label{H0_variance}
    \tilde\sigma^2 = \left\langle\left(H^{0}_{nn}\right)^2\right\rangle = \sigma^2\left(1-\Lambda^2\right),
\ee
\be
    \label{P0}
    P^{(0}(\boldsymbol{E})=\prod_{i=1}^N\frac{e^{-E_i^2/2\tilde\sigma^2}}{\sqrt{2\pi\tilde\sigma^2}}.
\ee

\subsection{Derivation of the ratio distribution for the transition from Poisson to GUE\label{RatioAnalyt}}

Starting from~\eqref{PDF}, we derive a Wigner-surmise like expression for the distribution $P^{0\to 2}(r)$, abbreviated as $P(r)$ in the following, of the ratios $r_i=\frac{e_{i+1}-e_i}{e_i-e_{i-1}}$ of the sorted eigenvales $e_i\leq e_{i+1},i=1,\dots\, ,N$ by restricting to $N=3$,
\be
    P(r)=3!\int_{-\infty}^{\infty}de_2\int_{-\infty}^{e_2}de_1\int_{e_2}^{\infty}de_3P(e_1,e_2,e_3)\delta\left(r-\frac{e_3-e_2}{e_2-e_1}\right).
\ee
We perform a variable transformation $[e_1,e_2,e_3]\rightarrow [\tilde e_1=e_1/(\sqrt{2}\Lambda) ,\tilde e_2 =e_2/(\sqrt{2}\Lambda) ,\tilde e_3=e_3/(\sqrt{2}\Lambda)]$, similarly $[E_1,E_2,E_3]\rightarrow [E_1/(\sqrt{2}\Lambda) ,E_2/(\sqrt{2}\Lambda) ,E_3/(\sqrt{2}\Lambda)]$, and then $[\tilde e_1,\tilde e_2,\tilde e_3]\longrightarrow [x=(\tilde e_2-\tilde e_1),z,y=(\tilde e_3-\tilde e_2)]$ yielding
\begin{align}
    \label{PR1}
    P(r) =&3!\left(\sqrt{\frac{2\Lambda^2}{\pi}}\right)^3\int_{-\infty}^{\infty}dE_1\int_{-\infty}^{\infty}dE_2\int_{-\infty}^{\infty}dE_3\frac{P_3^{(0)}(\sqrt{2}\Lambda\boldsymbol{E})}{\Delta(\boldsymbol{E})}e^{-(E_1^2+E_2^2+E_3^2)} \\
    \times&\int_0^\infty dx\int_0^\infty dy xy(x+y) \notag
    \exp\left[-\left(x^2+y^2\right)-2\left(xE_1-yE_3\right)\right]\delta\left(r-\frac{y}{x}\right) \\
    \times&\int_0^\infty dz\exp\left[-3z^2+2z\left(x-y+E_1+E_2+E_3\right)\right] \notag,
\end{align}
with $\Delta(\boldsymbol{E})=(E_3-E_2)(E_3-E_1)(E_2-E_1)$.
The integrations over $y$ and $z$ lead to
\begin{align}
    \label{PR2}
    P(r) =&3!\frac{r(r+1)}{\sqrt{3}}\frac{2\Lambda^2}{\pi}\int_{-\infty}^{\infty}dE_1\int_{-\infty}^{\infty}dE_2\int_{-\infty}^{\infty}dE_3\frac{P_3^{(0)}(\sqrt{2}\Lambda\boldsymbol{E})}{\Delta(\boldsymbol{E})}e^{-(E_1^2+E_2^2+E_3^2)}e^{(E_1+E_2+E_3)^2/3}\\
    \times&\int_0^\infty dx x^4 \notag
    \exp\left[-\frac{2}{3}(1+r+r^2)x^2+\frac{2x}{3}\left(-2E_1+E_2+E_3\right)+r\left(2E_3-E_1-E_2\right)\right].
\end{align}
Next, we insert for $P_3^{(0)}(\boldsymbol{E})$~\eqref{P0}, perform a variable change from $[E_1,E_2,E_3]\rightarrow [u=(E_2-E_1),w,v=(E_3-E_2)]$ and introduce the notation $\alpha^2=\frac{\Lambda^2}{1-\Lambda^2}=\lambda^2$, so that the integrals over $\{E_i\}$ become,
\begin{align}
    I=&\int_{-\infty}^{\infty}\frac{du}{u}\int_{-\infty}^{\infty}\frac{dv}{v}\frac{1}{u+v}\exp\left\{-(\alpha^2+1)(u^2+v^2)+\frac{(v-u)^2}{3}+\frac{2x}{3}\left[2u+v+r(2v+u)\right]\right\}\\
    \times &\int_{-\infty}^{\infty}dwe^{\left[-3\alpha^2w^2+2w(u-v)\alpha^2\right]}\left[\frac{1}{\sqrt{2\pi(1-\Lambda^2)}}\right]^3.
\end{align}
Performing the integral over $w$ yields
\begin{align}
    P(r)=&2r(r+1)\frac{\alpha^2}{\pi^2}\int_{-\infty}^{\infty}\frac{du}{u}\int_{-\infty}^{\infty}\frac{dv}{v}\frac{1}{u+v}\exp\left[-\frac{2}{3}(\alpha^2+1)(u^2+v^2+uv)\right]\\
    \times &\int_0^\infty dxx^4\exp\left[-Rx^2+2xF\right],
\end{align}
where we introduced the notations
\be
    R=\frac{2}{3}(1+r+r^2),\, F(u,v)=\frac{1}{3}[2u+v+r(u+2v)].
\ee
Integration over $x$ leaves us with a double integral,
\be
    \label{PR3}
    P(r)=2r(r+1)\frac{\alpha^2}{\pi^2}\int_{-\infty}^{\infty}\frac{du}{u}\int_{-\infty}^{\infty}\frac{dv}{v}\frac{1}{u+v}\exp\left[-\frac{2}{3}(\alpha^2+1)(u^2+v^2+uv)\right]g(u,v)
\ee
with
\be
\label{Guv}
    g(u,v)=\frac{1}{2R}\left(\frac{F}{R}\right)^3+\frac{5}{4R^2}\frac{F}{R}
    +e^{\frac{F^2}{R}}\frac{1}{2}\sqrt{\frac{\pi}{R}}\left[1+\Phi\left(\frac{F}{\sqrt{R}}\right)\right]\left\{\frac{3}{4R^2}+\frac{3}{R}\left(\frac{F}{R}\right)^2+\left(\frac{F}{R}\right)^4\right\}.
\ee
Here, $\Phi(x)$ denotes the error function.
Due to the symmetry properties of the integrand, terms with $g(u,v)=g(-u,-v)$ and $g(u,-v)=g(-u,v)$ cancel each other, so that the first term in the square bracket in~\eqref{Guv} vanishes upon integration.

Before we continue with the integration we consider the limit $\alpha\to 0$.
For this we introduce the variable transformation $[u,v]\to [\tilde u=\alpha u,\tilde v=\alpha v]$, resulting with $\tilde F(\tilde u,\tilde v)=\alpha F(u,v)$ in
\begin{align}
    & P(r)=\frac{2r(r+1)}{\pi^2}\int_{-\infty}^{\infty}\frac{du}{u}\int_{-\infty}^{\infty}\frac{dv}{v}\frac{1}{u+v}\exp\left[-\frac{2}{3}\left(1+\frac{1}{\alpha^2}\right)(u^2+v^2+uv)\right]\\
    & \times\left\{\frac{1}{2R}\left(\frac{\tilde F}{R}\right)^3+\alpha^2\frac{5}{4R^2}\frac{\tilde F}{R}
    +e^{\frac{\tilde F^2}{\alpha^2 R}}\frac{1}{2}\sqrt{\frac{\pi}{R}}\Phi\left(\frac{\tilde F}{\alpha\sqrt{R}}\right)\left[\alpha^3\frac{3}{4R^2}+\frac{3}{R}\alpha\left(\frac{\tilde F}{R}\right)^2+\frac{1}{\alpha}\left(\frac{\tilde F}{R}\right)^4\right]\right\}.
\end{align}
In the limit $\alpha\to 0$ only the last term in the curly bracket remains and we obtain with
\be
    \label{Deltafct}
    \frac{1}{\alpha}\exp\left[-\frac{1}{\alpha^2}\left(\frac{2}{3}(u^2+v^2+uv)-\frac{\tilde F^2}{R}\right)\right]\xrightarrow{\alpha\to 0}\sqrt{\pi}\delta\left(\sqrt{\frac{2}{3}(u^2+v^2+uv)-\frac{\tilde F^2}{R}}\right),
\ee
\begin{align}
    & P(r)\xrightarrow{\alpha\to 0}\sqrt{\pi}\frac{2r(r+1)}{\pi^2}\int_{-\infty}^{\infty}\frac{du}{u}\int_{-\infty}^{\infty}\frac{dv}{v}\frac{1}{u+v}\exp\left[-\frac{2}{3}(u^2+v^2+uv)\right]\\
    & \times\sqrt{3R}\delta\left(v-ur\right)\frac{1}{2}\sqrt{\frac{\pi}{R}}\frac{\tilde F}{\vert\tilde F\vert}\left(\frac{\tilde F}{R}\right)^4\\
    & =\frac{\sqrt{3}}{\pi}\int_{0}^{\infty}du 2ue^{-Ru^2}\\
    & =\frac{\sqrt{3}}{R\pi}=\frac{3\sqrt{3}}{2\pi(1+r+r^2)},
    \label{al0}
\end{align}
which is the Wigner-surmise like result for the ratio distribution of the Gaussian-distributed elements $E_i$ with arbitrary variance $\sigma^2$ of the diagonal matrix $\hat H_0$ in~\eqref{RPHUE}.
It differs from the result for Poissonian random numbers,
\be
    \label{PRPoi}
    P^{\rm Poi}(r)=\frac{1}{(1+r)^2}.
\ee

Next we compute the integral in~\eqref{PR3} for $\alpha > 0$.
In order to get rid of the factor $1/(u+v)$ in~\eqref{PR3}, we use
\be
    \frac{F}{uv(u+v)}=\frac{(2+r)}{v(u+v)}+\frac{(1+2r)}{u(u+v)},
\ee
and define new integration variables $[\tilde u=(u+v),\tilde v=v]$ for the first term and $[\tilde u=u, \tilde v= (u+v)]$ for the second one.
Introducing polar coordinates $[\tilde u=\rho\cos\varphi,\tilde v=\rho\sin\varphi]$ and the notations
\begin{align}
	& F_1(\varphi)=\frac{1}{3\sqrt{R}}\left\vert(2+r)\cos(\varphi/2)+(r-1)\sin(\varphi/2)\right\vert\label{Fi}\\
    & F_2(\varphi)=\frac{1}{3\sqrt{R}}\left\vert(r-1)\cos(\varphi/2)+(1+2r)\sin(\varphi/2)\right\vert\nonumber
\end{align}
leads to
\be
    P(r)=4\frac{r(r+1)}{R^3}\frac{\alpha^2}{\pi^2}\int_{-\pi}^{\pi}\frac{d\varphi}{\sin\varphi}\int_{0}^{\infty}d\rho\exp\left[-\frac{2}{3}(\alpha^2+1)\rho^2\left(1-\frac{\sin\varphi}{2}\right)\right]h(\varphi,\rho)
\ee
with
\begin{align}
    \label{I1}
    h(\varphi,\rho)&=\frac{f(\varphi)}{2}\rho +\frac{5}{4}\frac{1+r}{\rho}\\
    \label{I2}
    &+e^{F_1^2\rho^2}\frac{\sqrt{\pi}}{2}\Phi\left(F_1\rho\right)\left[\frac{3}{4}\frac{1}{F_1}\frac{1}{\rho^2}+3F_1+\left(F_1\right)^3\rho^2\right]\frac{2+r}{3}\\
    \label{I3}
    &+e^{F_2^2\rho^2}\frac{\sqrt{\pi}}{2}\Phi\left(F_2\rho\right)\left[\frac{3}{4}\frac{1}{F_2}\frac{1}{\rho^2}+3F_2+\left(F_2\right)^3\rho^2\right]\frac{1+2r}{3}\,
\end{align}
and
\begin{align}
    f(\varphi)&=\left[\frac{2+r}{3}F_1^2+\frac{1+2r}{3}F_2^2\right]\\
    &=\frac{1}{6R}\left[\left(2r+\frac{1}{2}\right)R+1+\frac{(2+r)(1+2r)(1-r)}{3}\cos\varphi+[(2-r)R-2]\sin\varphi\right]\\
    &:=\tilde a(r)+\tilde b(r)\cos\varphi +\tilde c(r)\sin\varphi\, .
\end{align}
The first integral in~\eqref{I1} equals
\begin{align}
    &\int_{-\pi}^{\pi}\frac{d\varphi}{\sin\varphi}\int_{0}^{\infty}d\rho\rho\exp\left[-\frac{2}{3}(\alpha^2+1)\rho^2\left(1-\frac{\sin\varphi}{2}\right)\right]f(\varphi)\\
    &=\frac{3}{4}\frac{1}{\alpha^2+1}\int_{-\pi}^{\pi}\frac{d\varphi}{\sin\varphi}\frac{f(\varphi)}{1-\frac{\sin\varphi}{2}}\\
    &=\frac{3}{2}\frac{1}{\alpha^2+1}\frac{\pi}{\sqrt{3}}\tilde a(r)+2\tilde c(r)\\
    &=\frac{\pi}{8}\frac{\sqrt{3}}{\alpha^2+1}[3R-2]\,.
\end{align}
For the second one in~\eqref{I1} we obtain
\begin{align}
    &\int_{-\pi}^{\pi}\frac{d\varphi}{\sin\varphi}\int_{0}^{\infty}\frac{d\rho}{\rho}\exp\left[-\frac{2}{3}(\alpha^2+1)\rho^2\left(1-\frac{\sin\varphi}{2}\right)\right]\\
    =&\int_0^{\pi}\frac{d\varphi}{\sin\varphi}\int_{0}^{\infty}\frac{d\rho}{\rho}\exp\left[-\frac{2}{3}(\alpha^2+1)\rho\right]\sinh\left[\frac{2}{3}(\alpha^2+1)\rho\frac{\sin\varphi}{2}\right]\\
    =&\int_0^{\pi/2}\frac{d\varphi}{\sin\varphi}\ln\left[\frac{1+\frac{\sin\varphi}{2}}{1-\frac{\sin\varphi}{2}}\right]\\
    =&\frac{\pi^2}{6}
\end{align}
We introduce the notations
\be
\label{Ai}
    A_i(\varphi)=\frac{2}{3}(\alpha^2+1)\left(1-\frac{\sin\varphi}{2}\right)-F_i^2(\varphi),\, i=1,2,
\ee
that can be brought to the forms
\begin{align}
    &A_1(\varphi)=\frac{1}{3R}\left[\alpha^2R\left(2-\sin\varphi\right)+(1+r)^2\left(\sin(\varphi/2)-\frac{r}{1+r}\cos(\varphi/2)\right)^2\right]\\
    &A_2(\varphi)=\frac{1}{3R}\left[\alpha^2R\left(2-\sin\varphi\right)+\left(\sin(\varphi/2)-(1+r)\cos(\varphi/2)\right)^2\right]\, ,
\end{align}
implying, that $A_i(\varphi)$ vanishes only for $\alpha =0$, namely, for $i=1$ at $\tan(\varphi/2)=\frac{r}{r+1}$ and for $i=2$ at $\tan(\varphi/2)=r+1$, respectively.
Consequently, in the limit $\alpha\to 0$, where only the integrals over the last terms in Eqs.~(\ref{I2}) and~(\ref{I3}) survive, the integrals over $\varphi$ only contribute at these values, and the result~\eqref{al0} is recovered.

\begin{figure}
    \centering
    \includegraphics[width=0.6\columnwidth]{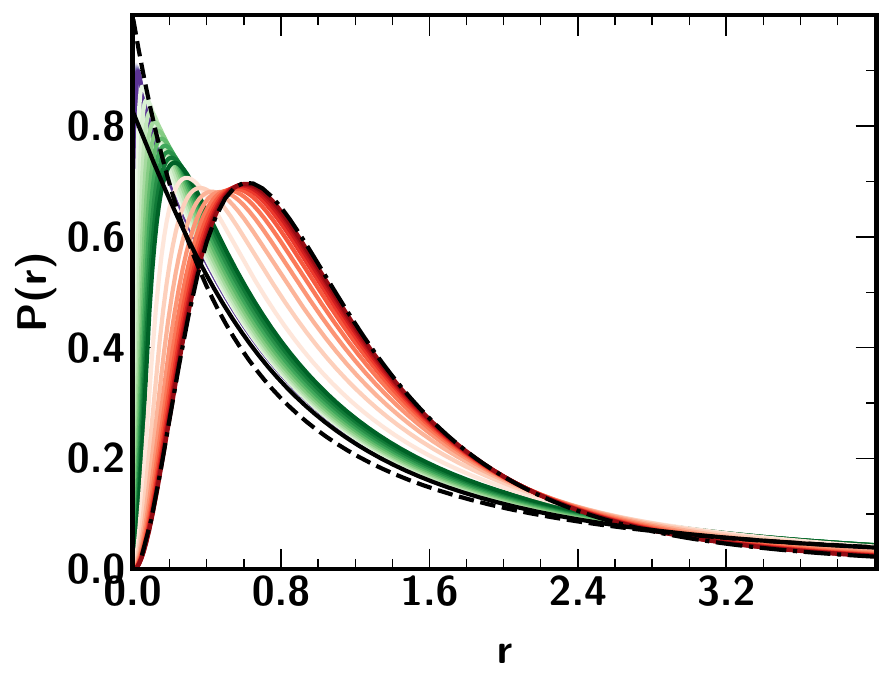}
	\caption{
	    Examples for the Wigner-surmise like result~\eqref{PRAnalyticGUE} for the ratio distribution of the eigenvalues of $\hat H^{0\to 2}$. For light to dark purple $\Lambda=0.0055,0.006,\dots0.01$, for light to dark green $\Lambda=0.02,0.03,\dots 0.1$, for light to dark red $\Lambda=0.15,0.2,\dots 0.6$. The corresponding values of $\alpha$ are obtained from the relation $\alpha^2=\frac{\Lambda^2}{1-\Lambda^2}$. We confirmed only analytically that with decreasing $\Lambda$ the limiting result~\eqref{al0} is approached (see main text). With increasing $\Lambda$ a transition from the result~\eqref{al0} to~\eqref{PRGUE} takes place. The analytical results~\eqref{al0},~\eqref{PRPoi} and~\eqref{PRAnalyticGUE} are shown as black solid, dashed and dash-dotted lines, respectively.}
    \label{RatioAnal}
\end{figure}

Performing integration by parts in the first integral of Eqs.~(\ref{I2}) and~(\ref{I3}) yields
\ba
    &&\frac{\sqrt{\pi}}{2}\int_{-\pi}^{\pi}\frac{d\varphi}{\sin\varphi}\frac{1}{F_i(\varphi)}\int_0^\infty\frac{d\rho}{\rho^2}e^{-A_i(\varphi)\rho^2}\Phi(F_i(\varphi)\rho)\\
    =&&-\sqrt{\pi}\int_{-\pi}^{\pi}\frac{d\varphi}{\sin\varphi}\frac{1}{F_i(\varphi)}A_i(\varphi)\int_0^\infty d\rho e^{-A_i(\varphi)\rho^2}\Phi(F_i(\varphi)\rho)\\
    +&&\int_{-\pi}^{\pi}\frac{d\varphi}{\sin\varphi}\int_0^\infty\frac{d\rho}{\rho}\exp\left[-\frac{2}{3}(\alpha^2+1)\rho^2\left(1-\frac{\sin\varphi}{2}\right)\right]\\
    =&&-\frac{3\pi}{8}\int_{-\pi}^{\pi}\frac{d\varphi}{\sin\varphi}X_i\left[1-\frac{2}{\pi}\arctan(X_i)\right]+\frac{\pi^2}{6},
\ea
where we introduced the notation
\be
\label{Xi}
    X_i=\frac{\sqrt{A_i(\varphi)}}{F_i(\varphi)}\, .
\ee
In the remaining integrals, integration over $\rho$ can be performed leaving the integrals over $\varphi$.
The final result reads
\ba
    \label{PRAnalyticGUE}
    P(r)=&&\frac{r(r+1)}{R^3}\frac{1}{2\pi}\left\{\sqrt{3}\frac{\alpha^2}{\alpha^2+1}\frac{(3R-2)}{2R}\right.\\
    +&&(2+r)\alpha^2\int_{-\pi}^{\pi}\frac{d\varphi}{\sin\varphi}\left(-X_1+\frac{2}{X_1}+\frac{1}{3X_1^3}\right)\left[1-\frac{2}{\pi}\arctan(X_1)\right]\nonumber\\
    +&&(1+2r)\alpha^2\int_{-\pi}^{\pi}\frac{d\varphi}{\sin\varphi}\left(-X_2+\frac{2}{X_2}+\frac{1}{3X_2^3}\right)\left[1-\frac{2}{\pi}\arctan(X_2)\right]\nonumber\\
    +&&\frac{2}{\pi}\alpha^2\left.\int_{-\pi}^{\pi}\frac{d\varphi}{\sin\varphi}\left[\frac{2+r}{3}\frac{1}{X_1^2(1+X_1^2)}+\frac{1+2r}{3}\frac{1}{X_2^2(1+X_2^2)}\right]\right\}\nonumber
\ea
The last integral can be further evaluated,
\begin{align}
    &\frac{2}{\pi}\alpha^2\int_{-\pi}^{\pi}\frac{d\varphi}{\sin\varphi}\left[\frac{2+r}{3}\frac{1}{X_1^2(1+X_1^2)}+\frac{1+2r}{3}\frac{1}{X_2^2(1+X_2^2)}\right]\\
    =&\frac{2}{\pi}\alpha^2\int_{-\pi}^{\pi}\frac{d\varphi}{\sin\varphi}\left[\frac{2+r}{3}\frac{1}{X_1^2}+\frac{1+2r}{3}\frac{1}{X_2^2}\right]-\sqrt{3}\frac{\alpha^2}{\alpha^2+1}\frac{(3R-2)}{2R}\, .
\end{align}
In the limit $\alpha\to\infty$ we have $\alpha^2\frac{1}{X_i^2(1+X_i^2)}\simeq\frac{\alpha^2}{X_i^4}\xrightarrow{\alpha\to\infty} 0$.
Using, that for $x > 1$ $\arctan(x)=\frac{\pi}{2}-{\rm arccot}\left(\frac{1}{x}\right)$, we obtain for the remaining integrals
\begin{align}
    &(2+r)\alpha^2\int_{-\pi}^{\pi}\frac{d\varphi}{\sin\varphi}\left(-X_1+\frac{2}{X_1}+\frac{1}{3X_1^3}\right)\left[1-\frac{2}{\pi}\arctan(X_1)\right]\\
    +&(1+2r)\alpha^2\int_{-\pi}^{\pi}\frac{d\varphi}{\sin\varphi}\left(-X_2+\frac{2}{X_2}+\frac{1}{3X_2^3}\right)\left[1-\frac{2}{\pi}\arctan(X_2)\right]\\
    &\simeq (2+r)\alpha^2\frac{7}{3}\int_{-\pi}^{\pi}\frac{d\varphi}{\sin\varphi}\frac{1}{X_1^2}
    +(1+2r)\alpha^2\frac{7}{3}\int_{-\pi}^{\pi}\frac{d\varphi}{\sin\varphi}\frac{1}{X_2^2}\\
    &\simeq21\int_{-\pi}^{\pi}\frac{d\varphi}{\sin\varphi}\frac{f(\varphi)}{2-\sin\varphi}=7\sqrt{3}\frac{\alpha^2}{\alpha^2+1}\frac{(3R-2)}{2R}
\end{align}
yielding
\be
    P(r)\xrightarrow{\alpha\to\infty}\frac{r(r+1)}{R^3}\frac{1}{2\pi}8\sqrt{3}\frac{(3R-2)}{2R}=\frac{81\sqrt{3}}{4\pi}\frac{\left[r(1+r)\right]^2}{(1+r+r^2)^4}\, ,
    \label{PRGUE}
\ee
which is the Wigner-surmise like result for the ratio distribution of the GUE.
In~\reffig{RatioAnal} we show the analytical result for the ratio distribution~\eqref{PRAnalyticGUE} for varying $\Lambda$. With increasing $\Lambda$ a transition from the result~\eqref{al0} for the eigenvalues of a $3\times 3$-dimensional diagonal matrix with Gaussian distributed entries for $N=3$ to~\eqref{PRGUE} for the surmise-like ratio distribution of the GUE takes place.

\subsection{Analytical results for long-range correlation functions for the transition from Poisson to GUE}
\label{Analytic}

In Ref.~\onlinecite{Frahm1998} an exact analytical expression was derived for $Y_2^{0\to 2}(r)$ based on the graded eigenvalue method,
%\ba/
%Y_2^{0\to 2}(r)&&=\frac{1}{2(\pi r)^2}\left[1-e^{-2\frac{r^2}{\tilde\alpha^2}}\cos(2\pi r)\right]-\frac{1}{(\pi\tilde\alpha)^2}
%\label{Y2Anal}\\
%&&+\frac{1}{\pi}\int_0^\infty \rho d\rho e^{-\frac{\rho^2}{2c}}\int_0^\pi d\phi\cos(\phi)\left[\Re (A)+\Re (B)\right]\nonumber\\
%A&&=\frac{e^{i\phi}\left[1-\frac{\rho}{\kappa}\sin\phi\right]}{1+i\frac{\rho e^{i\phi}}{2\kappa}}\exp\left[-i\frac{\rho ^2}{2c\kappa}\frac{1}{1-\frac{\rho}{\kappa}\sin\phi}\right]\nonumber\\
%B&&=\frac{e^{-i\phi}\left[1+\frac{\rho}{\kappa}\sin\phi\right]}{1+i\frac{\rho e^{-i\phi}}{2\kappa}}\exp\left[-i\frac{\rho^2}{2c\kappa}\frac{1}{1+\frac{\rho}{\kappa}\sin\phi}\right]\nonumber ,\\
%\kappa &&=\frac{r}{\pi\tilde\alpha ^2},\, c=\frac{1}{(\pi\tilde\alpha)^2}.\nonumber
%\ea
\ba
    Y_2^{0\to 2}(r)&&=\frac{1}{2(\pi r)^2}\left[1-e^{-2\frac{r^2}{\tilde\alpha^2}}\cos(2\pi r)\right]-\frac{1}{(\pi\tilde\alpha)^2}
    \label{Y2Anal}
    +\frac{1}{\pi}\int_0^\infty \rho d\rho e^{-\frac{\rho^2}{2c}}\int_0^\pi d\phi\cos(\phi)\left[\Re (A)+\Re (B)\right]\\
    A&&=\frac{e^{i\phi}\left[1-\frac{\rho}{\kappa}\sin\phi\right]}{1+i\frac{\rho e^{i\phi}}{2\kappa}}\exp\left[-i\frac{\rho ^2}{2c\kappa}\frac{1}{1-\frac{\rho}{\kappa}\sin\phi}\right]\nonumber,
    B=\frac{e^{-i\phi}\left[1+\frac{\rho}{\kappa}\sin\phi\right]}{1+i\frac{\rho e^{-i\phi}}{2\kappa}}\exp\left[-i\frac{\rho^2}{2c\kappa}\frac{1}{1+\frac{\rho}{\kappa}\sin\phi}\right]\nonumber ,\\
    \kappa &&=\frac{r}{\pi\tilde\alpha ^2},\, c=\frac{1}{(\pi\tilde\alpha)^2}.\nonumber
\ea
The number variance is deduced from~\eqref{Y2Anal} via the relation
\be
    \label{Sigma2Anal}
    \Sigma_{0\to 2}^2(L)=L-2\int_0^L(L-r)Y_2^{0\to 2}(r)dr\, . 
\ee 

\begin{figure}
    \centering
    \includegraphics[width=0.4\columnwidth]{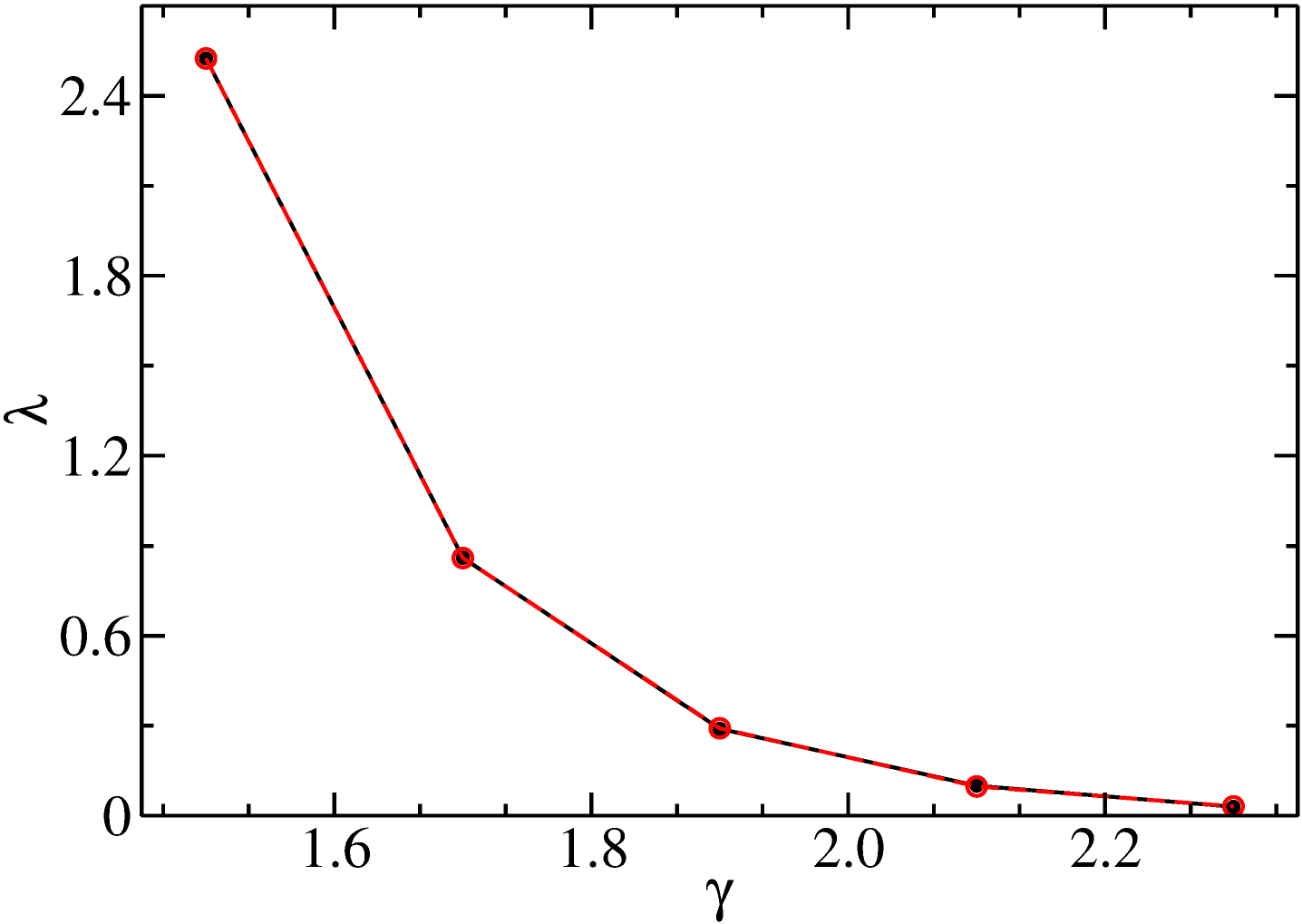}
    \includegraphics[width=0.4\columnwidth]{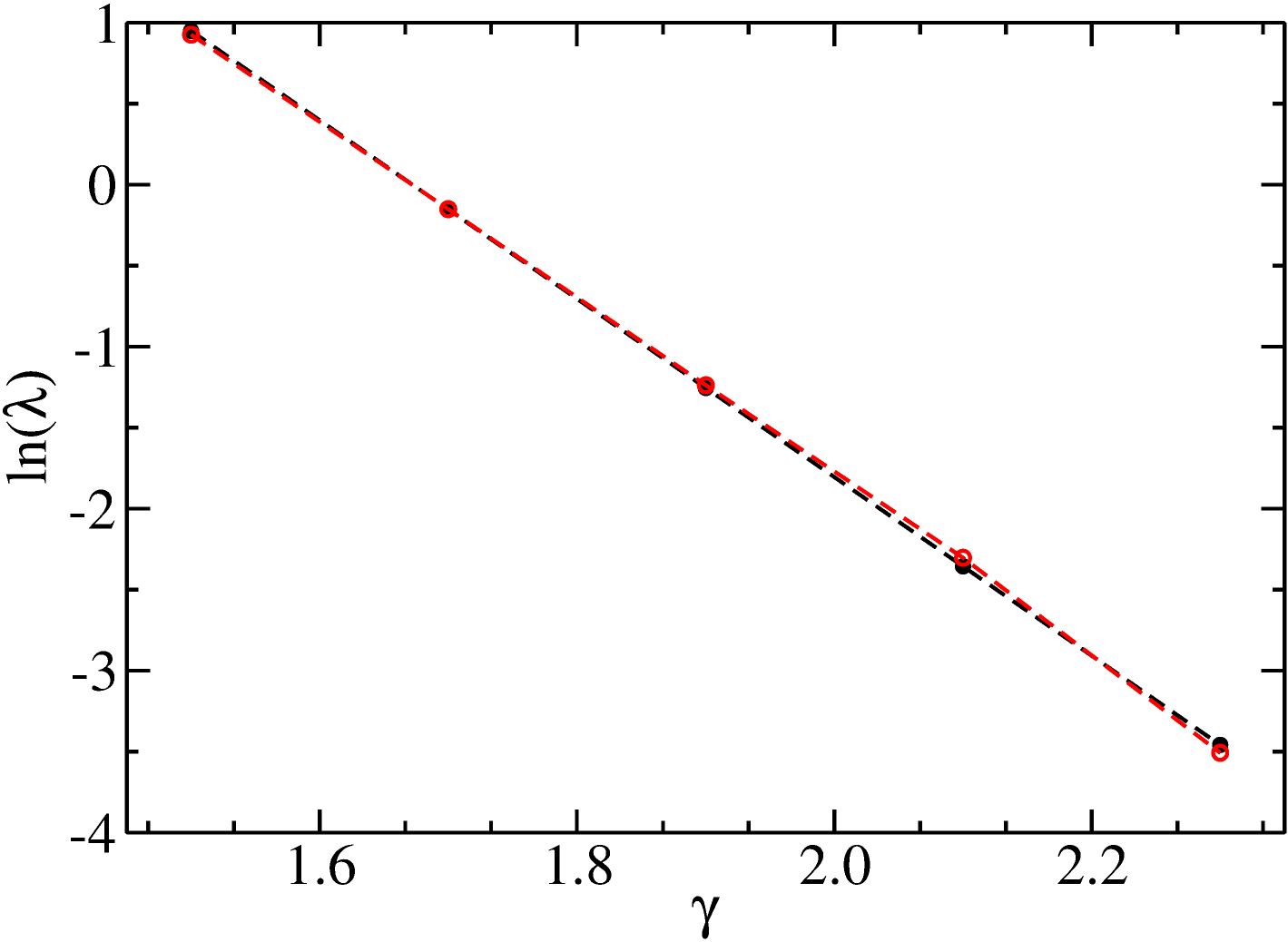}
	\caption{
	    Left: Values of $\lambda$ (black) obtained from the fit of the analytical result for $\Sigma^2(L)$ to the numerical results as function of $\gamma$. A fit of $AN^{-B\gamma}$ to $\lambda(\gamma)$ shown in red yields $A=8034.46$ and $B=0.49$.
	Right: A linear fit (red) $\ln(\lambda)\approx a-b\cdot\gamma\cdot {\rm ln}N$ to $\ln[\lambda(\gamma)]$ (black) yields $a=9.2122$ and $b=0.5$.
    }
    \label{fig:LamGamGUE}
\end{figure}

In Ref.~\onlinecite{Kunz1998} an exact analytical result was obtained for the form factor,
%\ba
%&&K^{0\to 2}(\tilde\tau)=1+\frac{2}{\gamma}I_1(\gamma)\exp{\left[-\pi\tilde\alpha^2\tilde\tau-\frac{\tilde\alpha^2\tilde\tau^2}{2}\right]}
%\label{B2Anal}\\
%&&-\frac{\tilde\tau}{2\pi}\gamma\int_1^\infty dt(t^2-1)I_1(\gamma t)\exp\left[-t^2\frac{\tilde\alpha^2\tilde\tau^2}{2}-\pi\tilde\alpha^2\tilde\tau\right],
%\nonumber\\
%&&\gamma =\sqrt{2\pi}\tilde\alpha^2\tilde\tau^{3/2},
%\ea
\ba
    &&K^{0\to 2}(\tilde\tau)=1+\frac{2}{\xi}I_1(\xi)\exp{\left[-\pi\tilde\alpha^2\tilde\tau-\frac{\tilde\alpha^2\tilde\tau^2}{2}\right]}
    \label{B2Anal}
    -\frac{\tilde\tau}{2\pi}\xi\int_1^\infty dt(t^2-1)I_1(\xi t)\exp\left[-t^2\frac{\tilde\alpha^2\tilde\tau^2}{2}-\pi\tilde\alpha^2\tilde\tau\right],
    \nonumber\\
    &&\xi =\sqrt{2\pi}\tilde\alpha^2\tilde\tau^{3/2},
\ea
which was rederived in Ref.~\onlinecite{kravtsov2015random}.
For the evaluation of the integral for values of $\tau\gtrsim \tau_{min}$, with $\tau_{min}$ denoting the value of $\tau$ at the minimum of $K(\tau)$, we performed a transformation of the integration variable $t$ to $t=\sqrt{1+x}$ as in Ref.~\onlinecite{kravtsov2015random}.

Note, that there are discrepancies in the scales of $\tilde\alpha$ and $\tilde\tau$ between Refs.~\cite{Kunz1998} and~\cite{Frahm1998}. These are due to differing definitions of the $N$-dependent scale $\Gamma_N$ in~\eqref{RPH}. We fixed this by computing the spectral form factor from the Fourier transform of the analytical result for the two-point cluster function~\cite{Frahm1998} given in~\eqref{Y2Anal} (right panel of~\reffig{fig:K}) and comparing it to the analytical result for $K(\tau)$~\cite{Kunz1998} given in~\eqref{B2Anal} (left panel of~\reffig{fig:K}). Furthermore, we compared the resulting values of $\tilde\alpha$ and $\tilde\tau$ with those obtained from the fits of the  Wigner-surmise like analytical result, $P_{0\to 2}(s)$, to the nearest-neighbor spacing distributions obtained for the gRP model, shown in~\reffig{fig:LamGam}, yielding $\tilde\alpha=\frac{\pi}{\sqrt{2}}\lambda$ and $\tilde\tau=\frac{\tau}{2\pi}$.

We performed random-matrix simulations for values of $\gamma$ varying from $0.9\leq\gamma\leq 2.5$ in the RP model~\eqref{RPH} and determined the corresponding values of $\lambda$ by fitting the analytical expression~\eqref{Sigma2Anal} deduced from~\eqref{Y2Anal} to the numerical ones. The resulting values are shown in~\reffig{fig:LamGamGUE}. They agree well with those shown in~\reffig{fig:LamGam} obtained from a fit of the distribution $P_{0\to 2}(s)$ given in~\eqref{PSGUE} to the numerical results for the random-matrix obtained for the gRP model with $\beta=2$. A fit to $AN^{B\gamma}$ yields $B\simeq 0.5$ as expected from the definitions of the generalized and original RP model, Eqs.~\eqref{eq:HgRP} with~\eqref{eq:gRP_variances} and~\eqref{RPH}.

\section{Additional numerical results}
\begin{figure}
    \centering
    \includegraphics[width=0.49\columnwidth]{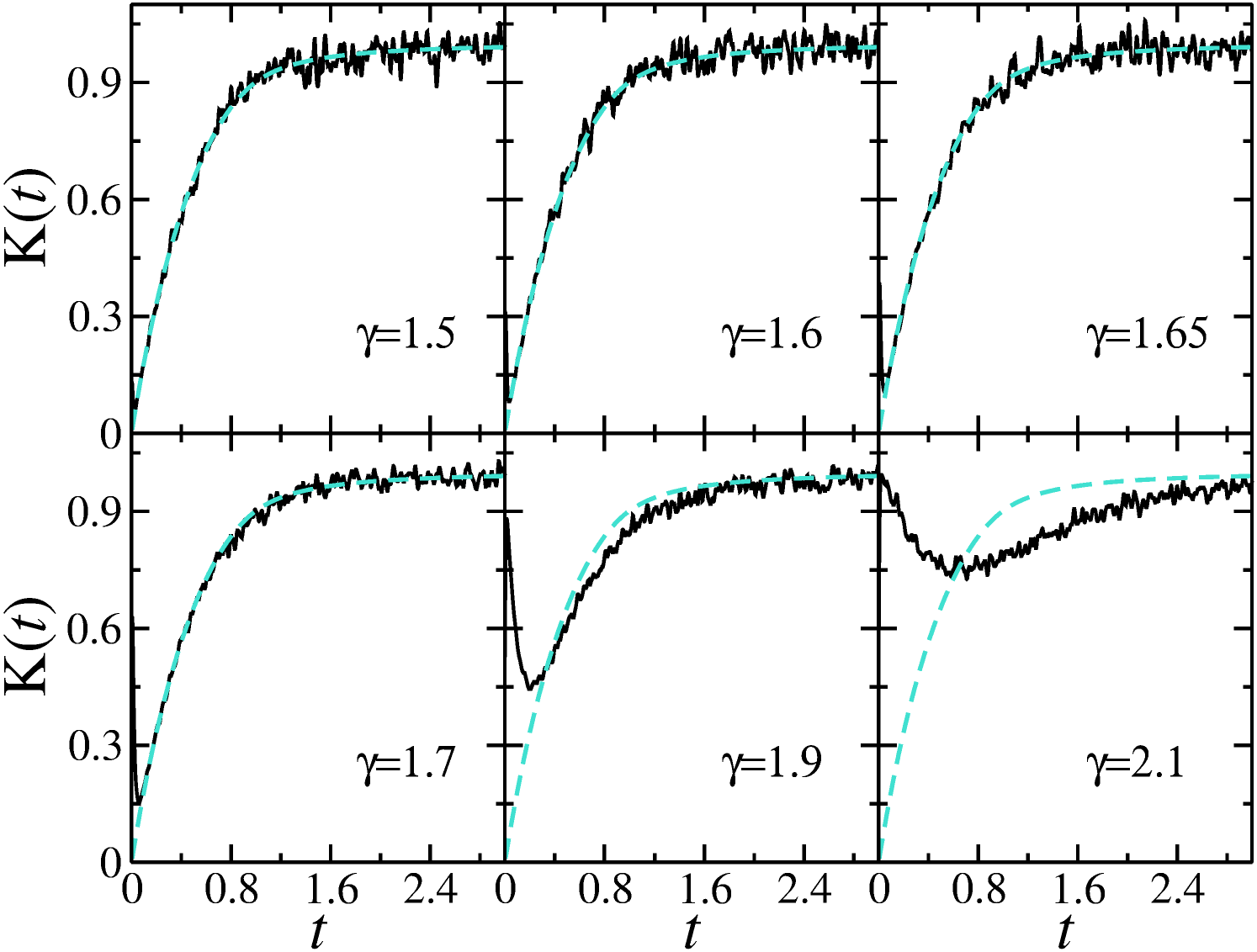}
    \includegraphics[width=0.49\columnwidth]{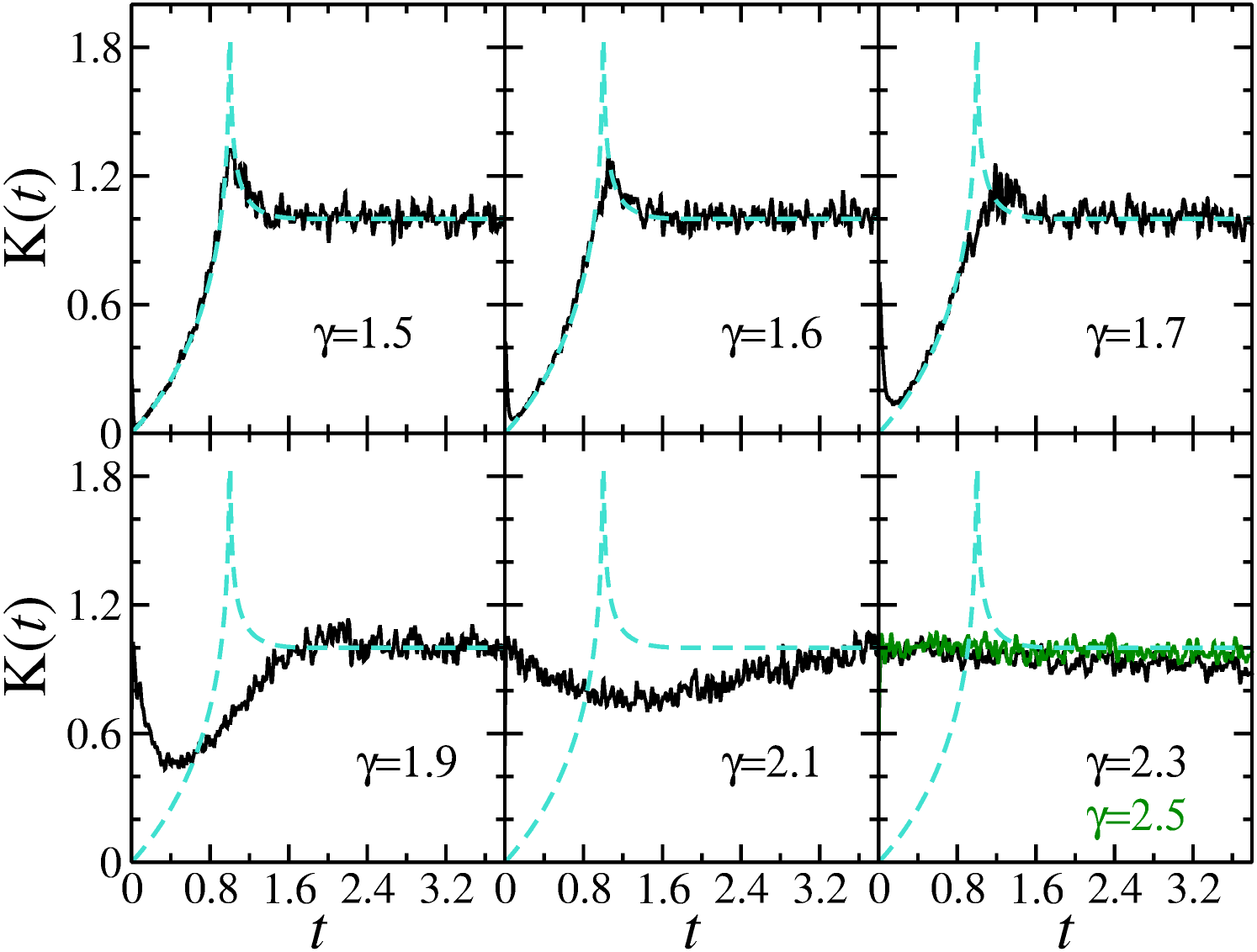}
    \caption{
	    Left: Form factor obtained from the random-matrix simulations for the gRP Hamiltonian~\eqref{eq:HgRP} (black) for the transition from Poisson to GOE for various values of $\gamma$.
        The turquoise line show the analytical curve for the GOE.
	    Right:  Form factor obtained from the random-matrix simulations for the gRP Hamiltonian~\eqref{eq:HgRP} (black) for the transition from Poisson to GSE for various values of $\gamma$.
        The turquoise line shows the analytical curve for the GSE.
    }
	\label{fig:K_GOE_GSE}
 \end{figure}

\begin{figure}
    \centering
    \includegraphics[width=0.6\columnwidth]{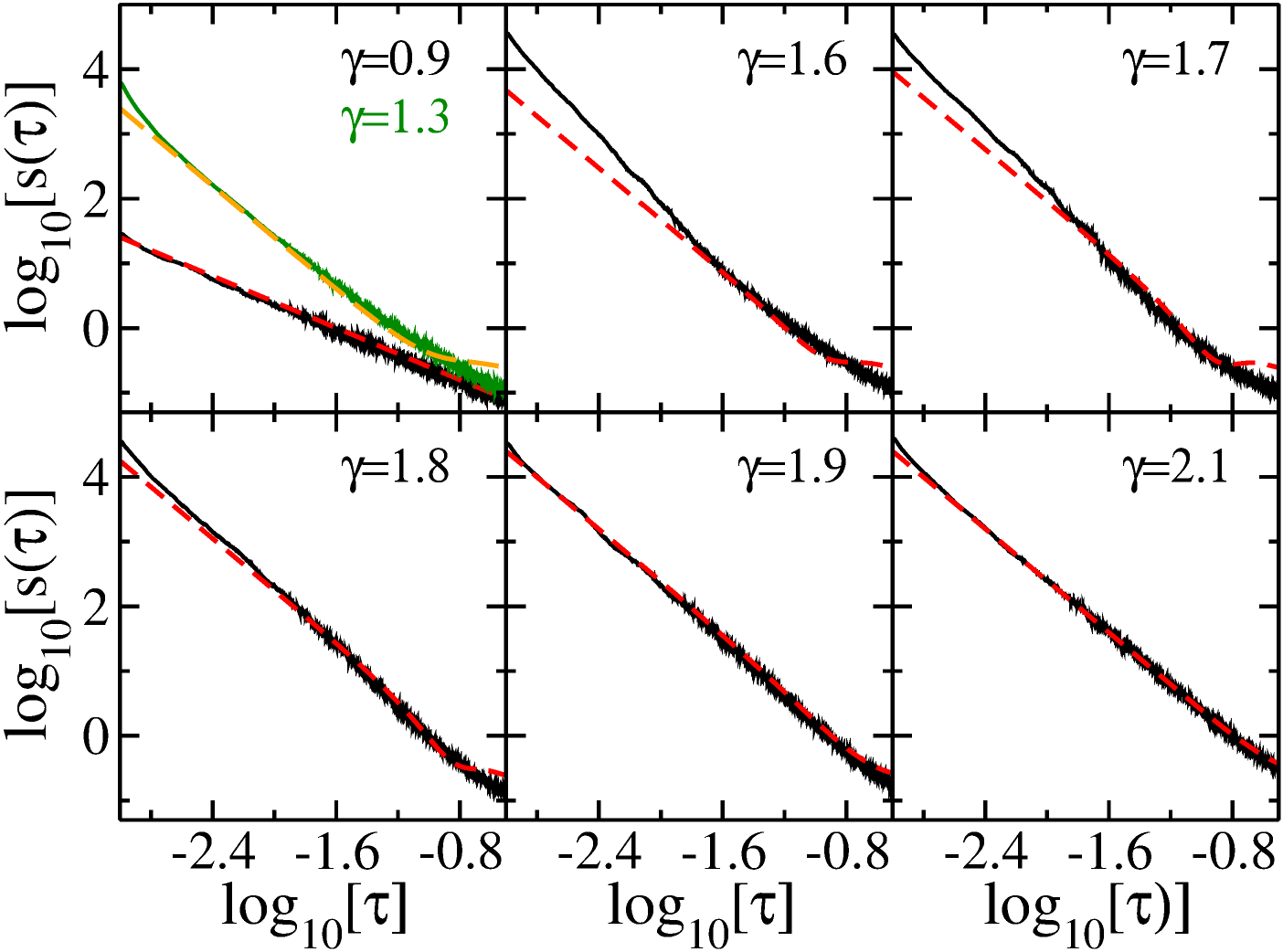}
	\caption{
	    Comparison of the power spectrum for $\beta =2$ (black and green lines) obtained from random-matrix simulatons for the Hamiltonian~\eqref{eq:HgRP} for various values of $\gamma$ with an analytical approximation in terms of the spectral form factor (red and orange dashed lines). We find clear deviations in the range $1.3\lesssim\gamma\lesssim 1.8$ where the exponent $\mu$ shown in $\reffig{fig:PowerSp}$ exhibits a drastic change.   
    }
    \label{fig:Power_GUE_Approx}
\end{figure}

The asymptotic behavior of the power spectrum for $\tau\ll 1$ is compared with approximate analytical results in terms of the spectral form factor in~\reffig{fig:Power_GUE_Approx}. In the intermediate region $1.3\lesssim\gamma\lesssim 1.8$ the approximation does not apply. In~\reffig{fig:K_GOE_GSE} the spectral form factor is depicted for six values of $\gamma$ for the GOE and the GSE. The turquoise dashed lines show the analytical result for the corresponding WD ensemble.

\end{appendix}	

\end{document}